\documentclass[showpacs,showkeys,amssymb,onecolumn]{revtex4}
\usepackage{graphicx}
\usepackage{dcolumn}
\usepackage{multirow}
\usepackage{bm}
\begin{document}

\title{Chiral Structure of Scalar and Pseudoscalar Mesons}

\author{Hua-Xing Chen}
\email{hxchen@buaa.edu.cn}
\affiliation{School of Physics and Nuclear Energy Engineering and International Research Center for Nuclei and Particles in the Cosmos, Beihang University, Beijing 100191, China}

\begin{abstract}
We systematically study the chiral structure of local tetraquark currents of flavor singlet and $J^P=0^+$. We also investigate their chiral partners, including scalar and pseudoscalar tetraquark currents of flavor singlet, octet, $\mathbf{10}$, $\mathbf{\overline {10}}$ and $\mathbf{27}$. We study their chiral transformation properties. Particularly, we use the tetraquark currents belonging to the ``non-exotic'' $[(\mathbf{\bar 3},\mathbf{3}) \oplus (\mathbf{3},\mathbf{\bar 3})]$ chiral multiplets to calculate the masses of light scalar mesons through QCD sum rule. The two-point correlation functions are calculated including all terms and only the connected parts~\cite{Weinberg:2013cfa,coleman,page}. The results are consistent with the experimental values.
\end{abstract}

\maketitle

\section{Introduction}

The quark model is very successful in explaining the hadron spectrum with simply using quark-antiquark mesons and three-quark baryons~\cite{GellMann:1964nj,GellMann:1962xb,Slansky:1981yr,Chirstman2011,de Swart:1963gc}. However, there are always multi-quark components in the Fock-space expansion of hadron states~\cite{Prelovsek:2005du,Fariborz:2009cq,Yndurain:2007qm}. Hence, it is useful to properly include these multi-quark components if we want to use Quantum Chromodynamics (QCD), the theory of strong interactions, to investigate hadrons in an exact way. Besides these ``exotic'' components, multi-quark states themselves are also important in order to understand the low-energy behavior of QCD. These subjects have been studied for more than thirties years by lots of theoretical and experimental physicists~\cite{Beringer:1900zz,Jaffe:2004ph,Weinstein:1982gc,Weinstein:1990gu,Brodsky:1977bs,Aerts:1979hn,Lipkin:1986dw,Choe:1997wz,Alexandrou:2001ip,Close:2002zu,Oka:2004xh,Ding:2006ya,Zhu:2007wz,Vijande:2007ix,Xie:2011uw,MartinezTorres:2008gy,MartinezTorres:2011vh,Xiao:2011rc,Lee:2009rt,Lipkin:2003zk,Ping:2009zzb,Sun:2012sy,Jaffe:1976ig,Jaffe:1976ih,Friedmann:2009mx,Dai:2009zz,Zhang:2007zzg,Giacosa:2006tf,Fariborz:2005gm,Napsuciale:2004au}. Particulary, the light scalar mesons are good candidates due to their tetraquark (or molecular) components.

The light scalar mesons $f_0(500)$ (or $\sigma(500)$), $\kappa(800)$, $a_0(980)$ and $f_0(980)$ compose a flavor nonet whose masses are all below 1 GeV~\cite{Beringer:1900zz}. Although such mesons have been intensively studied for many years, their nature is still not fully understood~\cite{Aitala:2000xu,Ablikim:2004qna}. In the conventional quark model, light scalar mesons have a $\bar q q$ configuration of $^3P_0$. However, because of their internal $p$-wave orbital excitation, their masses should exceed 1 GeV and the ordering should be $m_\sigma \sim m_{a_0} < m_\kappa < m_{f_0}$~\cite{Hatsuda:1994pi}, which is inconsistent with the experiments. In chiral models, light scalar mesons are very important because they are chiral partners of the Nambu-Goldstone bosons, $\pi$, $K$, $\eta$ and $\eta^\prime$~\cite{Hatsuda:1994pi}. Their masses are expected to be less than those of the quark model because of their collective nature. Light scalar mesons are also considered as tetraquark states or molecular states or containing large tetraquark components~\cite{Jaffe:1976ig,Jaffe:1976ih,Friedmann:2009mx,Zhang:2007zzg,Dai:2009zz,Chen:2007xr,Brito:2004tv}. Considering the diquark (antidiquark) inside has strong attraction, their masses are expected to be less than 1 GeV and the ordering is expected to be $m_\sigma <  m_\kappa < m_{f_0, a_0}$, which is consistent with the experiments.

To study the multi-quark components of the light scalar mesons, we can use group theoretical methods, which have been applied to study quark-antiquark mesons and three-quark baryons~\cite{Weinberg:1969hw,Weinberg:1990xn,Leinweber:1994nm,Leinweber:1995ie,Ioffe:1981kw,Ioffe:1982ce,Chung:1981cc,Espriu:1983hu,Cohen:1996sb,Chen:2008qv,Nagata:2007di,Cohen:2002st,Jido:2001nt,Jido:1999hd,Chen:2012ut}. T.~D.~Cohen and X.~D.~Ji studied the chiral structure of meson currents constructed using one quark and one antiquark fields and baryon currents constructed using three quark fields~\cite{Cohen:1996sb}. In this paper we shall follow their approaches and study the chiral structure of local scalar and pseudoscalar tetraquark currents. These tetraquark currents can be used in the QCD sum rule analyses~\cite{Brito:2004tv,Narison:2002pw,Jiao:2009ra,Du:2012pn} as well as the Lattice QCD calculations~\cite{Prelovsek:2005du,Cardoso:2011fq,Okiharu:2004ve,Prelovsek:2010kg,Loan:2008sd,Prelovsek:2008rf,Mathur:2006bs,McNeile:2006nv}.

In our previous references, we have applied the method of the QCD sum rule to calculate masses of light scalar mesons using local tetraquark currents~\cite{Chen:2006hy,Chen:2007xr,Chen:2008ej,Chen:2008qw}. We systematically classified the scalar tetraquark currents and found that there are altogether as many as five independent scalar local tetraquark currents for each flavor structure. Therefore, right currents should be used in order to study light scalar mesons. This is also closely related to the internal structure of light scalar mesons. A similar question for the baryon case has been studied in Refs.~\cite{Ioffe:1981kw,Ioffe:1982ce,Chung:1981cc,Espriu:1983hu} where there are three independent local baryon fields of flavor octet. Previously we chose some mixed currents which provided good QCD sum rule results~\cite{Chen:2007xr}. Although we did not know the relation of these currents with the internal structure of light scalar mesons at that time, we found that studying the chiral structure of scalar tetraquark currents can be useful to answer this question.

In this paper we shall try to answer this question (which currents should be used in order to study light scalar mesons). We shall systematically study the chiral structure of light scalar mesons through local tetraquark currents which belong to the ``non-exotic'' $[(\mathbf{\bar 3},\mathbf{ 3}) \oplus (\mathbf{3},\mathbf{\bar 3})]$ chiral multiplets. This chiral multiplet only contains flavor singlet and octet mesons, and it does not contain any meson having exotic flavor structure. Since there are no experimental signals observing scalar mesons having exotic flavors, we assume that all the nine light scalar mesons (or their dominant components) belong to this multiplet. Moreover, these nine light scalar mesons can together compose one $[(\mathbf{\bar 3},\mathbf{ 3}) \oplus (\mathbf{3},\mathbf{\bar 3})]$ chiral multiplet. To do a systematical study, we shall investigate both scalar and pseudoscalar tetraquark currents, since they are chiral partners. We shall also investigate tetraquark currents of flavor singlet, octet, $\mathbf{10}$, $\mathbf{\overline {10}}$ and $\mathbf{27}$, which can be useful for further studies. We shall use the left handed quark field $L_A^a \equiv q_{LA}^a = {1 - \gamma_5 \over 2} q_A^a$ and the right handed quark field $R_A^a \equiv q_{RA}^a = {1 + \gamma_5 \over 2} q_A^a$ to rewrite these currents. After making proper combinations we can clearly see their chiral structures.

In this paper we shall use the method of QCD sum rule to calculate the masses of light scalar mesons through local scalar tetraquark currents belonging to the ``non-exotic'' $[(\mathbf{\bar 3},\mathbf{ 3}) \oplus (\mathbf{3},\mathbf{\bar 3})]$ chiral multiplets. One tetraquark current can be always written as a combination of meson-meson currents through Fierz transformation (${\cal Q}(x)=\sum_{ij}C_{ij}{\cal B}_i(x){\cal B}_j(x)$), and so the two-point correlation function contain two parts: the disconnected parts
\begin{eqnarray}
\left\langle {\cal Q}(x){\cal Q}(y)\right\rangle_{\rm disconn} &=&
\sum_{ijkl}C_{ij}C_{kl}\left\langle {\cal B}_i(x){\cal B}_k(y)\right\rangle_0\left\langle {\cal B}_j(x){\cal B}_l(y)\right\rangle_0 \, ,
\end{eqnarray}
and the connected parts
\begin{eqnarray}
\left\langle {\cal Q}(x){\cal Q}(y)\right\rangle_{\rm conn} &=&
\sum_{ijkl}C_{ij}C_{kl} \left\langle {\cal B}_i(x){\cal B}_j(x) {\cal B}_k(y){\cal B}_l(y)\right\rangle_{0,{\rm conn}} \, .
\end{eqnarray}
In this paper we shall use both of them to perform QCD sum rule analysis. However, as suggested by S.~Weinberg in his recent reference~\cite{Weinberg:2013cfa,coleman,page} using the large $N_c$ approximation: ``A one tetraquark pole can only appear in the final, connected, term'', we shall also use (only) the connected parts to perform QCD sum rule analysis.

This paper is organized as follows. In Sec.~\ref{sec:currents} we investigate local tetraquark currents of flavor singlet and $J^P=0^+$, and others are listed in Appendix~\ref{app:othercurrents}. In Sec.~\ref{sec:chiral} we study their chiral transformation properties, and results are partly listed in Appendix~\ref{app:othertransform}. In Sec.~\ref{sec:sumrule} we use the method of QCD sum rule to study the light scalar mesons through local scalar tetraquark currents belonging to the ``non-exotic'' $[(\mathbf{\bar 3},\mathbf{ 3}) \oplus (\mathbf{3},\mathbf{\bar 3})]$ chiral multiplets. However, the results depend much on the threshold value $s_0$ suggesting a large contribution from the meson-meson continuum, and so in Sec.~\ref{sec:connected} we use only the connected parts of the two-point correlation function to perform the QCD sum rule analyses. Sec.~\ref{sec:summary} is a summary.

\section{Scalar Tetraquark Currents of Flavor Singlet}
\label{sec:currents}

We write the flavor structure of tetraquarks, and study local tetraquark currents of flavor singlet and $J^P=0^+$:
\begin{eqnarray}\label{eq:flavor}
\mathbf{3}\otimes\mathbf{3} \otimes \mathbf{\bar {3}}\otimes\mathbf{\bar{3}}
&=& \big( \mathbf{\bar 3}\oplus\mathbf{6} \big) \otimes \big( \mathbf{3}\oplus\mathbf{\bar6} \big)
\\ \nonumber &=& \big( \mathbf{\bar 3}\otimes\mathbf{3} \big) \oplus \big( \mathbf{\bar 3}\otimes\mathbf{\bar 6} \big)
\oplus \big( \mathbf{6}\otimes\mathbf{3} \big) \oplus \big( \mathbf{6}\otimes\mathbf{\bar 6} \big)
\\ \nonumber &=& \big( \mathbf{1}\oplus \mathbf{8}\big) \oplus \big( \mathbf{8}\oplus\mathbf{\overline{10}} \big)
\oplus \big( \mathbf{8}\oplus \mathbf{10}\big) \oplus \big( \mathbf{1}\oplus \mathbf{8}\oplus\mathbf{27} \big) \, .
\end{eqnarray}
There are two possibilities to construct a flavor single tetraquark current: both of the diquark and antidiquark have the antisymmetric flavor structure
$\mathbf{\bar 3}_F(qq) \otimes \mathbf{3}_F(\bar q \bar q) \rightarrow \mathbf{1}_F$, or have the symmetric flavor structure $\mathbf{6}_F(qq) \otimes \mathbf{\bar6}_F(\bar q \bar q) \rightarrow \mathbf{1}_F$. Together with five sets of Dirac matrices, $1$, $\gamma_5$, $\gamma_\mu$, $\gamma_\mu \gamma_5$ and $\sigma_{\mu \nu}$, we find the following ten independent local tetraquark currents of flavor singlet and $J^P = 0^+$:
\begin{eqnarray}
\label{eq:current11}
\nonumber \eta_1^{\rm S,\mathbb{S}} &=& q_A^{aT} \mathbb{C} \gamma_5 q_B^b ( \bar{q}_A^a \gamma_5 \mathbb{C} \bar{q}_B^{bT} - \bar{q}_A^b \gamma_5 \mathbb{C} \bar{q}_B^{aT} ) \, ,
\\ \nonumber \eta_2^{\rm S,\mathbb{S}} &=& q_A^{aT} \mathbb{C} q_B^b (\bar{q}_A^a \mathbb{C} \bar{q}_B^{bT}-\bar{q}_A^b \mathbb{C} \bar{q}_B^{aT}) \, ,
\\ \nonumber \eta_3^{\rm S,\mathbb{S}} &=& q_A^{aT} \mathbb{C} \sigma_{\mu\nu} q_B^b ( \bar{q}_A^a \sigma^{\mu\nu} \mathbb{C} \bar{q}_B^{bT} + \bar{q}_A^b \sigma^{\mu\nu} \mathbb{C} \bar{q}_B^{aT} ) \, ,
\\ \nonumber \eta_4^{\rm S,\mathbb{S}} &=& q_A^{aT} \mathbb{C} \gamma_5 q_B^b ( \bar{q}_A^a \gamma_5 \mathbb{C} \bar{q}_B^{bT} + \bar{q}_A^b \gamma_5 \mathbb{C} \bar{q}_B^{aT} ) \, ,
\\ \eta_5^{\rm S,\mathbb{S}} &=& q_A^{aT} \mathbb{C} q_B^b (\bar{q}_A^a \mathbb{C} \bar{q}_B^{bT} + \bar{q}_A^b \mathbb{C} \bar{q}_B^{aT}) \, ,
\\ \nonumber \eta_6^{\rm S,\mathbb{S}} &=& q_A^{aT} \mathbb{C} \sigma_{\mu\nu} q_B^b ( \bar{q}_A^a \sigma^{\mu\nu} \mathbb{C} \bar{q}_B^{bT} - \bar{q}_A^b \sigma^{\mu\nu} \mathbb{C} \bar{q}_B^{aT} ) \, ,
\\ \nonumber \eta_7^{\rm S,\mathbb{S}} &=& q_A^{aT} \mathbb{C} \gamma_\mu \gamma_5 q_B^b ( \bar{q}_A^a \gamma^\mu \gamma_5 \mathbb{C} \bar{q}_B^{bT} - \bar{q}_A^b\gamma^\mu \gamma_5 \mathbb{C} \bar{q}_B^{aT}) \, ,
\\ \nonumber \eta_8^{\rm S,\mathbb{S}} &=& q_A^{aT} \mathbb{C} \gamma_\mu q_B^b (\bar{q}_A^a \gamma^\mu \mathbb{C} \bar{q}_B^{bT}+\bar{q}_A^b \gamma^\mu \mathbb{C} \bar{q}_B^{aT}) \, ,
\\ \nonumber \eta_9^{\rm S,\mathbb{S}} &=& q_A^{aT} \mathbb{C} \gamma_\mu \gamma_5 q_B^b ( \bar{q}_A^a \gamma^\mu \gamma_5 \mathbb{C} \bar{q}_B^{bT} + \bar{q}_A^b \gamma^\mu \gamma_5 \mathbb{C} \bar{q}_B^{aT}) \, ,
\\ \nonumber \eta_{10}^{\rm S,\mathbb{S}} &=& q_A^{aT} \mathbb{C} \gamma_\mu q_B^b (\bar{q}_A^a \gamma^\mu \mathbb{C} \bar{q}_B^{bT} - \bar{q}_A^b \gamma^\mu \mathbb{C} \bar{q}_B^{aT}) \, .
\end{eqnarray}
In these expressions the summation is taken over repeated indices ($a$, $b$, $\cdots$ for color indices, $A$, $B$, $\cdots$ for flavor indices, and $\mu$, $\nu$, $\cdots$ for Lorentz indices). The two superscripts S and $\mathbb{S}$ denote scalar ($J^P = 0^+$) and flavor singlet, respectively. In this paper we also need to use the following notations: $\mathbb{C}$ is the charge-conjugation operator; $\epsilon^{ABC}$ is the totally anti-symmetric tensor; $S_P^{ABC}$ ($P=1\cdots10$) are the normalized totally symmetric matrices; $\bf \lambda_N$ ($N=1\cdots8$) are the Gell-Mann matrices; $S^{ABCD}_{U}$ ($U=1\cdots27$) are the matrices for the $\mathbf{27}$ flavor representation, as defined in Ref.~\cite{Chen:2012ut}.

Among the ten currents, five current ($\eta_{1,2,3,7,8}^{\rm S,\mathbb{S}}$) contain diquarks and antidiquarks both having the antisymmetric flavor structure $\mathbf{\bar 3} \otimes \mathbf{3}$, and the rest ($\eta_{4,5,6,9,10}^{\rm S,\mathbb{S}}$) contain diquarks and antidiquarks both having the symmetric flavor structure $\mathbf{6} \otimes \mathbf{\bar6}$. We note that after fixing the flavor and Lorentz structure of the internal diquarks and antidiquarks, their color structure is also fixed through Pauli's exclusion principle, as shown in Table~\ref{tab:singletscalar}.

The chiral structure of tetraquarks is more complicated than their flavor structure:
\begin{eqnarray}
\Big( ({\mathbf 3},{\mathbf 1})\oplus({\mathbf 1},{\mathbf 3}) \Big)^2 \otimes \Big( (\mathbf{ \bar {3}},{\mathbf 1})\oplus({\mathbf 1},\mathbf {\bar {3}}) \Big)^2
&=& \Big( (\mathbf {\bar 3},{\mathbf 1})\oplus({\mathbf 1},\mathbf {\bar 3}) \oplus ({\mathbf 6},{\mathbf 1})\oplus({\mathbf 1},{\mathbf 6}) \oplus ({\mathbf 3},{\mathbf 3}) \oplus ({\mathbf 3},{\mathbf 3}) \Big)
\otimes
\\ \nonumber && \Big( ({\mathbf 3},{\mathbf 1})\oplus({\mathbf 1},{\mathbf 3}) \oplus (\mathbf {\bar 6},{\mathbf 1})\oplus({\mathbf 1},\mathbf {\bar 6}) \oplus (\mathbf {\bar 3},\mathbf {\bar 3}) \oplus (\mathbf {\bar 3},\mathbf {\bar 3}) \Big) \, .
\end{eqnarray}
The full (expanded) expressions are shown in Ref.~\cite{Chen:2012ut}. Among them, the following multiplets contain flavor singlet tetraquarks currents: $[({\mathbf 1},{\mathbf 1})]$, $[(\mathbf {\bar 3},{\mathbf 3}) \oplus ({\mathbf 3},\mathbf {\bar 3})]$, $[({\mathbf 6},\mathbf {\bar 6}) \oplus (\mathbf {\bar 6},{\mathbf 6})]$ and $[({\mathbf 8},{\mathbf 8}) \oplus ({\mathbf 8},{\mathbf 8})]$, as well as their mirror multiples. Tetraquarks of all flavors can be chiral partners of flavor singlet tetraquarks because of the exotic $[({\mathbf 8},{\mathbf 8}) \oplus ({\mathbf 8},{\mathbf 8})]$ chiral multiplet.

To clearly see the chiral structure of Eqs.~(\ref{eq:current11}), we use the left-handed quark field $L_A^a \equiv q_{LA}^a = {1 - \gamma_5 \over 2} q_A^a$ and the right-handed quark field $R_A^a \equiv q_{RA}^a = {1 + \gamma_5 \over 2} q_A^a$ to rewrite these currents and then combine them properly:
\begin{eqnarray}
\nonumber \eta_1^{\rm S,\mathbb{S}} - \eta_2^{\rm S,\mathbb{S}} &=& - 2 L_A^{aT} \mathbb{C} L_B^b (\bar{L}_A^a \mathbb{C} \bar{L}_B^{bT} - \bar{L}_A^b \mathbb{C} \bar{L}_B^{aT})
- 2 R_A^{aT} \mathbb{C} R_B^b (\bar{R}_A^a \mathbb{C} \bar{R}_B^{bT} - \bar{R}_A^b \mathbb{C} \bar{R}_B^{aT}) \, ,
\\ \nonumber \eta_1^{\rm S,\mathbb{S}} + \eta_2^{\rm S,\mathbb{S}} &=& 2 L_A^{aT} \mathbb{C} L_B^b (\bar{R}_A^a \mathbb{C} \bar{R}_B^{bT} - \bar{R}_A^b \mathbb{C} \bar{R}_B^{aT})
+ 2 R_A^{aT} \mathbb{C} R_B^b (\bar{L}_A^a \mathbb{C} \bar{L}_B^{bT} - \bar{L}_A^b \mathbb{C} \bar{L}_B^{aT}) \, ,
\\ \nonumber \eta_4^{\rm S,\mathbb{S}} - \eta_5^{\rm S,\mathbb{S}} &=& - 2 L_A^{aT} \mathbb{C} L_B^b (\bar{L}_A^a \mathbb{C} \bar{L}_B^{bT} + \bar{L}_A^b \mathbb{C} \bar{L}_B^{aT})
- 2 R_A^{aT} \mathbb{C} R_B^b (\bar{R}_A^a \mathbb{C} \bar{R}_B^{bT} + \bar{R}_A^b \mathbb{C} \bar{R}_B^{aT}) \, ,
\\ \nonumber \eta_4^{\rm S,\mathbb{S}} + \eta_5^{\rm S,\mathbb{S}} &=& 2 L_A^{aT} \mathbb{C} L_B^b (\bar{R}_A^a \mathbb{C} \bar{R}_B^{bT} + \bar{R}_A^b \mathbb{C} \bar{R}_B^{aT})
+ 2 R_A^{aT} \mathbb{C} R_B^b (\bar{L}_A^a \mathbb{C} \bar{L}_B^{bT} + \bar{L}_A^b \mathbb{C} \bar{L}_B^{aT}) \, ,
\\ \eta_3^{\rm S,\mathbb{S}} &=& L_A^{aT} \mathbb{C} \sigma_{\mu\nu} L_B^b (\bar{R}_A^a \sigma_{\mu\nu} \mathbb{C} \bar{R}_B^{bT}
+ \bar{R}_A^b \sigma_{\mu\nu} \mathbb{C} \bar{R}_B^{aT})
+ R_A^{aT} \mathbb{C} \sigma_{\mu\nu} R_B^b (\bar{L}_A^a \sigma_{\mu\nu} \mathbb{C} \bar{L}_B^{bT} + \bar{L}_A^b \sigma_{\mu\nu} \mathbb{C} \bar{L}_B^{aT}) \, ,
\\ \nonumber \eta_6^{\rm S,\mathbb{S}} &=& L_A^{aT} \mathbb{C} \sigma_{\mu\nu} L_B^b (\bar{R}_A^a \sigma_{\mu\nu} \mathbb{C} \bar{R}_B^{bT} - \bar{R}_A^b \sigma_{\mu\nu} \mathbb{C} \bar{R}_B^{aT})
+ R_A^{aT} \mathbb{C} \sigma_{\mu\nu} R_B^b (\bar{L}_A^a \sigma_{\mu\nu} \mathbb{C} \bar{L}_B^{bT} - \bar{L}_A^b \sigma_{\mu\nu} \mathbb{C} \bar{L}_B^{aT}) \, ,
\\ \nonumber \eta_7^{\rm S,\mathbb{S}} - \eta_{10}^{\rm S,\mathbb{S}} &=& - 4 L_A^{aT} \mathbb{C} \gamma_\mu R_B^b ( \bar{L}_A^a \gamma^\mu \mathbb{C} \bar{R}_B^{bT} - \bar{L}_A^b \gamma^\mu \mathbb{C} \bar{R}_B^{aT}) \, ,
\\ \nonumber 2 \eta_7^{\rm S,\mathbb{S}} + \eta_{10}^{\rm S,\mathbb{S}} &=& 3 \lambda_N^{CA} \lambda_N^{DB} L_A^{aT} \mathbb{C} \gamma_\mu R_B^b ( \bar{L}_C^a \gamma^\mu \mathbb{C} \bar{R}_D^{bT} - \bar{L}_C^b \gamma^\mu \mathbb{C} \bar{R}_D^{aT}) \, ,
\\ \nonumber \eta_8^{\rm S,\mathbb{S}} - \eta_9^{\rm S,\mathbb{S}} &=& 4 L_A^{aT} \mathbb{C} \gamma_\mu R_B^b ( \bar{L}_A^a \gamma^\mu \mathbb{C} \bar{R}_B^{bT} + \bar{L}_A^b \gamma^\mu \mathbb{C} \bar{R}_B^{aT}) \, ,
\\ \nonumber 2 \eta_8^{\rm S,\mathbb{S}} + \eta_9^{\rm S,\mathbb{S}} &=& - 3 \lambda_N^{CA} \lambda_N^{DB} L_A^{aT} \mathbb{C} \gamma_\mu R_B^b ( \bar{L}_C^a \gamma^\mu \mathbb{C} \bar{R}_D^{bT} + \bar{L}_C^b \gamma^\mu \mathbb{C} \bar{R}_D^{aT}) \, ,
\end{eqnarray}
from which we can quickly find out their chiral structure (representations). For example, $\eta_{1+2}^{\rm S,\mathbb{S}} \equiv \eta_{1}^{\rm S,\mathbb{S}} + \eta_{2}^{\rm S,\mathbb{S}}$ partly contains
two left-handed quarks that have an antisymmetric flavor structure and two right-handed antiquarks that also have an antisymmetric flavor structure; therefore, this part has the chiral representation $( \mathbf {\bar 3}, {\mathbf 3})$, and its full chiral representation is just $[( \mathbf {\bar 3}, {\mathbf 3}) + ({\mathbf 3}, \mathbf {\bar 3})]$.

The results are listed in Table~\ref{tab:singletscalar}. There are four chiral singlets $[({\mathbf 1}, {\mathbf 1})]$, two $[( \mathbf {\bar 3}, {\mathbf 3}) + ({\mathbf 3}, \mathbf {\bar 3})]$ chiral multiplets, two $[({\mathbf 6}, \mathbf {\bar 6}) + (\mathbf {\bar 6}, {\mathbf 6})]$ chiral multiplets and two $[({\mathbf 8}, {\mathbf 8}) + ({\mathbf 8}, {\mathbf 8})]$ chiral multiplets. The $[( \mathbf {\bar 3}, {\mathbf 3}) + ({\mathbf 3}, \mathbf {\bar 3})]$ chiral representation contains the Nambu-Goldstone bosons, $\pi$, $K$, $\eta$ and $\eta^\prime$ mesons, and it does not contain any meson having exotic flavor structure. Considering $\sigma$ and $\pi$ are believed to be chiral partners in chiral models, we assume that light scalar mesons belong to this ``non-exotic'' multiplet, and we shall concentrate on it in our subsequent analysis. Based on this assumption, we do not need to study other possible tetraquark states having exotic flavor structure. Moreover, the current $\eta_3^{\rm S,\mathbb{S}}$ has the symmetric color structure ${\mathbf 6} \otimes \mathbf {\bar 6}$, where color interactions between quarks and antiquarks are repulsive. Therefore, it is questionable to use this current, but we shall still use it to perform the QCD sum rule analysis for comparison. We note that mixed currents used in Ref.~\cite{Chen:2007xr} belong to the mixing of $[({\mathbf 1},{\mathbf 1})]$ and exotic $[({\mathbf 8},{\mathbf 8}) \oplus ({\mathbf 8},{\mathbf 8})]$ multiplets.

\begin{table}[hbt]
\renewcommand{\arraystretch}{1.5}
\begin{center}
\caption{Flavor singlet tetraquark currents of $J^P = 0^+$, showing their chiral representations and chirality. The second and third columns show the flavor and color structures of the diquark and antidiquark inside, respectively.}
\begin{tabular}{c c c c c}
\hline\hline
Currents & Flavor & Color & Representations & Chirality
\\ \hline \hline
$\eta_{1}^{\rm S,\mathbb{S}} - \eta_{2}^{\rm S,\mathbb{S}}$ & $\mathbf{\bar 3} \otimes \mathbf{3}$ & $\mathbf{\bar 3} \otimes \mathbf{3}$ & \multirow{2}{*}{$[({\mathbf 1}, {\mathbf 1})]$} & \multirow{2}{*}{$L L \bar L \bar L+ R R \bar R \bar R$}
\\ \cline{1-3} $\eta_{4}^{\rm S,\mathbb{S}} - \eta_{5}^{\rm S,\mathbb{S}}$ & $\mathbf{6} \otimes \mathbf{\bar 6}$ & $\mathbf{6} \otimes \mathbf{\bar 6}$ & &
\\ \hline
$\eta_{1}^{\rm S,\mathbb{S}} + \eta_{2}^{\rm S,\mathbb{S}}$ & $\mathbf{\bar 3} \otimes \mathbf{3}$ & $\mathbf{\bar 3} \otimes \mathbf{3}$ & $[(\bar {\mathbf 3}, {\mathbf 3}) + ({\mathbf 3}, \bar {\mathbf 3})]$ & $L L \bar R \bar R + R R \bar L \bar L$
\\ \hline
$\eta_{4}^{\rm S,\mathbb{S}} + \eta_{5}^{\rm S,\mathbb{S}}$ & $\mathbf{6} \otimes \mathbf{\bar 6}$ & $\mathbf{6} \otimes \mathbf{\bar 6}$ & $[({\mathbf 6}, \bar {\mathbf 6}) + (\bar {\mathbf 6}, {\mathbf 6})]$ & $L L \bar R \bar R + R R \bar L \bar L$
\\ \hline
$\eta_3^{\rm S,\mathbb{S}}$ & $\mathbf{\bar 3} \otimes \mathbf{3}$ & $\mathbf{6} \otimes \mathbf{\bar 6}$ & $[(\bar {\mathbf 3}, {\mathbf 3}) + ({\mathbf 3}, \bar {\mathbf 3})]$ & $L L \bar R \bar R + R R \bar L \bar L$
\\ \hline
$\eta_6^{\rm S,\mathbb{S}}$ & $\mathbf{6} \otimes \mathbf{\bar 6}$ & $\mathbf{\bar 3} \otimes \mathbf{3}$ & $[({\mathbf 6}, \bar {\mathbf 6}) + (\bar {\mathbf 6}, {\mathbf 6})]$ & $L L \bar R \bar R + R R \bar L \bar L$
\\ \hline
$\eta_{7}^{\rm S,\mathbb{S}} - \eta_{10}^{\rm S,\mathbb{S}}$ & mixed & $\mathbf{\bar 3} \otimes \mathbf{3}$ & \multirow{2}{*}{$[({\mathbf 1}, {\mathbf 1})]$} & \multirow{2}{*}{$L R \bar L \bar R + R L \bar R \bar L$}
\\ \cline{1-3} $\eta_{8}^{\rm S,\mathbb{S}} - \eta_{9}^{\rm S,\mathbb{S}}$ & mixed & $\mathbf{6} \otimes \mathbf{\bar 6}$ & &
\\ \hline
$2\eta_{7}^{\rm S,\mathbb{S}} + \eta_{10}^{\rm S,\mathbb{S}}$ & mixed & $\mathbf{\bar 3} \otimes \mathbf{3}$ & \multirow{2}{*}{$[({\mathbf 8}, {\mathbf 8}) + ({\mathbf 8}, {\mathbf 8})]$} & \multirow{2}{*}{$L R \bar L \bar R + R L \bar R \bar L$}
\\ \cline{1-3} $2\eta_{8}^{\rm S,\mathbb{S}} + \eta_{9}^{\rm S,\mathbb{S}}$ & mixed & $\mathbf{6} \otimes \mathbf{\bar 6}$
\\ \hline\hline
\end{tabular}
\label{tab:singletscalar}
\end{center}
\renewcommand{\arraystretch}{1}
\end{table}

To fully study this multiplet, the chiral partners of Eqs.~(\ref{eq:current11}) are also studied, i.e., the scalar and pseudoscalar tetraquark currents of flavor singlet, octet, $\mathbf{10}$, $\mathbf{\overline {10}}$ and $\mathbf{27}$. The results are shown in Appendix.~\ref{app:othercurrents}. The conventional pseudoscalar and scalar mesons made by one $\bar q q$ pair can also belong to the $[(\mathbf{3},\mathbf{\bar{3}}) \oplus (\mathbf{\bar{3}},\mathbf{3})]$ chiral multiplet. However, all the scalar tetraquark currents inside this multiplet have the $q_L q_L \bar q_R \bar q_R + q_R q_R \bar q_L \bar q_L$ chirality, and so they are not direct chiral partners of these $\bar q q$ mesons addressed by chiral singlet quark-antiquark pairs, which have the $( \bar q_L q_R + \bar q_R q_L) \otimes (\bar q_L q_L + \bar q_R q_R)$ chirality (``chiral'' Fock-space expansion), unless these two types of chirality mix with each other. Similarly, all the pseudoscalar tetraquark currents inside this multiplet have the same $q_L q_L \bar q_R \bar q_R + q_R q_R \bar q_L \bar q_L$ chirality, and so they are not (direct) terms in the ``chiral'' Fock-space expansion of the $\bar q q$ pseudoscalar mesons ($\pi$, etc).

\section{Chiral Transformations}
\label{sec:chiral}

We can study their chiral transformation properties to verify which currents are chiral partners. Under the $\bf{U(1)_{V}}$, $\bf{U(1)_{A}}$, $\bf{SU(3)_V}$ and $\bf{SU(3)_A}$ chiral transformations, the quark field, $q= q_L + q_R$, transforms as
\begin{eqnarray}
\nonumber
\bf{U(1)_{V}} &:& q \to \exp(i a^0) q  = q + \delta q \, ,
\\
\bf{SU(3)_V} &:& q \to \exp (i \vec \lambda \cdot \vec a ){q} = q + \delta^{\vec{a}} q \, ,
\\ \nonumber
\bf{U(1)_{A}} &:& q \to \exp(i \gamma_5 b^0) q = q + \delta_5 q \, ,
\\ \nonumber
\bf{SU(3)_A} &:& q \to \exp (i \gamma_{5} \vec \lambda \cdot \vec b){q} = q + \delta_5^{\vec{b}} q \, ,
\end{eqnarray}
where $\vec \lambda$ are the eight Gell-Mann matrices; $a^0$ is an infinitesimal parameter for the $\bf{U(1)_{V}}$ transformation, $\vec{a}$ the octet of $\bf{SU(3)_V}$ group parameters, $b^0$ an infinitesimal parameter for the $\bf{U(1)_{A}}$ transformation, and $\vec{b}$ the octet of the chiral transformations.

The chiral transformation equations for these tetraquark currents can be calculated straightforwardly, and we only show the final results. The local scalar and pseudoscalar tetraquark currents have been classified in Sec.~\ref{sec:currents} and Sec.~\ref{app:othercurrents}. We find that there are four $[({\mathbf 1},{\mathbf 1})]$ chiral multiplets:
\begin{eqnarray}
\nonumber && \big ( \eta_{1}^{\rm S, \mathbb{S}} - \eta_{2}^{\rm S, \mathbb{S}}, \eta_{1}^{\rm PS, \mathbb{S}} - \eta_{2}^{\rm PS, \mathbb{S}} \big ) \, ,
\\ && \big ( \eta_{4}^{\rm S, \mathbb{S}} - \eta_{5}^{\rm S, \mathbb{S}}, \eta_{4}^{\rm PS, \mathbb{S}} - \eta_{5}^{\rm PS, \mathbb{S}}\big ) \, ,
\\ \nonumber && \big ( \eta_{7}^{\rm S, \mathbb{S}} - \eta_{10}^{\rm S, \mathbb{S}}, \eta_{7}^{\rm PS, \mathbb{S}} - \eta_{10}^{\rm PS, \mathbb{S}} = 0 \big ) \, ,
\\ \nonumber && \big ( \eta_{8}^{\rm S, \mathbb{S}} - \eta_{9}^{\rm S, \mathbb{S}}, \eta_{8}^{\rm PS, \mathbb{S}} - \eta_{9}^{\rm PS, \mathbb{S}}  = 0 \big ) \, ;
\end{eqnarray}
there are two $[({\mathbf 3},\mathbf {\bar 3}) \oplus (\mathbf {\bar 3},{\mathbf 3})]$ chiral multiplets (or mirror multiplets):
\begin{eqnarray}
\nonumber && \big ( \eta_{1}^{\rm S, \mathbb{S}} + \eta_{2}^{\rm S, \mathbb{S}}, \eta_{1}^{\rm PS, \mathbb{S}} + \eta_{2}^{\rm PS, \mathbb{S}}, \eta_{1,N}^{\rm S, \mathbb{O}} + \eta_{2,N}^{\rm S, \mathbb{O}}, \eta_{1,N}^{\rm PS, \mathbb{O}} + \eta_{2,N}^{\rm PS, \mathbb{O}} \big ) \, ,
\\ && \big ( \eta_{3}^{\rm S, \mathbb{S}}, \eta_{3}^{\rm PS, \mathbb{S}}, \eta_{3,N}^{\rm S, \mathbb{O}}, \eta_{3,N}^{\rm PS, \mathbb{O}} \big )\, ;
\end{eqnarray}
there are two $[({\mathbf 6},\mathbf {\bar 6}) \oplus (\mathbf {\bar 6},{\mathbf 6})]$ chiral multiplets (or mirror multiplets):
\begin{eqnarray}
\nonumber && \big ( \eta_{4}^{\rm S, \mathbb{S}} + \eta_{5}^{\rm S, \mathbb{S}}, \eta_{4}^{\rm PS, \mathbb{S}} + \eta_{5}^{\rm PS, \mathbb{S}}, \eta_{4,N}^{\rm S, \mathbb{O}} + \eta_{5,N}^{\rm S, \mathbb{O}}, \eta_{4,N}^{\rm PS, \mathbb{O}} + \eta_{5,N}^{\rm PS, \mathbb{O}}, \eta_{1,U}^{\rm S, \mathbb{TS}} + \eta_{2,U}^{\rm S, \mathbb{TS}}, \eta_{1,U}^{\rm PS, \mathbb{TS}} + \eta_{2,U}^{\rm PS, \mathbb{TS}} \big ) \, ,
\\ && \big ( \eta_{6}^{\rm S, \mathbb{S}}, \eta_{6}^{\rm PS, \mathbb{S}}, \eta_{6,N}^{\rm S, \mathbb{O}}, \eta_{6,N}^{\rm PS, \mathbb{O}}, \eta_{3,U}^{\rm S, \mathbb{TS}}, \eta_{3,U}^{\rm PS, \mathbb{TS}} \big ) \, ;
\end{eqnarray}
there are two $[({\mathbf 8},{\mathbf 8}) \oplus ({\mathbf 8},{\mathbf 8})]$ chiral multiplets:
\begin{eqnarray}
\nonumber && \big ( 2 \eta_{7}^{\rm S, \mathbb{S}} + \eta_{10}^{\rm S, \mathbb{S}}, 2 \eta_{7}^{\rm PS, \mathbb{S}} + \eta_{10}^{\rm PS, \mathbb{S}} = 0, 5 \eta_{7,N}^{\rm S, \mathbb{O}} + \eta_{10,N}^{\rm S, \mathbb{O}}, \eta_{7,N}^{\rm PS, \mathbb{O}} + \eta_{10,N}^{\rm PS, \mathbb{O}}, \eta_{5,U}^{\rm S, \mathbb{TS}}, \eta_{1,P}^{\rm PS, \mathbb{D}}, \eta_{2,P}^{\rm PS, \bar \mathbb{D}} \big ) \, ,
\\ && \big ( 2 \eta_{8}^{\rm S, \mathbb{S}} + \eta_{9}^{\rm S, \mathbb{S}}, 2 \eta_{8}^{\rm PS, \mathbb{S}} + \eta_{9}^{\rm PS, \mathbb{S}} = 0, 5 \eta_{8,N}^{\rm S, \mathbb{O}} + \eta_{9,N}^{\rm S, \mathbb{O}}, \eta_{8,N}^{\rm PS, \mathbb{O}} + \eta_{9,N}^{\rm PS, \mathbb{O}}, \eta_{4,U}^{\rm S, \mathbb{TS}}, \eta_{2,P}^{\rm PS, \mathbb{D}}, \eta_{1,P}^{\rm PS, \bar \mathbb{D}} \big ) \, ;
\end{eqnarray}
there are four $[({\mathbf 8},{\mathbf 1}) \oplus ({\mathbf 1},{\mathbf 8})]$ chiral multiplets (or mirror multiplets):
\begin{eqnarray}
\nonumber && \big ( \eta_{1,N}^{\rm S, \mathbb{O}} - \eta_{2,N}^{\rm S, \mathbb{O}}, \eta_{1,N}^{\rm PS, \mathbb{O}} - \eta_{2,N}^{\rm PS, \mathbb{O}} \big ) \, ,
\\ && \big ( \eta_{4,N}^{\rm S, \mathbb{O}} - \eta_{5,N}^{\rm S, \mathbb{O}}, \eta_{4,N}^{\rm PS, \mathbb{O}} - \eta_{5,N}^{\rm PS, \mathbb{O}} \big ) \, ,
\\ \nonumber && \big ( \eta_{7,N}^{\rm S, \mathbb{O}} - \eta_{10,N}^{\rm S, \mathbb{O}}, \eta_{7,N}^{\rm PS, \mathbb{O}} - \eta_{10,N}^{\rm PS, \mathbb{O}} \big ) \, ,
\\ \nonumber && \big ( \eta_{8,N}^{\rm S, \mathbb{O}} - \eta_{9,N}^{\rm S, \mathbb{O}}, \eta_{8,N}^{\rm PS, \mathbb{O}} - \eta_{9,N}^{\rm PS, \mathbb{O}} \big ) \, ;
\end{eqnarray}
there is only one $[(\mathbf{27},{\mathbf 1}) \oplus ({\mathbf 1},\mathbf{27})]$ chiral multiplet:
\begin{eqnarray}
\big ( \eta_{1,U}^{\rm S, \mathbb{TS}} - \eta_{2,U}^{\rm S, \mathbb{TS}}, \eta_{1,U}^{\rm PS, \mathbb{TS}} - \eta_{2,U}^{\rm PS, \mathbb{TS}} \big ) \, .
\end{eqnarray}
Their chiral transformation properties are shown in Appendix.~\ref{app:othertransform}, except those for the two $[(\mathbf{\bar 3},\mathbf{3}) \oplus (\mathbf{3},\mathbf{\bar 3})]$ chiral multiplets, which we show here. We use $( \eta_{(\mathbf{\bar 3},\mathbf{3})}^{\rm S, \mathbb{S}}, \eta_{( \mathbf{\bar 3},\mathbf{3})}^{\rm PS, \mathbb{S}}, \eta_{(\mathbf{\bar 3},\mathbf{3}),N}^{\rm S, \mathbb{O}}, \eta_{(\mathbf{\bar 3},\mathbf{3}),N}^{\rm PS, \mathbb{O}} )$ to denote these two multiplets, $\big ( \eta_{1}^{\rm S, \mathbb{S}} + \eta_{2}^{\rm S, \mathbb{S}}, \eta_{1}^{\rm PS, \mathbb{S}} + \eta_{2}^{\rm PS, \mathbb{S}}, \eta_{1,N}^{\rm S, \mathbb{O}} + \eta_{2,N}^{\rm S, \mathbb{O}}, \eta_{1,N}^{\rm PS, \mathbb{O}} + \eta_{2,N}^{\rm PS, \mathbb{O}} \big )$ and
$\big ( \eta_{3}^{\rm S, \mathbb{S}}, \eta_{3}^{\rm PS, \mathbb{S}}, \eta_{3,N}^{\rm S, \mathbb{O}}, \eta_{3,N}^{\rm PS, \mathbb{O}} \big )$, and they have the same chiral transformation properties:
\begin{eqnarray}
\nonumber \delta_5 \eta_{(\mathbf{\bar 3},\mathbf{3})}^{\rm S(PS), \mathbb{S}} &=& 4 i b \eta_{(\mathbf{\bar 3},\mathbf{3})}^{\rm PS(S), \mathbb{S}} \, ,
\\ \nonumber \delta^{\vec a} \eta_{(\mathbf{\bar 3},\mathbf{3})}^{\rm S(PS), \mathbb{S}} &=& 0 \, ,
\\ \delta_5^{\vec b} \eta_{(\mathbf{\bar 3},\mathbf{3})}^{\rm S(PS), \mathbb{S}} &=& 4 i b^N \eta_{(\mathbf{\bar 3},\mathbf{3}),N}^{\rm PS(S), \mathbb{O}} \, ,
\\ \nonumber \delta_5 \eta_{(\mathbf{\bar 3},\mathbf{3}),N}^{\rm S(PS), \mathbb{O}} &=& 4 i b \eta_{(\mathbf{\bar 3},\mathbf{3}),N}^{\rm PS(S), \mathbb{O}} \, ,
\\ \nonumber \delta^{\vec a} \eta_{(\mathbf{\bar 3},\mathbf{3}),N}^{\rm S(PS), \mathbb{O}} &=& 2 a^N f_{NMO} \eta_{(\mathbf{\bar 3},\mathbf{3}),O}^{\rm S(PS), \mathbb{O}} \, ,
\\ \nonumber \delta_5^{\vec b} \eta_{(\mathbf{\bar 3},\mathbf{3}),N}^{\rm S(PS), \mathbb{O}} &=& {2\over3} i b^M \eta_{(\mathbf{\bar 3},\mathbf{3})}^{\rm PS(S), \mathbb{S}} - 2 i b^N d_{NMO} \eta_{(\mathbf{\bar 3},\mathbf{3}),O}^{\rm PS(S), \mathbb{O}} \, .
\end{eqnarray}
These chiral transformation equations can be compared to those calculated in Ref.~\cite{Chen:2012ut} which have the same chirality and chiral representation, but in Ref.~\cite{Chen:2012ut} only the flavor structure is taken into account. The ${\bf SU}(3)_{\bf V}$ and ${\bf SU}(3)_{\bf A}$ equations are similar to those of $\bar q q$ mesons as well as baryons belonging to the same chiral multiplet~\cite{Chen:2008qv,Chen:2012ut}, suggesting chiral transformation properties are closely related to chiral representations; while ${\bf U}(1)_{\bf A}$ equations are different, which may be reasons for the $U(1)_A$ anomaly.

The following formula obtained from Ref.~\cite{Chen:2012ut} is used in the calculations:
\begin{eqnarray}
\lambda^N_{CA} \lambda^M_{DB} &=& -{1\over12}\delta^{NM} \delta_{CA} \delta_{DB} + {1\over4}\delta^{NM} \delta_{DA} \delta_{CB}
\\ \nonumber && - {2\over5} d_{NMO} \delta_{CA} \lambda^O_{DB} + ( {3\over5} d_{NMO} + {i\over3} f_{NMO} ) \delta_{DA} \lambda^O_{CB}
\\ \nonumber && + ( {3\over5} d_{NMO} - {i\over3} f_{NMO} ) \delta_{CB} \lambda^O_{DA} - {2\over5} d_{NMO} \delta_{DB} \lambda^O_{CA}
\\ \nonumber && + ({\bf T}_{8\times10}^N)^*_{MP} \epsilon_{ABE} S^{{\bf 10}, P}_{CDE} + ({\bf T}_{8\times10}^N)_{MP} \epsilon_{CDE} S^{{\bf 10}, P}_{ABE}
\\ \nonumber && + ({\bf T}_{8\times27}^N)_{MU} S^{{\bf 27}, U}_{CD;AB} \, ,
\end{eqnarray}
as well as several other formulae:
\begin{eqnarray}
\epsilon_{ABE} S^{{\bf 10},P}_{CGE} \lambda^N_{DG} &=& {\bf 0}^{NP} \times \delta_{CA} \delta_{DB} + {\bf 0}^{NP} \times \delta_{DA} \delta_{CB}
\\ \nonumber && - {1 \over 3}({\bf T}_{8\times10}^N)^T_{PO} \delta_{CA} \lambda^O_{DB} + {2 \over 3}({\bf T}_{8\times10}^N)^T_{PO} \delta_{DA} \lambda^O_{CB}
\\ \nonumber && + {1 \over 3}({\bf T}_{8\times10}^N)^T_{PO} \delta_{CB} \lambda^O_{DA} - {2 \over 3}({\bf T}_{8\times10}^N)^T_{PO} \delta_{DB} \lambda^O_{CA}
\\ \nonumber && + ({\bf T}_{10\times10}^N)^T_{PQ} \epsilon_{ABE} S^{{\bf 10}, P}_{CDE} + {\bf 0}^{NPQ} \times \epsilon_{CDE} S^{{\bf 10}, P}_{ABE}
\\ \nonumber && + {\bf 0}^{NPU} \times S^{{\bf 27}, U}_{CD;AB} \, ,
\\ S^{{\bf 10},P}_{AGE} \epsilon_{CDE} \lambda^N_{GB} &=& {\bf 0}^{NP} \times \delta_{CA} \delta_{DB} + {\bf 0}^{NP} \times \delta_{DA} \delta_{CB}
\\ \nonumber && - {1 \over 3}({\bf T}_{8\times10}^N)^\dagger_{PO} \delta_{CA} \lambda^O_{DB} + {1 \over 3}({\bf T}_{8\times10}^N)^\dagger_{PO} \delta_{DA} \lambda^O_{CB}
\\ \nonumber && + {2 \over 3}({\bf T}_{8\times10}^N)^\dagger_{PO} \delta_{CB} \lambda^O_{DA} - {2 \over 3}({\bf T}_{8\times10}^N)^\dagger_{PO} \delta_{DB} \lambda^O_{CA}
\\ \nonumber && + {\bf 0}^{NPQ} \epsilon_{ABE} S^{{\bf 10}, P}_{CDE} + ({\bf T}_{10\times10}^N)_{PQ}\times \epsilon_{CDE} S^{{\bf 10}, P}_{ABE}
\\ \nonumber && + {\bf 0}^{NPU} \times S^{{\bf 27}, U}_{CD;AB} \, ,
\\ \epsilon_{AGE} S^{{\bf 10},P}_{CDE} \lambda^N_{GB} &=& {\bf 0}^{NP} \times \delta_{CA} \delta_{DB} + {\bf 0}^{NP} \times \delta_{DA} \delta_{CB}
\\ \nonumber && + {1\over15} ({\bf T}_{8\times10}^N)^T_{PO} \delta_{CA} \lambda^O_{DB} + {1\over15} ({\bf T}_{8\times10}^N)^T_{PO} \delta_{DA} \lambda^O_{CB}
\\ \nonumber && - {4\over15} ({\bf T}_{8\times10}^N)^T_{PO} \delta_{CB} \lambda^O_{DA} - {4\over15} ({\bf T}_{8\times10}^N)^T_{PO} \delta_{DB} \lambda^O_{CA}
\\ \nonumber && - {1\over2} ({\bf T}_{10\times10}^N)^T_{PQ} \epsilon_{ABE} S^{{\bf 10}, P}_{CDE} + {\bf 0}^{NPQ} \times \epsilon_{CDE} S^{{\bf 10}, P}_{ABE}
\\ \nonumber && + ({\bf T^B}_{10\times27})^N_{PU} S^{{\bf 27}, U}_{CD;AB} \, ,
\\ S^{{\bf 10},P}_{ABE} \epsilon_{CGE} \lambda^N_{DG} &=& {\bf 0}^{NP} \times \delta_{CA} \delta_{DB} + {\bf 0}^{NP} \times \delta_{DA} \delta_{CB}
\\ \nonumber && + {1\over15} ({\bf T}_{8\times10}^N)^\dagger_{PO} \delta_{CA} \lambda^O_{DB} - {4\over15} ({\bf T}_{8\times10}^N)^\dagger_{PO} \delta_{DA} \lambda^O_{CB}
\\ \nonumber && + {1\over15} ({\bf T}_{8\times10}^N)^\dagger_{PO} \delta_{CB} \lambda^O_{DA} - {4\over15} ({\bf T}_{8\times10}^N)^\dagger_{PO} \delta_{DB} \lambda^O_{CA}
\\ \nonumber && + {\bf 0}^{NPQ} \epsilon_{ABE} S^{{\bf 10}, P}_{CDE} - {1\over2} ({\bf T}_{10\times10}^N)_{PQ} \times \epsilon_{CDE} S^{{\bf 10}, P}_{ABE}
\\ \nonumber && + ({\bf T^A}_{10\times27})^N_{PU} S^{{\bf 27}, U}_{CD;AB} \, ,
\\ S^{{\bf 27},U}_{CD;AG} \lambda^N_{GB} &=& {\bf 0}^{NU} \times \delta_{CA} \delta_{DB} + {\bf 0}^{NU} \times \delta_{DA} \delta_{CB}
\\ \nonumber && - {1\over10} ({\bf T}_{8\times27}^N)^\dagger_{UO} \delta_{CA} \lambda^O_{DB} - {1\over10} ({\bf T}_{8\times27}^N)^\dagger_{UO} \delta_{DA} \lambda^O_{CB}
\\ \nonumber && + {2\over5} ({\bf T}_{8\times27}^N)^\dagger_{UO} \delta_{CB} \lambda^O_{DA} + {2\over5} ({\bf T}_{8\times27}^N)^\dagger_{UO} \delta_{DB} \lambda^O_{CA}
\\ \nonumber && + {3\over2}({\bf T^B}_{10\times27})^{N\dagger}_{UP} \times \epsilon_{ABE} S^{{\bf 10}, P}_{CDE} + {\bf 0}^{NUP} \times \epsilon_{CDE} S^{{\bf 10}, P}_{ABE}
\\ \nonumber && + ({\bf T^A}_{27\times27})^N_{UV} S^{{\bf 27}, U}_{CD;AB} \, ,
\\ S^{{\bf 27},U}_{CG;AB} \lambda^N_{DG} &=& {\bf 0}^{NU} \times \delta_{CA} \delta_{DB} + {\bf 0}^{NU} \times \delta_{DA} \delta_{CB}
\\ \nonumber && - {1\over10} ({\bf T}_{8\times27}^N)^\dagger_{UO} \delta_{CA} \lambda^O_{DB} + {2\over5} ({\bf T}_{8\times27}^N)^\dagger_{UO} \delta_{DA} \lambda^O_{CB}
\\ \nonumber && - {1\over10} ({\bf T}_{8\times27}^N)^\dagger_{UO} \delta_{CB} \lambda^O_{DA} + {2\over5} ({\bf T}_{8\times27}^N)^\dagger_{UO} \delta_{DB} \lambda^O_{CA}
\\ \nonumber && + {\bf 0}^{NUP} \times \epsilon_{ABE} S^{{\bf 10}, P}_{CDE} + {3\over2}({\bf T^A}_{10\times27})^{N\dagger}_{UP} \times \epsilon_{CDE} S^{{\bf 10}, P}_{ABE}
\\ \nonumber && + ({\bf T^B}_{27\times27})^N_{UV} S^{{\bf 27}, U}_{CD;AB} \, .
\end{eqnarray}
The transition matrices ${\bf T}_{8\times10}^N$ and ${\bf T}_{8\times27}^N$ have been obtained and listed in Ref.~\cite{Chen:2012ut}. We list the transition matrices ${\bf T}_{10\times10}$, ${\bf T^A}_{10\times27}$ and ${\bf T^B}_{10\times27}$ in Appendix.~\ref{app:matrices}. However, the transition matrices ${\bf T^A}_{27\times27}$ and ${\bf T^B}_{27\times27}$ are omitted due to their long expressions.

\section{QCD Sum Rule Analysis}
\label{sec:sumrule}

For the past decades QCD sum rule has proven to be a powerful and successful non-perturbative method~\cite{Shifman:1978bx,Reinders:1984sr}. In sum rule analyses, we consider two-point correlation functions:
\begin{equation}
\Pi(q^2) \, \equiv \, i \int d^4x e^{iqx} \langle 0 | T J(x) J^\dagger (0) | 0 \rangle \, ,
\label{def:pi}
\end{equation}
where $J(x)$ is an interpolating field (current) coupling to a tetraquark state. Here we shall choose the tetraquark currents studied in Sec.~\ref{sec:currents} and Appendix.~\ref{app:othercurrents}. We compute $\Pi(q^2)$ in the operator product expansion (OPE) of QCD up to certain order in the expansion, which is then matched with a hadronic parametrization to extract information about hadron properties. At the hadron level, we express the correlation function in the form of the dispersion relation with a spectral function:
\begin{equation}
\Pi(q^2)={1\over\pi}\int^\infty_{s_<}\frac{{\rm Im} \Pi(s)}{s-q^2-i\varepsilon}ds \, ,
\label{eq:disper}
\end{equation}
where the integration starts from the mass square of all current quarks. The imaginal part of the two-point correlation function is
\begin{eqnarray}
{\rm Im} \Pi(s) & \equiv & \pi \sum_n\delta(s-M^2_n)\langle 0|\eta|n\rangle\langle n|{\eta^\dagger}|0\rangle \, .
\label{eq:rho}
\end{eqnarray}
For the second equation, as usual, we adopt a parametrization of one pole dominance for the ground state $Y$ $\big(\langle 0 | \eta(0) | Y \rangle \equiv f_Y$; where $f_Y$ is the decay constant$\big)$ and a continuum contribution. The sum rule analysis is then performed after the Borel transformation of the two expressions of the correlation function, (\ref{def:pi}) and (\ref{eq:disper})
\begin{equation}
\Pi^{(all)}(M_B^2)\equiv\mathcal{B}_{M_B^2}\Pi(p^2) = {1\over\pi} \int^\infty_{s_<} e^{-s/M_B^2} {\rm Im} \Pi(s) ds \, .
\label{eq:borel}
\end{equation}
Assuming the contribution from the continuum states can be approximated well by the spectral density of OPE above a threshold value $s_0$ (duality), we arrive at the sum rule equation
\begin{eqnarray}
f^2_Y e^{-M_Y^2/M_B^2} &=& \Pi(s_0, M_B^2) \equiv {1\over\pi} \int^{s_0}_{s_<} e^{-s/M_B^2} {\rm Im} \Pi(s) ds
\label{eq:fin} \, .
\end{eqnarray}
Differentiating Eq.~(\ref{eq:fin}) with respect to $1 / M_B^2$ and dividing it by Eq. (\ref{eq:fin}), finally we obtain
\begin{equation}
M^2_Y = \frac{\frac{\partial}{\partial(-1/M_B^2)}\Pi(s_0, M_B^2)}{\Pi(s_0, M_B^2)} \, .
\end{equation}

The tetraquark currents classified in Sec.~\ref{sec:currents} and Appendix.~\ref{app:othercurrents} can couple to mesons that belong to (or partly belong to) the same representation. Here, we assume that the scalar ones belonging to the ``non-exotic'' $[(\mathbf{\bar 3},\mathbf{ 3}) \oplus (\mathbf{3}, \mathbf{\bar 3})]$ chiral multiplets can couple to the light scalar mesons $f_0(500)$, $\kappa(800)$, $a_0(980)$ and $f_0(980)$. Using these currents, we can calculate the mass of the light scalar mesons through the method of QCD sum rule. In the calculations, we assume an ideal mixing. Hence, the mass of the $f_0(500)$ meson is calculated through tetraquark currents
\begin{eqnarray}
\label{eq:sigma}
\sigma_{1+2} &=& {1\over6}\big(\eta_{1}^{\rm S,\mathbb{S}} + \eta_{2}^{\rm S,\mathbb{S}}\big ) +{1\over\sqrt3}\big(\eta_{1,N=8}^{\rm S,\mathbb{O}} + \eta_{2,N=8}^{\rm S,\mathbb{O}}\big) \, ,
\\ \nonumber \sigma_{3} &=& {1\over6} \eta_{3}^{\rm S,\mathbb{S}} + {1\over\sqrt3} \eta_{3,N=8}^{\rm S,\mathbb{O}} \, ,
\end{eqnarray}
whose quark contents are $ud\bar u \bar d$; $\kappa^+(800)$ through
\begin{eqnarray}
\label{eq:kappa}
\kappa_{1+2} &=& {1\over2}\big(\eta_{1,N=4}^{\rm S,\mathbb{O}} + \eta_{2,N=4}^{\rm S,\mathbb{O}} - i \eta_{1,N=5}^{\rm S,\mathbb{O}} - i \eta_{2,N=5}^{\rm S,\mathbb{O}} \big ) \, ,
\\ \nonumber \kappa_{3} &=& {1\over2}\big(\eta_{3,N=4}^{\rm S,\mathbb{O}} - i \eta_{3,N=5}^{\rm S,\mathbb{O}} \big )\, ,
\end{eqnarray}
whose quark contents are $u d \bar d \bar s$; $a_0^+(980)$ through
\begin{eqnarray}
\label{eq:a0}
a0_{1+2} &=& {1\over2}\big(\eta_{1,N=1}^{\rm S,\mathbb{O}} + \eta_{2,N=1}^{\rm S,\mathbb{O}} - i \eta_{1,N=2}^{\rm S,\mathbb{O}} - i \eta_{2,N=2}^{\rm S,\mathbb{O}} \big ) \, ,
\\ \nonumber a0_{3} &=& {1\over2}\big(\eta_{3,N=1}^{\rm S,\mathbb{O}} - i \eta_{3,N=2}^{\rm S,\mathbb{O}} \big )\, ,
\end{eqnarray}
whose quark contents are $u s \bar d \bar s$; $f_0(980)$ through
\begin{eqnarray}
\label{eq:f0}
f0_{1+2} &=& {1\over3\sqrt2}\big(\eta_{1}^{\rm S,\mathbb{S}} + \eta_{2}^{\rm S,\mathbb{S}} \big ) - {1\over\sqrt6}\big( \eta_{1,N=8}^{\rm S,\mathbb{O}} + \eta_{2,N=8}^{\rm S,\mathbb{O}} \big ) \, ,
\\ \nonumber f0_{3} &=& {1\over3\sqrt2} \eta_{3}^{\rm S,\mathbb{S}} - {1\over\sqrt6} \eta_{3,N=8}^{\rm S,\mathbb{O}} \, ,
\end{eqnarray}
whose quark contents are $u s \bar u \bar s + d s \bar d \bar s$.

They lead to the following QCD sum rules where we have computed the operator product expansion up to the eighth dimension:
\begin{eqnarray}
\label{eq:sigma12sumrule}
f^2_{\sigma_{1+2}} e^{-M_{\sigma_{1+2}}^2/M_B^2} &=& \Pi^{\sigma_{1+2}}(s_0, M_B^2)
\\ \nonumber &=& \int^{s_0}_{s_<} e^{-s/M_B^2} ds \times \Big (
\frac{1}{30720 \pi ^6} s^4
+ \frac{\langle g^2 GG \rangle}{3072 \pi ^6}
s^2
\Big ) \, ,
\\ f^2_{\sigma_3} e^{-M_{\sigma_3}^2/M_B^2} &=& \Pi^{\sigma_3}(s_0, M_B^2)
\\ \nonumber &=& \int^{s_0}_{s_<} e^{-s/M_B^2} ds \times \Big ( \frac{1}{1280 \pi ^6} s^4
+ \frac{11 \langle g^2 GG \rangle}{768 \pi ^6}
s^2
\Big ) \, ,
\\ f^2_{\kappa_{1+2}} e^{-M_{\kappa_{1+2}}^2/M_B^2} &=& \Pi^{\kappa_{1+2}}(s_0, M_B^2)
\\ \nonumber &=& \int^{s_0}_{s_<} e^{-s/M_B^2} ds \times \Big (
\frac{1}{30720 \pi ^6} s^4
- \frac{m_s^2}{1536 \pi ^6} s^3
+ \left( \frac{\langle g^2 GG \rangle}{3072 \pi ^6} + \frac{ m_s \langle \bar s s \rangle}{192 \pi ^4} \right) s^2
\\ \nonumber && -\frac{ m_s^2 \langle g^2 GG \rangle}{1024 \pi ^6} s
+ \frac{ m_s \langle g^2 GG \rangle \langle \bar s s \rangle}{768 \pi ^4}
\Big ) \, ,
\\ f^2_{\kappa_3} e^{-M_{\kappa_3}^2/M_B^2} &=& \Pi^{\kappa_3}(s_0, M_B^2)
\\ \nonumber &=& \int^{s_0}_{s_<} e^{-s/M_B^2} ds \times \Big (
\frac{1}{1280 \pi ^6} s^4
- \frac{m_s^2}{64 \pi ^6} s^3
+ \left( \frac{11 \langle g^2 GG \rangle}{768 \pi ^6} + \frac{ m_s \langle \bar s s \rangle}{8 \pi ^4} \right) s^2
\\ \nonumber && -\frac{ 11 m_s^2 \langle g^2 GG \rangle}{256 \pi ^6} s
+ \frac{ 11 m_s \langle g^2 GG \rangle \langle \bar s s \rangle}{192 \pi ^4}
\Big ) \, ,
\\ f^2_{a0_{1+2}} e^{-M_{a0_{1+2}}^2/M_B^2} &=& \Pi^{a0_{1+2}}(s_0, M_B^2)
\\ \nonumber &=& \int^{s_0}_{s_<} e^{-s/M_B^2} ds \times \Big (
\frac{1}{30720 \pi ^6} s^4
- \frac{m_s^2}{768 \pi ^6} s^3
+ \left( \frac{\langle g^2 GG \rangle}{3072 \pi ^6} + \frac{ m_s \langle \bar s s \rangle}{96 \pi ^4} \right) s^2
\\ \nonumber && -\frac{ m_s^2 \langle g^2 GG \rangle}{512 \pi ^6} s
+ \frac{ m_s \langle g^2 GG \rangle \langle \bar s s \rangle}{384 \pi ^4} + \frac{ m_s^2 (4 \langle \bar q q \rangle^2 + \langle \bar s s \rangle^2)}{24 \pi ^2}
\Big ) \, ,
\\ f^2_{a0_3} e^{-M_{a0_3}^2/M_B^2} &=& \Pi^{a0_3}(s_0, M_B^2)
\label{eq:a03sumrule}
\\ \nonumber &=& \int^{s_0}_{s_<} e^{-s/M_B^2} ds \times \Big (
\frac{1}{1280 \pi ^6} s^4
- \frac{m_s^2}{32 \pi ^6} s^3
+ \left( \frac{11 \langle g^2 GG \rangle}{768 \pi ^6} + \frac{ m_s \langle \bar s s \rangle}{4 \pi ^4} \right) s^2
\\ \nonumber && -\frac{ 11 m_s^2 \langle g^2 GG \rangle}{128 \pi ^6} s
+ \frac{ 11 m_s \langle g^2 GG \rangle \langle \bar s s \rangle}{96 \pi ^4} + \frac{ m_s^2 (4 \langle \bar q q \rangle^2 + \langle \bar s s \rangle^2)}{ \pi ^2}
\Big ) \, .
\end{eqnarray}
In this expression we only show terms containing the $strange$ current quark mass up to $m_s^2$, while we keep all terms in the calculations. We also keep the terms containing the $up$ and $down$ current quark masses in the calculations, although they are quite small and give little contribution~\cite{ioffe}.
We note that we do not include high dimension terms which can be important, particulary the tree-level term $\alpha_s \langle \bar q q \rangle^4$~\cite{Oganesian:2005jm,Oganesian:2006ug}.

In Eqs.~(\ref{eq:sigma12sumrule})-(\ref{eq:a03sumrule}), many terms cancelled, including condensates $\langle \bar q q \rangle^2$ and $\langle \bar q q \rangle \langle g \bar q G q \rangle$, which are usually much larger than others. Moreover, Eq.~(\ref{eq:sigma12sumrule}) shows that effects of gluons are significant in the OPE of the $\sigma$ meson since the $up$ and $down$ current quark masses are quite small. The sum rules for $f_0(980)$ are the same as those for $a_0(980)$, and so we obtain the same mass for $a_0(980)$ and $f_0(980)$.

To perform the numerical analysis, we use the following values for the condensates and other parameters, which correspond to the energy scale of 1 GeV~\cite{Yang:1993bp,Hwang:1994vp,Beringer:1900zz,Narison:2002pw,Gimenez:2005nt,Jamin:2002ev,Ioffe:2002be,Ovchinnikov:1988gk,colangelo}:
\begin{eqnarray}
\nonumber &&\langle\bar qq \rangle=-(0.240 \pm 0.010)^3 \mbox{ GeV}^3\, ,
\\
\nonumber &&\langle\bar ss\rangle=-(0.8\pm 0.1)\times(0.240 \pm 0.010)^3 \mbox{
GeV}^3\, ,
\\
\nonumber &&\langle g_s^2GG\rangle =(0.48\pm 0.14) \mbox{ GeV}^4\, ,
\\ && m_u = 2.90 \pm 0.20 \mbox{ MeV}\, ,m_d = 6.35 \pm 0.27 \mbox{ MeV}\, ,
\\
\label{condensates} \nonumber &&m_s=125 \pm 20 \mbox{ MeV}\, ,
\\
\nonumber && \langle g_s\bar q\sigma G
q\rangle=-M_0^2\times\langle\bar qq\rangle\, ,
\\
\nonumber &&M_0^2=(0.8\pm0.2)\mbox{ GeV}^2\, .
\end{eqnarray}
As usual we assume the vacuum saturation for higher dimensional operators such as $\langle 0 | \bar q q \bar q q |0\rangle \sim \langle 0 | \bar q q |0\rangle \langle 0|\bar q q |0\rangle$. There is a minus sign in the definition of the mixed condensate
$\langle g_s\bar q\sigma G q\rangle$, which is different from that
used in some other QCD sum rule studies. This difference just comes
from the definition of coupling constant
$g_s$~\cite{Yang:1993bp,Hwang:1994vp}.

\begin{figure}[hbt]
\begin{center}
\scalebox{0.6}{\includegraphics{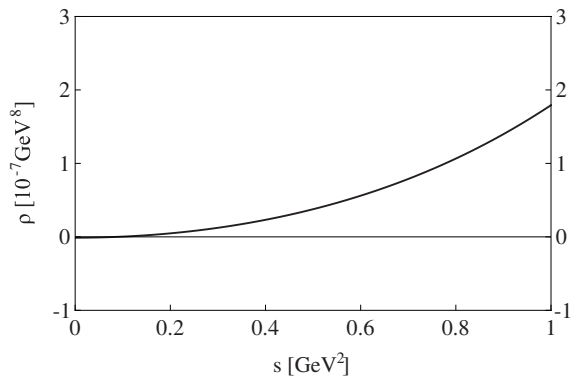}}
\caption{Spectral density for the current $\sigma_{1+2}$ as a function of the energy $s$.}
\label{fig:rhosigma12}
\end{center}
\end{figure}

\begin{figure}[hbt]
\begin{center}
\scalebox{0.6}{\includegraphics{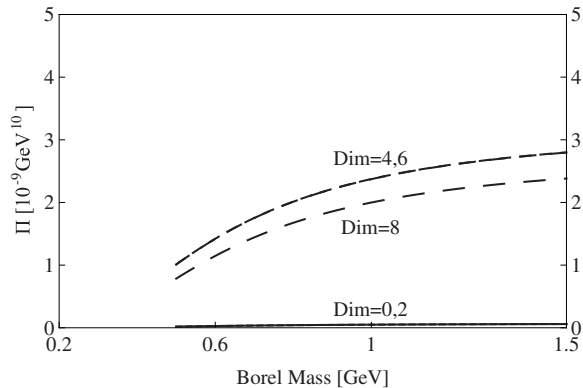}}
\caption{Various
contributions to the correlation function of the current $\sigma_{1+2}$ as functions of the Borel mass $M_B$ in units of GeV$^{10}$ at $s_0$ = 0.4 GeV$^2$. The labels indicate the dimension up
to which the OPE terms are included.} \label{fig:pi}
\end{center}
\end{figure}

We use the current $\sigma_{1+2}$ as an example. First we extract its spectral density from Eq.~(\ref{eq:sigma12sumrule}) and show it in Fig.~\ref{fig:rhosigma12} as a function of the energy $s$. It is almost positive definite, and so we can use it to perform QCD sum rule analyses. Then we need to study its OPE convergence. The Borel transformed correlation function of the current $\sigma_{1+2}$ is shown in Fig.~\ref{fig:pi}, when we take $s_0=0.4$ GeV$^2$. We can clearly see that the $D=4$ terms give large contributions, and the convergence is good in the region $M_B>0.5$ GeV, where OPEs are reliable. To fix the upper bound of the Borel window, we need to use the pole contribution, defined as the pole part divided by the sum of the pole and the continuum parts in the two-point correlation function Eq.~(\ref{def:pi}):
\begin{equation}\label{eq_pole}
\mbox{Pole contribution} \equiv \frac{ \int^{s_0}_0 e^{-s/M_B^2}
\rho(s)ds }{\int^\infty_0 e^{-s/M_B^2} \rho(s)ds}\, .
\end{equation}
It nearly vanishes for $\kappa(800)$ meson when using $\kappa_{1+2}$, as shown in Table~\ref{table_pole}. For the $f_0(500)$ meson it is also not large. This suggests that the two-meson continuum contributes significantly. Only for the $a_0(980)$ and $f_0(980)$ mesons it is acceptable. Mathematically, this is because the continuum term is growing as $s^4$ and the condensates $\langle \bar q q \rangle^2$ and $\langle \bar q q \rangle \langle g \bar q G q \rangle$ cancelled.

\begin{figure}[hbt]
\begin{center}
\scalebox{0.45}{\includegraphics{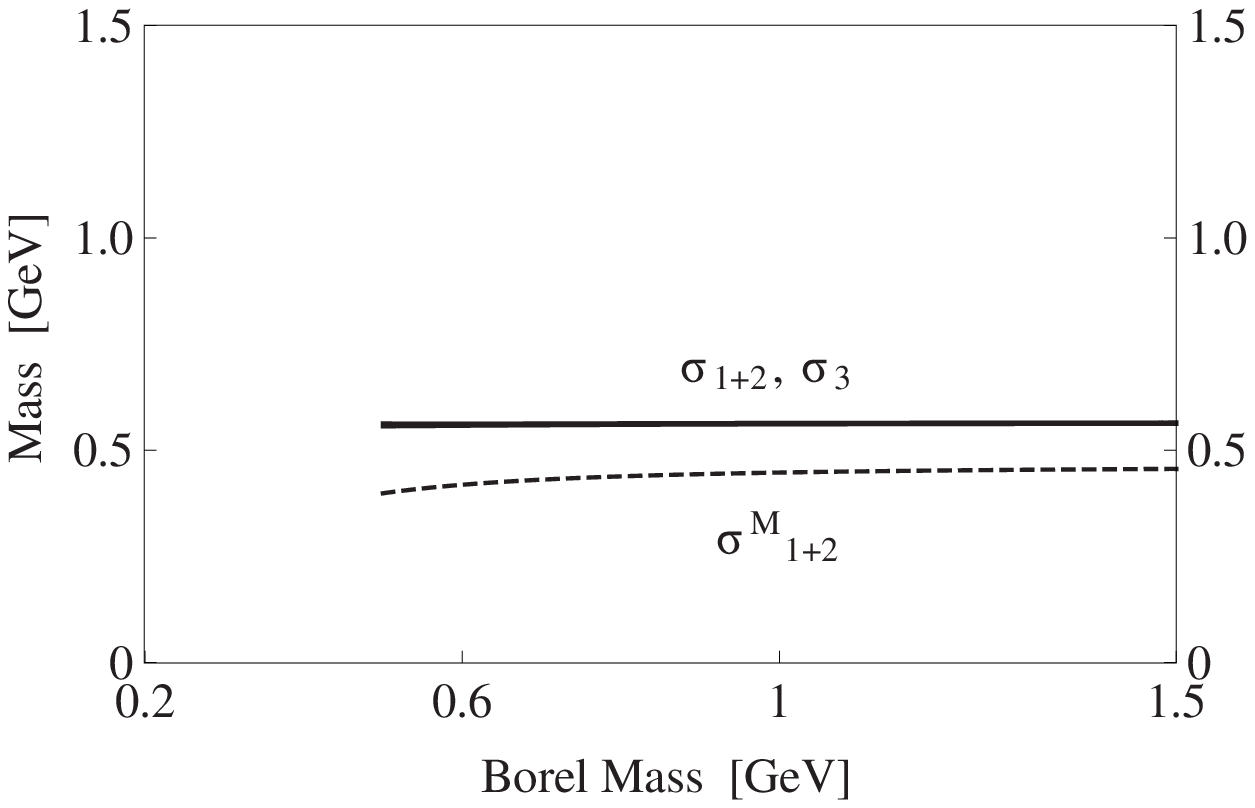}}
\scalebox{0.45}{\includegraphics{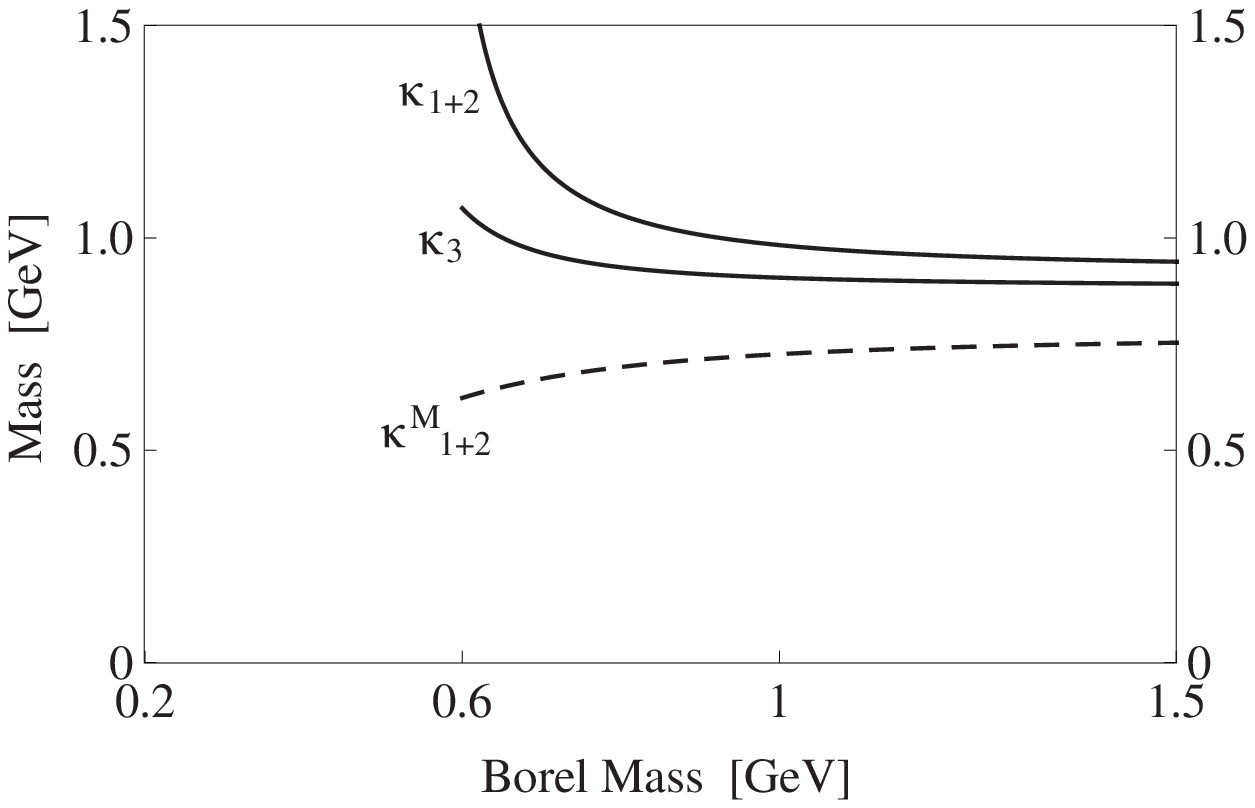}}
\scalebox{0.45}{\includegraphics{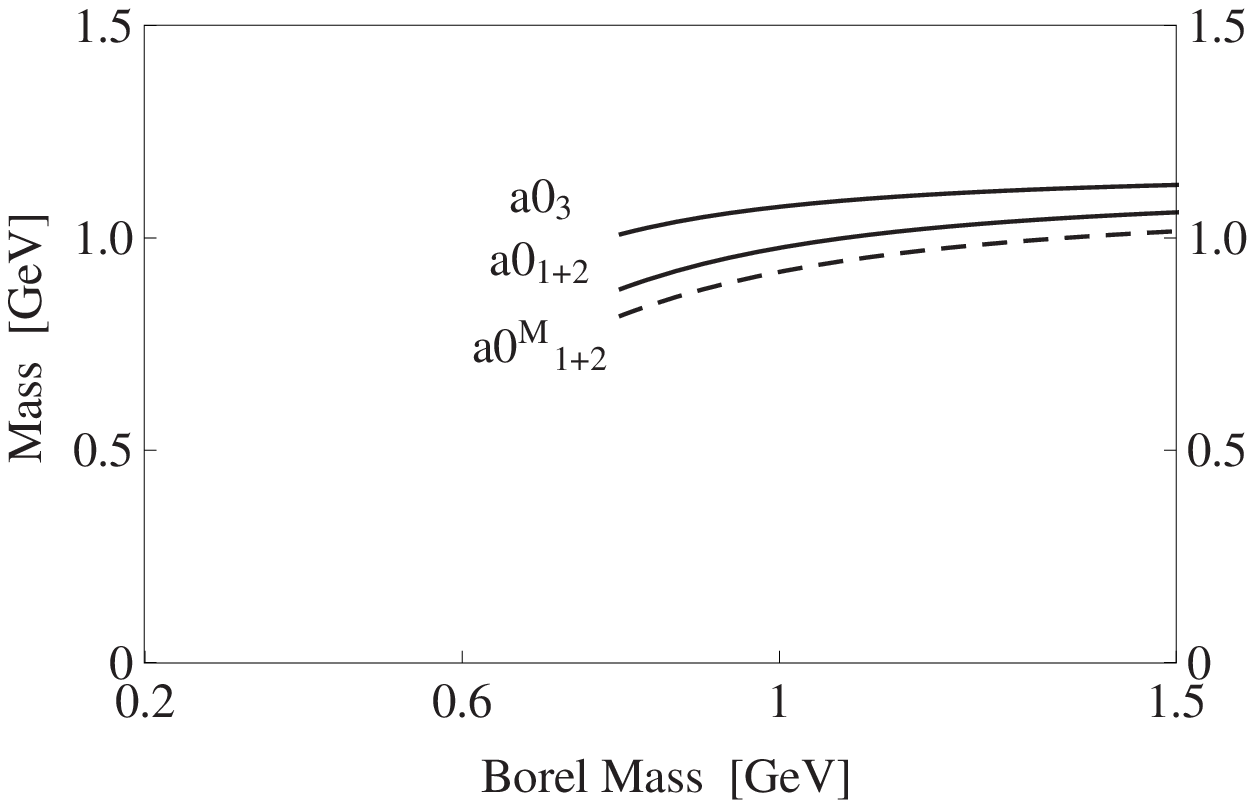}}
\caption{Mass of light scalar mesons $f_0(500)$, $\kappa(800)$ and $a_0(980)$ ($f_0(980)$) as functions of Borel mass $M_B$. The threshold values $s_0$ for $f_0(500)$, $\kappa(800)$ and $a_0(980)$ ($f_0(980)$) are chosen to be $0.40$, $0.90$, $1.80$ GeV$^2$, respectively. The solid curves are obtained using the tetraquark currents $\sigma_{1+2}$, $\sigma_{3}$, $\kappa_{1+2}$, $\kappa_{3}$, $a0_{1+2}$ and $a0_{3}$ (see definitions in Eqs.~(\ref{eq:sigma}), (\ref{eq:kappa}), (\ref{eq:a0}) and (\ref{eq:f0})), while the dashed curves are obtained using the modified currents $\sigma^{mod}_{1+2}$, $\kappa^{mod}_{1+2}$ and $a0^{mod}_{1+2}$ (see definitions in Eqs.~(\ref{eq:modification})). The results for $f_0(980)$ are the same as those for $a_0(980)$.}
\label{fig:mass}
\end{center}
\end{figure}

\begin{figure}[hbt]
\begin{center}
\scalebox{0.45}{\includegraphics{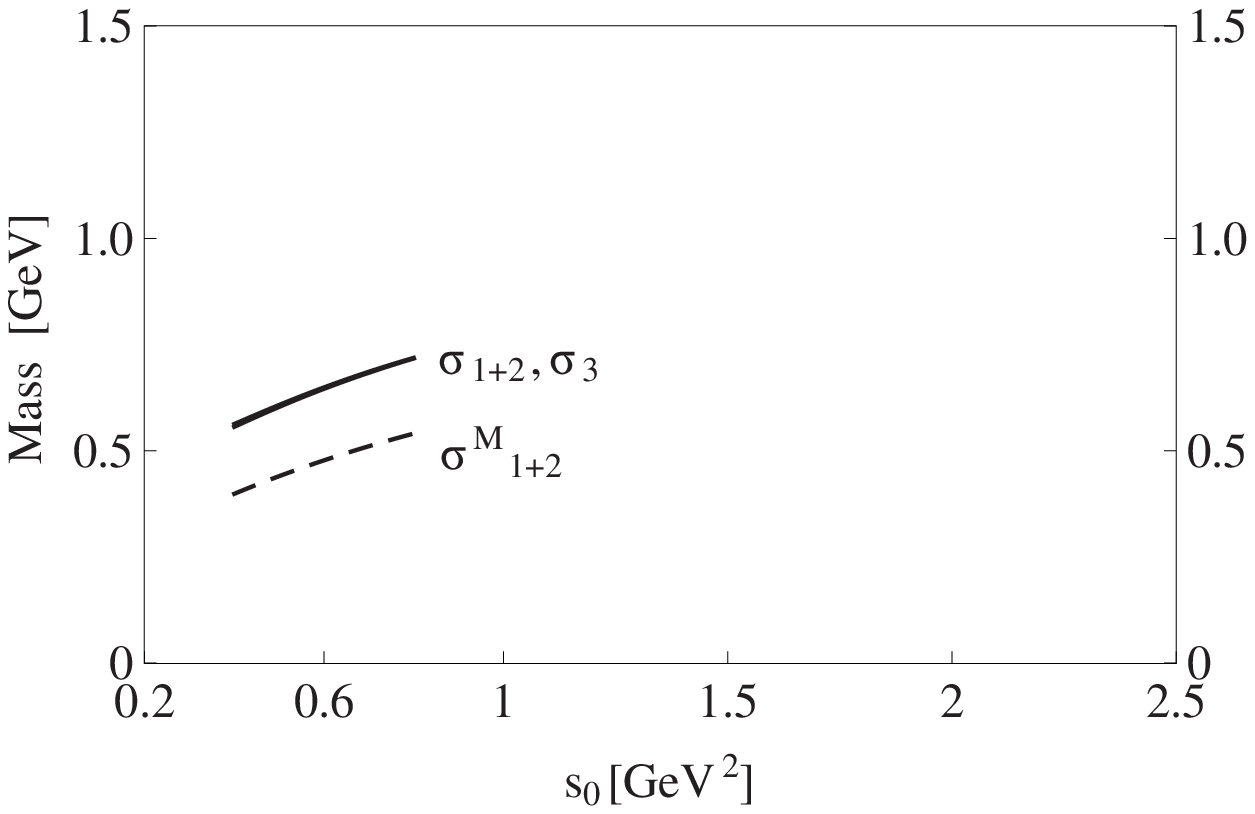}}
\scalebox{0.45}{\includegraphics{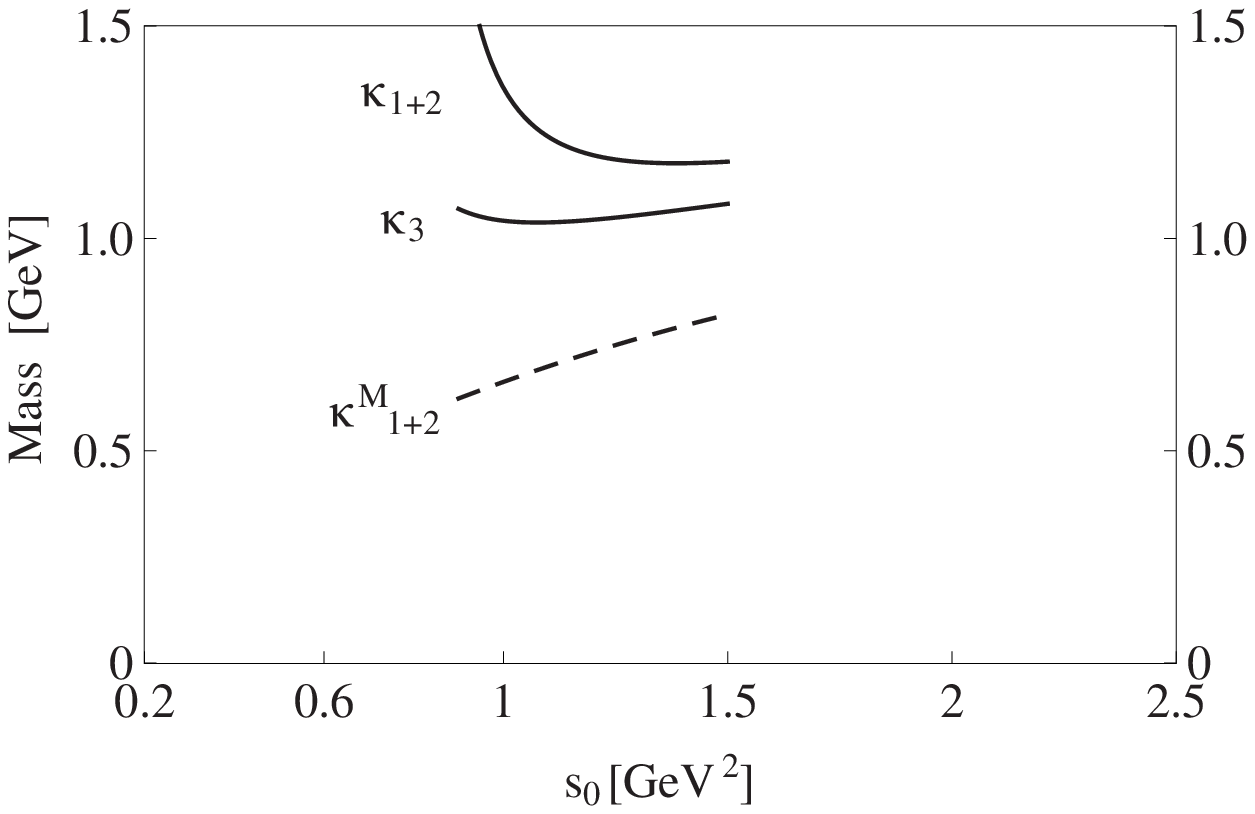}}
\scalebox{0.45}{\includegraphics{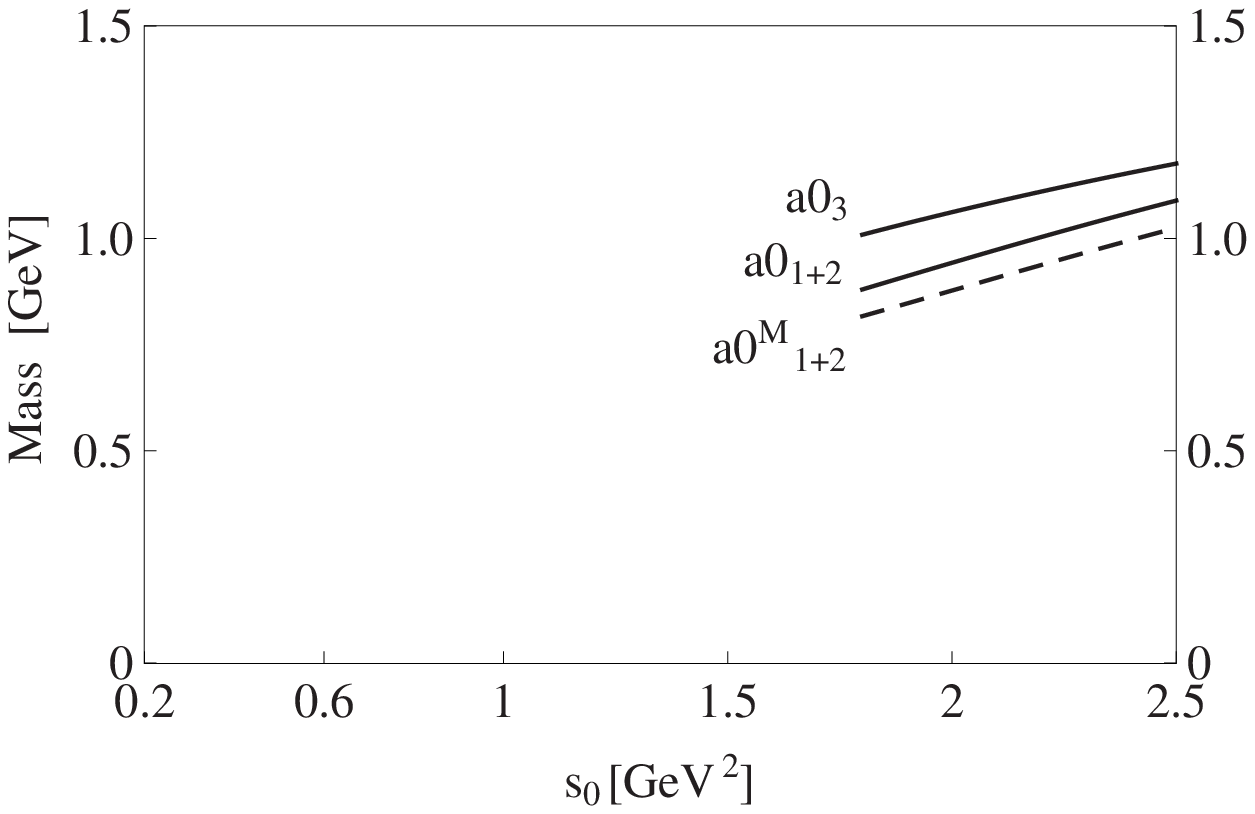}}
\caption{Mass of light scalar mesons $f_0(500)$, $\kappa(800)$ and $a_0(980)$ ($f_0(980)$) as functions of the threshold value $s_0$. The Borel Mass $M_B$ for $f_0(500)$, $\kappa(800)$ and $a_0(980)$ ($f_0(980)$) are chosen to be $0.50$, $0.60$, $0.80$ GeV$^2$, respectively. The solid curves are obtained using the tetraquark currents $\sigma_{1+2}$, $\sigma_{3}$, $\kappa_{1+2}$, $\kappa_{3}$, $a0_{1+2}$ and $a0_{3}$ (see definitions in Eqs.~(\ref{eq:sigma}), (\ref{eq:kappa}), (\ref{eq:a0}) and (\ref{eq:f0})), while the dashed curves are obtained using the modified currents $\sigma^{mod}_{1+2}$, $\kappa^{mod}_{1+2}$ and $a0^{mod}_{1+2}$ (see definitions in Eqs.~(\ref{eq:modification})). The results for $f_0(980)$ are the same as those for $a_0(980)$.}
\label{fig:masss0}
\end{center}
\end{figure}

We show masses of light scalar mesons as functions of the Borel mass $M_B$ and the threshold value $s_0$ in Fig.~\ref{fig:mass} and Fig.~\ref{fig:masss0}, using solid curves.  The masses of $\sigma$, $\kappa$ and $a_0$ ($f_0$) are around 600 MeV, 900 MeV and 1100 MeV, respectively. However, these results much depend on threshold value $s_0$, especially for $\sigma$ and $a_0$, once more suggesting the contribution of meson-meson continuum can not be neglected. In such cases, the use of local quark-hadron duality with one resonance approximation is not valid.

In order to use the quark-hadron duality and obtain more reliable QCD sum rules, we should try to increase the pole contribution. This can be done by slightly changing the mixing parameters of currents $\eta_{1+2}$, which have the antisymmetric color structure $\mathbf{\bar 3} \otimes \mathbf {3}$ and color interactions between quarks and antiquarks are repulsive (the details expressions are similar to Eqs.~(\ref{eq:sigma}), (\ref{eq:kappa}), (\ref{eq:a0}) and (\ref{eq:f0})):
\begin{eqnarray}
\label{eq:modification}
\nonumber \sigma_{1+2} \sim \eta^{{\rm S},{\mathbb S}({\mathbb O})}_{1} + \eta^{{\rm S},{\mathbb S}({\mathbb O})}_{2} &\longrightarrow& \sigma^{mod}_{1+2} \sim 0.99 \times \eta^{{\rm S},{\mathbb S}({\mathbb O})}_{1} + \eta^{{\rm S},{\mathbb S}({\mathbb O})}_{2} \, ,
\\ \kappa_{1+2} \sim \eta^{{\rm S},{\mathbb S}({\mathbb O})}_{1} + \eta^{{\rm S},{\mathbb S}({\mathbb O})}_{2} &\longrightarrow& \kappa^{mod}_{1+2} \sim 0.99 \times \eta^{{\rm S},{\mathbb S}({\mathbb O})}_{1} + \eta^{{\rm S},{\mathbb S}({\mathbb O})}_{2} \, ,
\\ \nonumber a0_{1+2} \sim \eta^{{\rm S},{\mathbb S}({\mathbb O})}_{1} + \eta^{{\rm S},{\mathbb S}({\mathbb O})}_{2} &\longrightarrow& a0^{mod}_{1+2} \sim 0.99 \times \eta^{{\rm S},{\mathbb S}({\mathbb O})}_{1} + \eta^{{\rm S},{\mathbb S}({\mathbb O})}_{2} \, ,
\\ \nonumber f0_{1+2} \sim \eta^{{\rm S},{\mathbb S}({\mathbb O})}_{1} + \eta^{{\rm S},{\mathbb S}({\mathbb O})}_{2} &\longrightarrow& f0^{mod}_{1+2} \sim 0.99 \times \eta^{{\rm S},{\mathbb S}({\mathbb O})}_{1} + \eta^{{\rm S},{\mathbb S}({\mathbb O})}_{2} \, ,
\end{eqnarray}
We note that doing this we introduce a few $[({\mathbf 8},{\mathbf 1}) \oplus ({\mathbf 1},{\mathbf 8})]$ components, which are still ``non-exotic''. Mathematically, the condensates $\langle \bar q q \rangle^2$ and $\langle \bar q q \rangle \langle g \bar q G q \rangle$ appear and contribute, although the mixing parameters are only slightly modified.

\begin{figure}[hbt]
\begin{center}
\scalebox{0.6}{\includegraphics{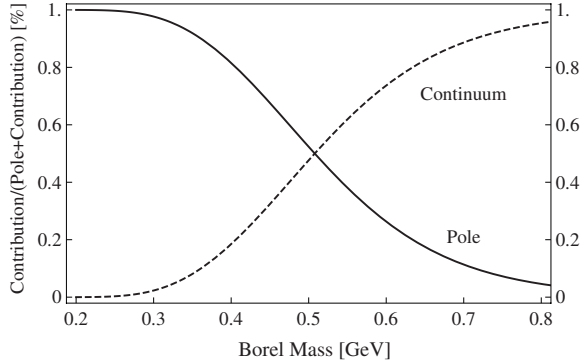}}
\caption{The solid curve shows the pole contribution and the dashed curve shows the continuum contribution ($= 1 - $pole contribution).} \label{fig:pole}
\end{center}
\end{figure}

Still we use the current $\sigma^{mod}_{1+2}$ as an example. The comparison between pole and continuum contributions for $s_0 = 0.4$ GeV$^2$ is shown in Fig.~\ref{fig:pole}~\cite{Matheus:2007ta,Zhang:2012gp,Zhang:2012in}. We find that the pole contribution is significantly increased to around 50\% when $M_B$ is around $0.5$ MeV, but it decreases very quickly as the Borel mass increases. Therefore, we obtain a very narrow Borel window around $M_B \sim 0.5$ GeV.

Using the modified currents listed in Eqs.~(\ref{eq:modification}) we calculate masses of light scalar mesons. The results are shown in Fig.~\ref{fig:mass} as functions of Borel Mass $M_B$, but using dashed curves. The masses of $\sigma$, $\kappa$ and $a_0$ ($f_0$) are around $500$ MeV, $700$ MeV and $900$ MeV, respectively, better consistent with the experimental results. The pole contributions are significantly increased to be around 50\% for $f_0(500)$ and $\kappa(800)$, as shown in Table~\ref{table_pole}. Since there are still 50\% continuum, and considering nearly all the continuum comes from the $\pi$-$\pi$ contribution, the $\sigma$ meson is probably still contributed significantly by its underlying $\pi$-$\pi$ continuum.

We have also studied the threshold value $s_0$ dependence. The results are shown in Fig.~\ref{fig:masss0}, using dashed lines. We can see that $s_0$ dependence is still significant suggesting the contribution of meson-meson continuum can not be neglected. To solve this problem, we shall use only the connected parts of the two-point correlation function to perform the QCD sum rule analysis in the next section~\cite{Weinberg:2013cfa,coleman,page}.

\begin{table}[hbt]
\caption{Pole contributions of various currents.}
\begin{center}
\begin{tabular}{c||ccc|ccc|ccc}
\hline\hline & $\sigma_{1+2}$ & $\sigma^{mod}_{1+2}$ & $\sigma_3$ & $\kappa_{1+2}$ & $\kappa^{mod}_{1+2}$ & $\kappa_3$ & $a0_{1+2}$ & $a0^{mod}_{1+2}$ & $a0_3$
\\ \hline\hline $M_B$ (GeV) & 0.5 & 0.5 & 0.5 & 0.6 & 0.6 & 0.6 & 0.8 & 0.8 & 0.8
\\ \hline $s_0$ (GeV$^2$) & 0.4 & 0.4 & 0.4 & 0.9 & 0.9 & 0.9 & 1.8 & 1.8 & 1.8
\\ \hline Pole (\%) & 15 & 52 & 16 & 5 & 45 & 18 & 35 & 39 & 39
\\ \hline\hline
\end{tabular}\label{table_pole}
\end{center}
\end{table}

To investigate this meson-meson continuum we simply use the theory of relativity to estimate how far at most the two final pseudoscalar mesons can travel away from each other in the lifetimes of the initial light scalar mesons. From this distance we shall clearly see the difficulty to separate the meson-meson continuum. Our assumptions are very simple and straightforward: the initial state, an unstable particle $X$, is at rest in the beginning; it has mass $m_X$ and decay width $\Gamma_X$; it decays into two particles $A$ and $B$, having masses $m_A$ and $m_B$, respectively; when $X$ is decaying into $A$ and $B$, the mass difference between the initial and final states, $m_X-m_A-m_B$, is totally and immediately transferred into kinetic energies of $A$ and $B$; this makes they have speeds $v_A$ and $v_B$, in opposite direction. This process can be easily described using the following equations:
\begin{eqnarray}
 {m_A c^2\over\sqrt{1-v_A^2/c^2}} + {m_B c^2\over\sqrt{1-v_B^2/c^2}} &=& m_X c^2\, ,
 \label{eq:motion1}
\\ {m_A\over\sqrt{1-v_A^2/c^2}} v_A + {m_B\over\sqrt{1-v_B^2/c^2}} v_B &=& 0 \, .
\label{eq:motion2}
\end{eqnarray}
Here $c$ is the speed of light. The quantity $d_X\equiv{\hbar \over \Gamma_X} (|v_A| + |v_B|)$  is just the farthest distance that $A$ and $B$ can travel away from each other in the half-life of $X$. We can use the uncertainty principle $\Delta x \Delta p \geq \hbar /2$ to estimate the theoretical uncertainty of $d_X$:
\begin{eqnarray}
\Delta d_X = {\hbar \over 2} {1 \over {m_A\over\sqrt{1-v_A^2/c^2}} |v_A| + {m_B\over\sqrt{1-v_B^2/c^2}} |v_B|} \, ,
\label{def:Deltadx}
\end{eqnarray}
Using these equations we obtain: $d_{f_0(500)\rightarrow \pi \pi} = 0.41-0.85$ fm, $d_{\kappa(800)\rightarrow \pi K} = 0.19-0.37$ fm, $d_{a_0(980)\rightarrow \pi \eta} = 2.76-5.69$ fm and $d_{f_0(980)\rightarrow \pi \pi} = 3.79-9.50$ fm. The relevant theoretical error bars from the uncertainty principle are $\Delta d_{f_0(500)} = 0.34$ fm, $\Delta d_{\kappa(800)} = 0.76$ fm, $\Delta d_{a_0(980)} = 0.16$ fm and $\Delta d_{f_0(980)} = 0.11$ fm. Moreover, in order to obtain these results we have assumed that the mass difference is totally and immediately transferred into kinetic energies, and so the actual distance that the two final states travel away from each other in the half-life of the initial unstable hadron can be even smaller. Therefore, in the cases of $f_0(500)$ and $\kappa(800)$, if the initial hadron is spherical in the beginning and the two final hadrons are both spherical in the end, the two final states may not separate geometrically even after the whole decay process. From this effect we clearly see the meson-meson continuum contributes much in the cases of $f_0(500)$ and $\kappa(800)$, and it is quite difficulty to separate the meson-meson continuum. We note that this distance can be estimated for other hadrons, and this problem is not only for light scalar mesons.

\section{QCD Sum Rule using only Connected Parts}
\label{sec:connected}

In this section we use only the connected parts of the two-point correlation function to perform the QCD sum rule analysis. Using the large $N_c$ approximation, S.~Weinberg suggested in his recent reference~\cite{Weinberg:2013cfa,coleman,page}: ``A one tetraquark pole can only appear in the final, connected, term'':
\begin{eqnarray}
\label{eq:weinberg}
\nonumber \left\langle {\cal Q}(x){\cal Q}(y)\right\rangle_0 &=&
\sum_{ijkl}C_{ij}C_{kl}\Bigg[\left\langle {\cal B}_i(x){\cal B}_k(y)\right\rangle_0\left\langle {\cal B}_j(x){\cal B}_l(y)\right\rangle_0
+ \left\langle {\cal B}_i(x){\cal B}_j(x) {\cal B}_k(y){\cal B}_l(y)\right\rangle_{{\rm conn}}\Bigg] \, ,
\end{eqnarray}
where ${\cal Q}(x)$ is a color-neutral operator, i.e., a tetraquark current. Using Fierz transformation, it can be written in the form (see Appendix.~\ref{app:mesoniccurrents} for details):
\begin{equation}
{\cal Q}(x)=\sum_{ij}C_{ij}{\cal B}_i(x){\cal B}_j(x) \, ,
\end{equation}
and ${\cal B}_i(x)$ are color-neutral quark bilinears:
\begin{equation}
{\cal B}_i(x)=\sum_a\overline{q^A_a (x)}\Gamma_i q^B_a(x) \, ,
\end{equation}

In the previous section we have included both the connected parts (the second term in Eq.~(\ref{eq:weinberg})) and the disconnected parts (the first term in Eq.~(\ref{eq:weinberg})) to perform the QCD sum rule analysis. In this section we shall use only the connected parts. We shall use the same tetraquark currents. Although these currents are constructed using diquark and antidiquark fields, we do not need to change them to meson-meson form. We can simply select the connected parts in the contracted two-point correlation function. Take the current
\begin{equation}
J(x) = [u_a^T C \gamma_5 d_b][\bar{u}_a \gamma_5 C \bar{d}_b^T - \bar{u}_b \gamma_5 C \bar{d}_a^T] \, ,
\end{equation}
as an example. We use $S_{ab}^q(x)$ to denote the quark propagator ($q=u$ for up quark, and $q=d$ for down quark), and the contracted two-point correlation function is
\begin{eqnarray}
\nonumber \langle0|J(x)J^\dagger(0)|0\rangle &=& {\rm Tr}[\mathbb{C} S^{uT}_{a1a2}(x) \mathbb{C} \gamma_5 S^d_{b1b2}(x) \gamma_5] \times {\rm Tr}[S^u_{a2a1}(-x)\gamma_5 \mathbb{C} S^{dT}_{b2b1}(-x) \mathbb{C} \gamma_5]
\\ \nonumber && - {\rm Tr}[\mathbb{C} S^{uT}_{a1a2}(x) \mathbb{C} \gamma_5 S^d_{b1b2}(x) \gamma_5] \times {\rm Tr}[S^u_{b2a1}(-x)\gamma_5 \mathbb{C} S^{dT}_{a2b1}(-x) \mathbb{C} \gamma_5]
\\ \nonumber && - {\rm Tr}[\mathbb{C} S^{uT}_{a1a2}(x) \mathbb{C} \gamma_5 S^d_{b1b2}(x) \gamma_5] \times {\rm Tr}[S^u_{a2b1}(-x)\gamma_5 \mathbb{C} S^{dT}_{b2a1}(-x) \mathbb{C} \gamma_5]
\\ \nonumber && + {\rm Tr}[\mathbb{C} S^{uT}_{a1a2}(x) \mathbb{C} \gamma_5 S^d_{b1b2}(x) \gamma_5] \times {\rm Tr}[S^u_{b2b1}(-x)\gamma_5 \mathbb{C} S^{dT}_{a2a1}(-x) \mathbb{C} \gamma_5] \, ,
\end{eqnarray}
where $a1,a2,b1,b2$ are color indices. Its connected parts are just
\begin{eqnarray}
\nonumber \langle0|J(x)J^\dagger(0)|0\rangle_{\rm conn} &=& - {\rm Tr}[\mathbb{C} S^{uT}_{a1a2}(x) \mathbb{C} \gamma_5 S^d_{b1b2}(x) \gamma_5] \times {\rm Tr}[S^u_{b2a1}(-x)\gamma_5 \mathbb{C} S^{dT}_{a2b1}(-x) \mathbb{C} \gamma_5]
\\ \nonumber && - {\rm Tr}[\mathbb{C} S^{uT}_{a1a2}(x) \mathbb{C} \gamma_5 S^d_{b1b2}(x) \gamma_5] \times {\rm Tr}[S^u_{a2b1}(-x)\gamma_5 \mathbb{C} S^{dT}_{b2a1}(-x) \mathbb{C} \gamma_5] \, .
\end{eqnarray}
The tetraquark currents Eq.~(\ref{eq:sigma}) to (\ref{eq:f0}) lead to the following ``connected'' spectral densities:
\begin{eqnarray}
\label{eq:sigma12sumruleconnect}
f^2_{\sigma_{1+2}} e^{-M_{\sigma_{1+2}}^2/M_B^2} &=& \Pi^{\sigma_{1+2}}(s_0, M_B^2)
\\ \nonumber &=& \int^{s_0}_{s_<} e^{-s/M_B^2} ds \times \Big (
- \frac{1}{61440 \pi^6} s^4
+ \frac{\langle g^2 GG \rangle}{3072 \pi ^6}
\Big ) \, ,
\label{eq:sigma3sumruleconnect}
\\ f^2_{\sigma_3} e^{-M_{\sigma_3}^2/M_B^2} &=& \Pi^{\sigma_3}(s_0, M_B^2)
\\ \nonumber &=& \int^{s_0}_{s_<} e^{-s/M_B^2} ds \times \Big (
\frac{1}{5120 \pi^6} s^4
+ \frac{5 \langle g^2 GG \rangle}{768 \pi ^6}
\Big ) \, ,
\label{eq:kappa12sumruleconnect}
\\ f^2_{\kappa_{1+2}} e^{-M_{\kappa_{1+2}}^2/M_B^2} &=& \Pi^{\kappa_{1+2}}(s_0, M_B^2)
\\ \nonumber &=& \int^{s_0}_{s_<} e^{-s/M_B^2} ds \times \Big (
- \frac{1}{61440 \pi ^6} s^4
+ \frac{m_s^2}{3072 \pi ^6} s^3
\\ \nonumber && + \left( \frac{\langle g^2 GG \rangle}{3072 \pi ^6} - \frac{ m_s \langle \bar s s \rangle}{384 \pi ^4} \right) s^2
-\frac{ m_s^2 \langle g^2 GG \rangle}{1024 \pi ^6} s
+ \frac{ m_s \langle g^2 GG \rangle \langle \bar s s \rangle}{768 \pi ^4}
\Big ) \, ,
\label{eq:kappa3sumruleconnect}
\\ f^2_{\kappa_3} e^{-M_{\kappa_3}^2/M_B^2} &=& \Pi^{\kappa_3}(s_0, M_B^2)
\\ \nonumber &=& \int^{s_0}_{s_<} e^{-s/M_B^2} ds \times \Big (
\frac{1}{5120 \pi ^6} s^4
- \frac{m_s^2}{256 \pi ^6} s^3
+ \left( \frac{5 \langle g^2 GG \rangle}{768 \pi ^6} + \frac{ m_s \langle \bar s s \rangle}{32 \pi ^4} \right) s^2
\\ \nonumber && -\frac{ 5 m_s^2 \langle g^2 GG \rangle}{256 \pi ^6} s
+ \frac{ 5 m_s \langle g^2 GG \rangle \langle \bar s s \rangle}{192 \pi ^4}
\Big ) \, ,
\label{eq:a012sumruleconnect}
\\ f^2_{a0_{1+2}} e^{-M_{a0_{1+2}}^2/M_B^2} &=& \Pi^{a0_{1+2}}(s_0, M_B^2)
\\ \nonumber &=& \int^{s_0}_{s_<} e^{-s/M_B^2} ds \times \Big (
- \frac{1}{61440 \pi^6} s^4
+ \frac{m_s^2}{1536 \pi^6} s^3
\\ \nonumber && + \left( \frac{\langle g^2 GG \rangle}{3072 \pi ^6} - \frac{ m_s \langle \bar s s \rangle}{192 \pi ^4} \right) s^2
-\frac{ m_s^2 \langle g^2 GG \rangle}{512 \pi ^6} s
\\ \nonumber && + \frac{ m_s \langle g^2 GG \rangle \langle \bar s s \rangle}{384 \pi ^4} - \frac{ m_s^2 (4 \langle \bar q q \rangle^2 + \langle \bar s s \rangle^2)}{48 \pi ^2}
\Big ) \, ,
\\ f^2_{a0_3} e^{-M_{a0_3}^2/M_B^2} &=& \Pi^{a0_3}(s_0, M_B^2)
\label{eq:a03sumruleconnect}
\\ \nonumber &=& \int^{s_0}_{s_<} e^{-s/M_B^2} ds \times \Big (
\frac{1}{5120 \pi ^6} s^4
- \frac{m_s^2}{128 \pi ^6} s^3
+ \left( \frac{5 \langle g^2 GG \rangle}{768 \pi ^6} + \frac{ m_s \langle \bar s s \rangle}{16 \pi ^4} \right) s^2
\\ \nonumber && -\frac{ 5 m_s^2 \langle g^2 GG \rangle}{128 \pi ^6} s
+ \frac{ 5 m_s \langle g^2 GG \rangle \langle \bar s s \rangle}{96 \pi ^4} + \frac{ m_s^2 (4 \langle \bar q q \rangle^2 + \langle \bar s s \rangle^2)}{ 4 \pi ^2}
\Big ) \, .
\end{eqnarray}
The sum rules Eqs.~(\ref{eq:sigma3sumruleconnect}), (\ref{eq:kappa3sumruleconnect}) and (\ref{eq:a03sumruleconnect}) using tetraquark currents $\sigma_3$, $\kappa_3$ and $a0_3$ ($f0_3$) do not change significantly, i.e., the connected and disconnected parts lead to similar results. This suggests that the meson-meson contribution is significant in both the connected and disconnected parts of these currents. We note that they have the symmetry color structure ${\mathbf 6} \otimes \mathbf {\bar 6}$, where color interactions between quarks and antiquarks are repulsive.

\begin{figure}[hbt]
\begin{center}
\scalebox{0.45}{\includegraphics{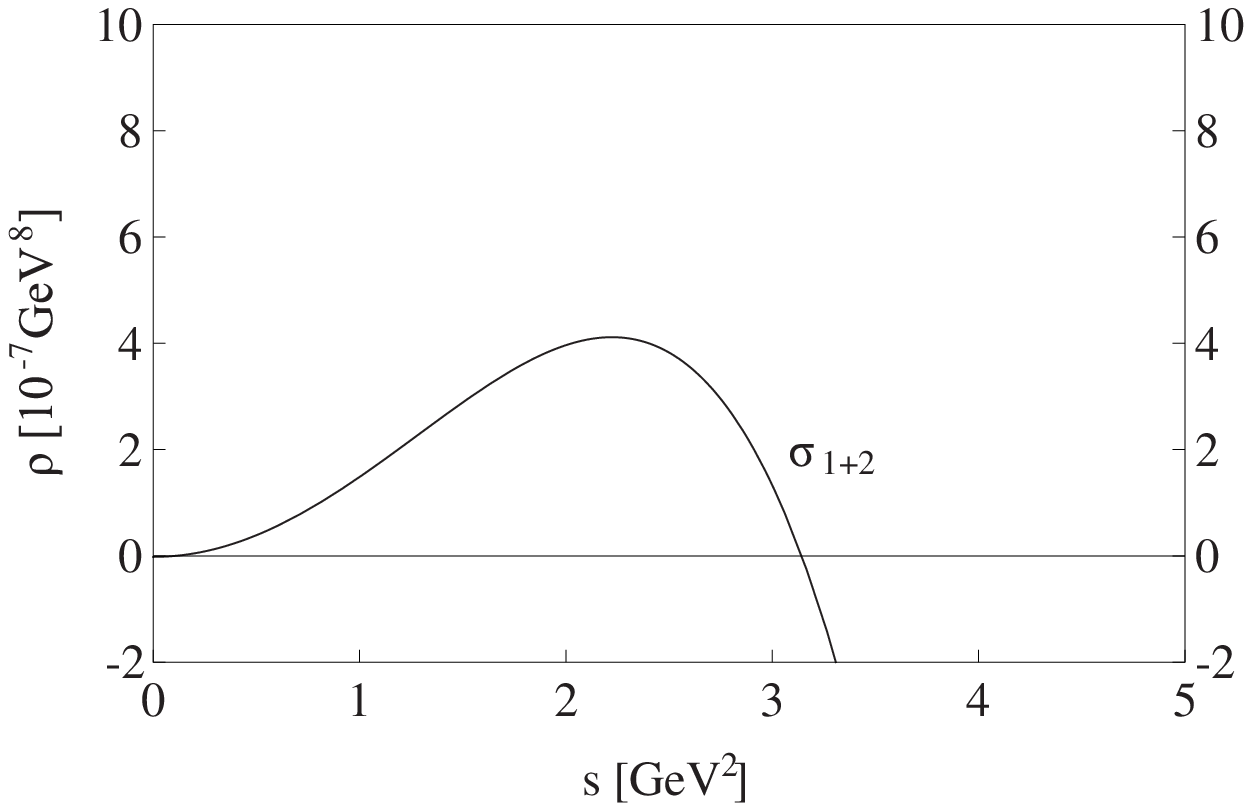}}
\scalebox{0.45}{\includegraphics{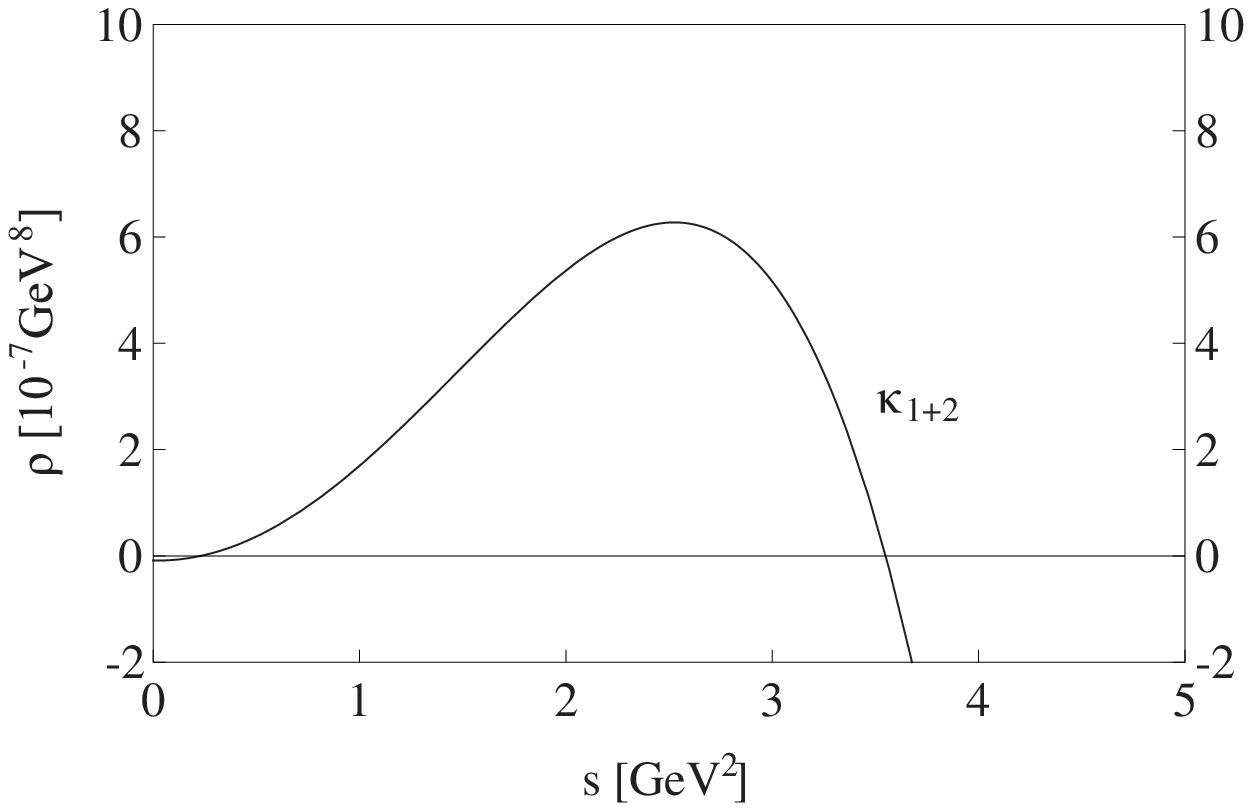}}
\scalebox{0.45}{\includegraphics{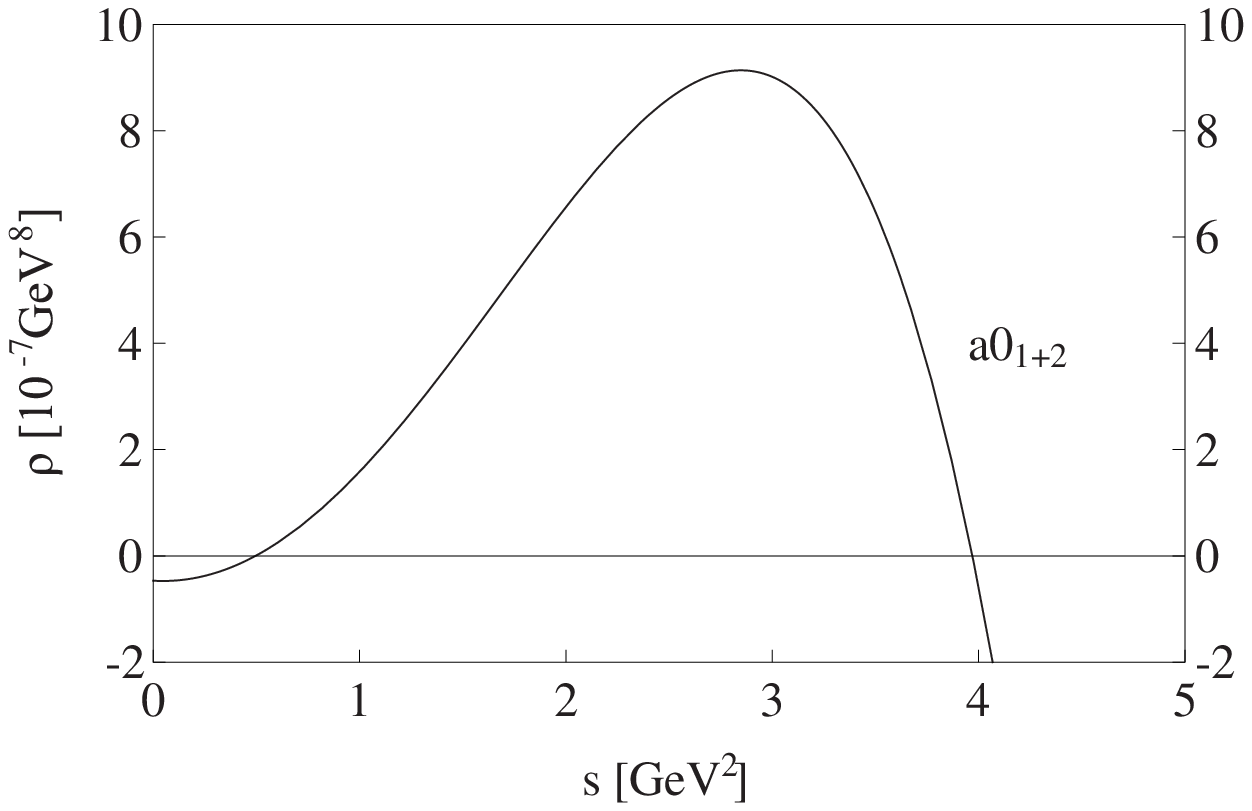}}
\caption{Spectral densities of light scalar mesons $f_0(500)$, $\kappa(800)$ and $a_0(980)$ ($f_0(980)$) as functions of the energy $s$, where only the connected parts are taken into account.}
\label{fig:rho}
\end{center}
\end{figure}

The sum rules Eqs.~(\ref{eq:sigma12sumruleconnect}), (\ref{eq:kappa12sumruleconnect}) and (\ref{eq:a012sumruleconnect}) do change significantly.Although the continuum term proportional to $s^4$ is negative, the spectral densities are positive in our working region $s \sim 1$ GeV$^2$, as shown in Fig.~\ref{fig:rho} for the spectral densities $\rho^{\sigma_{1+2}}$, $\rho^{\kappa_{1+2}}$ and $\rho^{a0_{1+2}}$. We note that the pole contribution is not well defined because these spectral densities are negative when $s$ is large.

\begin{figure}[hbt]
\begin{center}
\scalebox{0.6}{\includegraphics{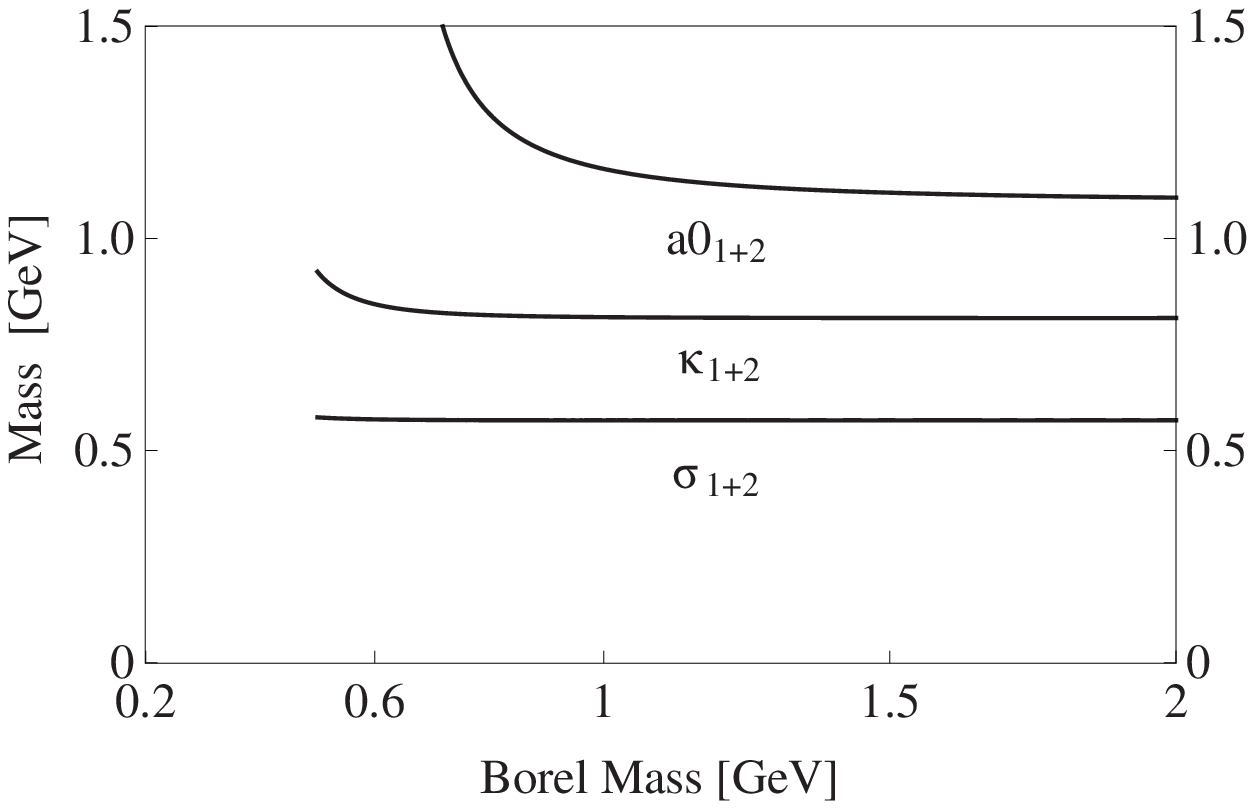}}
\scalebox{0.6}{\includegraphics{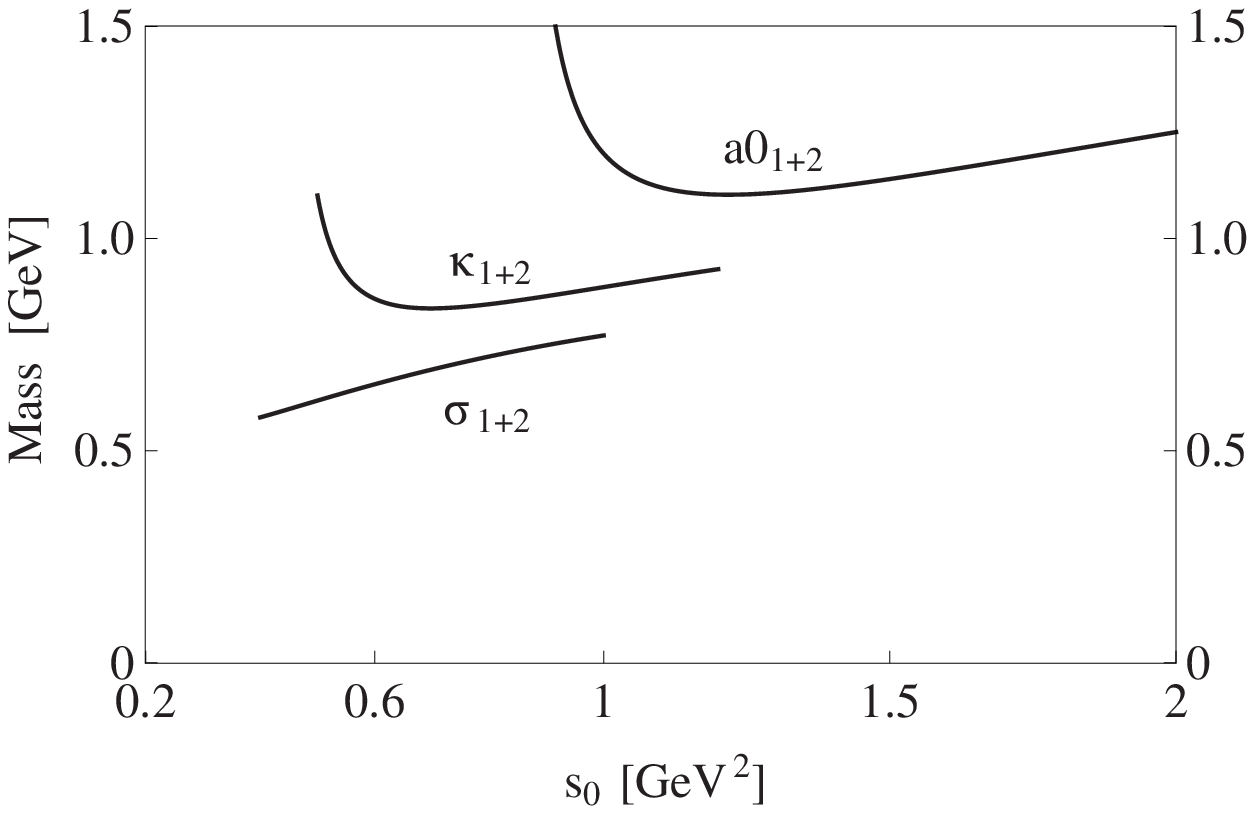}}
\caption{Masses of light scalar mesons $f_0(500)$, $\kappa(800)$ and $a_0(800)$ ($f_0(980)$) as functions of the Borel Mass $M_B$ and the threshold value $s_0$, where only the connected parts, Eq.~(\ref{eq:sigma12sumruleconnect}), are taken into account. The threshold value $s_0$ for the left figure is chosen to be 0.4 GeV$^2$, 0.8 GeV$^2$ and 1.3 GeV$^2$, while the Borel Mass for the right figure is chosen to be 0.5 GeV, 0.6 GeV and 1.5 GeV, for $f_0(500)$, $\kappa(800)$ and $a_0(800)$ ($f_0(980)$), respectively.}
\label{fig:massWB}
\end{center}
\end{figure}

Masses of light scalar mesons are calculated using only the connected parts, and the results are shown in Figs.~\ref{fig:massWB}, as functions of the Borel Mass $M_B$ and the threshold value $s_0$. We clearly see that the Borel Mass dependence is still not much; the mass of $\sigma$ still grows as the threshold value $s_0$ increases, suggesting that there is still much two-meson contribution (or related to its broad decay width); but the mass curves of $\kappa$ and $a_0$ have minimums around $s_0 = 0.8$ GeV$^2$ for $\kappa$ and $s_0 = 1.3$ GeV$^2$ for $a_0$, where the $s_0$ dependence is weak. We use these values as inputs, and calculate the masses of light scalar mesons.

Altogether there are two kinds of error bars: one is due to the two-meson continuum and the other is due to the $[({\mathbf 8},{\mathbf 1}) \oplus ({\mathbf 1},{\mathbf 8})]$ components. This makes our results have large error bars: the mass of $f_0(500)$ is around $400 \sim 600$ MeV, the mass of $\kappa(800)$ is around $700 \sim 900$ MeV, and masses of $a_0(980)$ and $f_0(980)$ are around $900 \sim 1100$ MeV.

\section{Summary}
\label{sec:summary}

We systematically studied the chiral structure of light scalar mesons using local scalar tetraquark currents that belong to the ``non-exotic'' $[( \mathbf{\bar 3},\mathbf{ 3}) \oplus (\mathbf{3},\mathbf{\bar 3})]$ chiral multiplets. This chiral representation only contains flavor singlet and octet mesons, and it does not contain any meson having exotic flavor structure. The nine light scalar mesons can just compose one $[(\mathbf{\bar 3},\mathbf{ 3}) \oplus (\mathbf{3},\mathbf{\bar 3})]$ chiral multiplet. To do a systematical study, we investigated both scalar and pseudoscalar tetraquark currents, since they are chiral partners. We also investigated tetraquark currents of flavor singlet, octet, $\mathbf{10}$, $\mathbf{\overline {10}}$ and $\mathbf{27}$, which can be useful for further studies. Then we used the left handed quark field $L_A^a \equiv q_{LA}^a = {1 - \gamma_5 \over 2} q_A^a$ and the right handed quark field $R_A^a \equiv q_{RA}^a = {1 + \gamma_5 \over 2} q_A^a$ to rewrite these currents. After making proper combinations we verified their chiral representations.

We then used the QCD sum rule to calculate their masses. The masses of $\sigma$, $\kappa$, $a_0$ and $f_0$ are around 600 MeV, 900 MeV, 1100 MeV and 1100 MeV, respectively, generally consistent with the experimental values. However, the pole contributions are very small. Then we introduced a few $[({\mathbf 8},{\mathbf 1}) \oplus ({\mathbf 1},{\mathbf 8})]$ components by slightly changing the mixing parameters from $\eta_{1} + \eta_{2} \longrightarrow 0.99 \times \eta_{1} + \eta_{2}$. The masses of $\sigma$, $\kappa$, $a_0$ and $f_0$ are now around $500$ MeV, $700$ MeV, $900$ MeV and $900$ MeV, respectively, better consistent with the experimental results. The pole contributions are significantly increased to be around 50\% for $f_0(500)$ and $\kappa(800)$. However, these results still depend much on the threshold value $s_0$.

To solve this problem, we use only the connected parts of the two-point correlation function to perform the QCD sum rule analysis.
We find that the results obtained using the tetraquark currents $\sigma_3$, $\kappa_3$, $a0_3$ and $f0_3$ (see Eqs.~(\ref{eq:sigma})-(\ref{eq:f0})) do not change significantly.
However, the results obtained using the tetraquark currents $\kappa_{1+2}$, $a0_{1+2}$ and $f0_{1+2}$ are improved: the mass curves of $\kappa$ and $a_0$ have minimums around $s_0 = 0.8$ GeV$^2$ for $\kappa$ and $s_0 = 1.3$ GeV$^2$ for $a_0$, where the $s_0$ dependence is weak. We use these values as inputs, and calculate the masses.

Altogether there are three kinds of error bars. The dominant one is due to the two-meson continuum. All light scalar mesons couple strongly to it, but we still do not know how to effectively separate them. The second one is due to the mixing of different chiral components. For example, we have included a few $[({\mathbf 8},{\mathbf 1}) \oplus ({\mathbf 1},{\mathbf 8})]$ components to make our results reliable, but we do not know how much it is contained in light scalar mesons. The third one comes from our QCD sum rule calculations that we did not include the high dimensional terms, such as $\alpha_s \langle \bar q q \rangle^4$. Consequently, we obtained masses of light scalar mesons with large error bars: the mass of $f_0(500)$ is around $400 \sim 600$ MeV, the mass of $\kappa(800)$ is around $700 \sim 900$ MeV, and the mass of $a_0(980)$ and $f_0(980)$ is around $900 \sim 1100$ MeV. We note that in Ref.~\cite{Chen:2007xr} we used the same method to calculate masses of $\bar q q$ scalar mesons, which are all above 1 GeV.

We have also used these pseudoscalar tetraquark currents to perform the QCD sum rule analyses. For example, the one containing quark contents $q s \bar q \bar s$ has a mass around 1.3-1.6 GeV. This is significantly larger than the masses of the $\eta$ and $\eta^\prime$ mesons, suggesting that the Nambu-Goldstone bosons, $\pi$, $K$, $\eta$ and $\eta^\prime$, are predominantly $\bar q q$ states. We note the finite decay width of light scalar mesons can be taken into account which does not change the final result significantly~\cite{Chen:2007xr}. We also note that the contribution of instanton has not been considered in this paper whose effects can be significant since light scalar mesons have the same quantum numbers as vacuum. There are many papers discussing this~\cite{Novikov:1984rf,Geshkenbein:1979vb,Schafer:2003nu,Lee:2006vk}.

We note that we can also use the Firez transformation to write tetraquark currents in a mesonic-mesonic form. Some relations are shown in Appendix.~\ref{app:mesoniccurrents}, and here we show one example:
\begin{eqnarray}
\eta_1^{\rm S,\mathbb{S}} + \eta_2^{\rm S,\mathbb{S}}
&=& - {1\over2}(\bar{q}_A^a \gamma_5 q_A^a)(\bar q_B^b \gamma_5 q_B^b) + {1\over2}(\bar{q}_A^a \gamma_5 q_B^a)(\bar{q}_B^b \gamma_5 q_A^b)
\\ \nonumber && - {1\over2}(\bar{q}_A^a q_A^a)(\bar q_B^b q_B^b) + {1\over2}(\bar{q}_A^a q_B^a)(\bar{q}_B^b q_A^b)
\\ \nonumber && + {1\over4}(\bar{q}_A^a \sigma_{\mu\nu} q_A^a)(\bar q_B^b \sigma^{\mu\nu} q_B^b) - {1\over4}(\bar{q}_A^a \sigma_{\mu\nu} q_B^a)(\bar{q}_B^b \sigma^{\mu\nu} q_A^b)
\\ \nonumber &=& - [(\bar q_{LA}^a q_{RA}^a)(\bar q_{LB}^b q_{RB}^b) + (\bar q_{RA}^a q_{LA}^a)(\bar q_{RB}^b q_{LB}^b)]
\\ \nonumber && + [(\bar{q}_{LA}^a q_{RB}^a)(\bar{q}_{LB}^b q_{RA}^b) + (\bar{q}_{RA}^a q_{LB}^a)(\bar{q}_{RB}^b q_{LA}^b)]
\\ \nonumber && + {1\over4}[ (\bar{q}_{LA}^a \sigma_{\mu\nu} q_{RA}^a)(\bar q_{LB}^b \sigma^{\mu\nu} q_{RB}^b) + (\bar{q}_{RA}^a \sigma_{\mu\nu} q_{LB}^a)(\bar{q}_{RB}^b \sigma^{\mu\nu} q_{LA}^b)]
\\ \nonumber && - {1\over4}[ (\bar{q}_{LA}^a \sigma_{\mu\nu} q_{RB}^a)(\bar q_{LB}^b \sigma^{\mu\nu} q_{RA}^b) + (\bar{q}_{RA}^a \sigma_{\mu\nu} q_{LA}^a)(\bar{q}_{RB}^b \sigma^{\mu\nu} q_{LB}^b)] \, .
\end{eqnarray}
Considering $\bar q_R q_L$ and $\bar q_L q_R$ both belong to $[(\mathbf{\bar 3},\mathbf{3}) \oplus (\mathbf{3},\mathbf{\bar 3})]$ representation (or its mirror), all local scalar tetraquark currents that belong to $[(\mathbf{\bar 3},\mathbf{3}) \oplus (\mathbf{3},\mathbf{\bar 3})]$ chiral multiplets are more similar to the combination of two $\bar q q$ mesons that both belong to this same representation. Consequently, the light scalar mesons are more similar to (like) tetraquarks or molecular states consisting two ``non-chiral-singlet'' $\bar q q$ mesons, unless different types of chirality mix with others.

The conventional pseudoscalar and scalar mesons made by one $\bar q q$ pair can also belong to the $[(\mathbf{3},\mathbf{\bar 3}) \oplus ( \mathbf{\bar 3},\mathbf{3})]$ chiral multiplet. However, all the scalar tetraquark currents inside this multiplet have the $q_L q_L \bar q_R \bar q_R + q_R q_R \bar q_L \bar q_L$ chirality, and so they are not direct chiral partners of these $\bar q q$ mesons addressed by chiral singlet quark-antiquark pairs, which have the $( \bar q_L q_R + \bar q_R q_L) \otimes (\bar q_L q_L + \bar q_R q_R)$ chirality (``chiral'' Fock-space expansion), unless these two types of chirality mix with each other; they are more similar to conventional $\bar q q$ mesons addressed by quark condensates, i.e., ${\rm mesons} (\bar q \Gamma q) \otimes {\rm condensates} \langle \bar q q \rangle  = ( \bar q_L q_R + \bar q_R q_L) \otimes (\bar q_L q_R + \bar q_R q_L)$.

Similarly, all the pseudoscalar tetraquark currents inside this multiplet also have the $q_L q_L \bar q_R \bar q_R + q_R q_R \bar q_L \bar q_L$ chirality, and so they are not (direct) terms in the ``chiral'' Fock-space expansion of the $\bar q q$ pseudoscalar mesons ($\pi$, etc). Therefore, in order to write the Fock-space expansion of the conventional pseudoscalar and scalar mesons, we probably need to study the mix of different types of chirality, which will be our next focus. In Ref.~\cite{Weinberg:2013cfa}, S.~Weinberg calculated the decay width of tetraquarks using the Large-N method. This can be done also using the method of QCD sum rule, which will be also our next focus.

\section*{Acknowledgments}

This work is partly supported by the National Natural Science Foundation of China under Grant No. 11205011, and the Fundamental Research Funds for the Central Universities.

\appendix
\section{Other Tetraquark Currents}
\label{app:othercurrents}

\subsection{Pseudoscalar Tetraquark Currents of Flavor Singlet}
\label{subsec:singletpseudoscalar}

In this subsection we study flavor singlet tetraquark currents of $J^P = 0^-$. There are altogether six independent pseudoscalar currents as listed in the following:
\begin{eqnarray}
\nonumber \eta_1^{\rm PS,\mathbb{S}} &=& q_A^{aT} \mathbb{C} q_B^b (\bar{q}_A^a \gamma_5 \mathbb{C} \bar{q}_B^{bT} - \bar{q}_A^b \gamma_5 \mathbb{C} \bar{q}_B^{aT}) \, ,
\\ \nonumber \eta_2^{\rm PS,\mathbb{S}} &=& q_A^{aT} \mathbb{C} \gamma_5 q_B^b (\bar{q}_A^a \mathbb{C} \bar{q}_B^{bT} - \bar{q}_A^b \mathbb{C} \bar{q}_B^{aT}) \, ,
\\ \eta_3^{\rm PS,\mathbb{S}} &=& q_A^{aT} \mathbb{C} \sigma_{\mu\nu} q_B^b (\bar{q}_A^a \sigma^{\mu\nu} \gamma_5 \mathbb{C} \bar{q}_B^{bT} + \bar{q}_A^b \sigma^{\mu\nu} \gamma_5 \mathbb{C} \bar{q}_B^{aT}) \, ,
\\ \nonumber \eta_4^{\rm PS,\mathbb{S}} &=& q_A^{aT} \mathbb{C} q_B^b (\bar{q}_A^a \gamma_5 \mathbb{C} \bar{q}_B^{bT} +\bar{q}_A^b \gamma_5 \mathbb{C} \bar{q}_B^{aT}) \, ,
\\ \nonumber \eta_5^{\rm PS,\mathbb{S}} &=& q_A^{aT} \mathbb{C} \gamma_5 q_B^b (\bar{q}_A^a \mathbb{C} \bar{q}_B^{bT} + \bar{q}_A^b \mathbb{C} \bar{q}_B^{aT}) \, ,
\\ \nonumber \eta_6^{\rm PS,\mathbb{S}} &=& q_A^{aT} \mathbb{C} \sigma_{\mu\nu} q_B^b (\bar{q}_A^a \sigma^{\mu\nu} \gamma_5 \mathbb{C} \bar{q}_B^{bT} - \bar{q}_A^b \sigma^{\mu\nu} \gamma_5 C
\bar{q}_B^{aT}) \, .
\end{eqnarray}
We note that we can prove
\begin{eqnarray}
\nonumber \eta_{7}^{\rm PS,\mathbb{S}} &=& ( q_A^{aT} \mathbb{C} \gamma_\mu q_B^b ) (\bar{q}_A^a \gamma^\mu \gamma_5 \mathbb{C} \bar{q}_B^{bT} - \bar{q}_A^b \gamma^\mu \gamma_5 \mathbb{C} \bar{q}_B^{aT} ) =0 \, ,
\\ \eta_{8}^{\rm PS,\mathbb{S}} &=& ( q_A^{aT} \mathbb{C} \gamma_\mu \gamma_5 q_B^b ) (\bar{q}_A^a \gamma^\mu \mathbb{C} \bar{q}_B^{bT} + \bar{q}_A^b \gamma^\mu \mathbb{C} \bar{q}_B^{aT} ) =0 \, ,
\\ \nonumber \eta_{9}^{\rm PS,\mathbb{S}} &=& ( q_A^{aT} \mathbb{C} \gamma_\mu q_B^b ) (\bar{q}_A^a \gamma^\mu \gamma_5 \mathbb{C} \bar{q}_B^{bT} + \bar{q}_A^b \gamma^\mu \gamma_5 \mathbb{C} \bar{q}_B^{aT} ) =0 \, ,
\\ \nonumber \eta_{10}^{\rm PS,\mathbb{S}} &=& ( q_A^{aT} \mathbb{C} \gamma_\mu \gamma_5 q_B^b ) (\bar{q}_A^a \gamma^\mu \mathbb{C} \bar{q}_B^{bT} - \bar{q}_A^b \gamma^\mu \mathbb{C} \bar{q}_B^{aT} ) =0 \, .
\end{eqnarray}
The two superscripts PS and $\mathbb{S}$ denote pseudoscalar ($J^P = 0^-$) and flavor singlet, respectively. $\eta_{1,2,3}^{\rm PS,\mathbb{S}}$ contain diquarks and antidiquarks having the antisymmetric flavor structure $\mathbf{\bar 3} \otimes \mathbf{3}$; $\eta_{4,5,6}^{\rm PS,\mathbb{S}}$ contain diquarks and antidiquarks having the symmetric flavor structure $\mathbf{6} \otimes \mathbf{\bar6}$. From the following combinations we can clearly see their chiral structure, where the left handed quark field $L_A^a \equiv q_{LA}^a = {1 - \gamma_5 \over 2} q_A^a$ and the right handed quark field $R_A^a \equiv q_{RA}^a = {1 + \gamma_5 \over 2} q_A^a$ are used:
\begin{eqnarray}
\nonumber \eta_1^{\rm PS,\mathbb{S}} - \eta_2^{\rm PS,\mathbb{S}} &=& 2 L_A^{aT} \mathbb{C} L_B^b (\bar{L}_A^a \mathbb{C} \bar{L}_B^{bT} - \bar{L}_A^b \mathbb{C} \bar{L}_B^{aT}) - 2 R_A^{aT} \mathbb{C} R_B^b (\bar{R}_A^a \mathbb{C} \bar{R}_B^{bT} - \bar{R}_A^b \mathbb{C} \bar{R}_B^{aT}) \, ,
\\ \nonumber \eta_1^{\rm PS,\mathbb{S}} + \eta_2^{\rm PS,\mathbb{S}} &=& - 2 L_A^{aT} \mathbb{C} L_B^b (\bar{R}_A^a \mathbb{C} \bar{R}_B^{bT} - \bar{R}_A^b \mathbb{C} \bar{R}_B^{aT}) + 2 R_A^{aT} \mathbb{C} R_B^b (\bar{L}_A^a \mathbb{C} \bar{L}_B^{bT} - \bar{L}_A^b \mathbb{C} \bar{L}_B^{aT}) \, ,
\\ \eta_4^{\rm PS,\mathbb{S}} - \eta_5^{\rm PS,\mathbb{S}} &=& 2 L_A^{aT} \mathbb{C} L_B^b (\bar{L}_A^a \mathbb{C} \bar{L}_B^{bT} + \bar{L}_A^b \mathbb{C} \bar{L}_B^{aT}) - 2 R_A^{aT} \mathbb{C} R_B^b (\bar{R}_A^a \mathbb{C} \bar{R}_B^{bT} + \bar{R}_A^b \mathbb{C} \bar{R}_B^{aT}) \, ,
\\ \nonumber \eta_4^{\rm PS,\mathbb{S}} + \eta_5^{\rm PS,\mathbb{S}} &=& - 2 L_A^{aT} \mathbb{C} L_B^b (\bar{R}_A^a \mathbb{C} \bar{R}_B^{bT} + \bar{R}_A^b \mathbb{C} \bar{R}_B^{aT}) + 2 R_A^{aT} \mathbb{C} R_B^b (\bar{L}_A^a \mathbb{C} \bar{L}_B^{bT} + \bar{L}_A^b \mathbb{C} \bar{L}_B^{aT}) \, ,
\\ \nonumber \eta_3^{\rm PS,\mathbb{S}} &=& - L_A^{aT} \mathbb{C} \sigma_{\mu\nu} L_B^b (\bar{R}_A^a \sigma_{\mu\nu} \mathbb{C} \bar{R}_B^{bT} + \bar{R}_A^b \sigma_{\mu\nu} \mathbb{C} \bar{R}_B^{aT})
+ R_A^{aT} \mathbb{C} \sigma_{\mu\nu} R_B^b (\bar{L}_A^a \sigma_{\mu\nu} \mathbb{C} \bar{L}_B^{bT} + \bar{L}_A^b \sigma_{\mu\nu} \mathbb{C} \bar{L}_B^{aT}) \, ,
\\ \nonumber \eta_6^{\rm PS,\mathbb{S}} &=& - L_A^{aT} \mathbb{C} \sigma_{\mu\nu} L_B^b (\bar{R}_A^a \sigma_{\mu\nu} \mathbb{C} \bar{R}_B^{bT} - \bar{R}_A^b \sigma_{\mu\nu} \mathbb{C} \bar{R}_B^{aT})
+ R_A^{aT} \mathbb{C} \sigma_{\mu\nu} R_B^b (\bar{L}_A^a \sigma_{\mu\nu} \mathbb{C} \bar{L}_B^{bT} - \bar{L}_A^b \sigma_{\mu\nu} \mathbb{C} \bar{L}_B^{aT}) \, .
\end{eqnarray}
We list their chirality and chiral representations in Table~\ref{tab:singletpseudoscalar}.
\begin{table}[!hbt]
\renewcommand{\arraystretch}{1.5}
\begin{center}
\caption{Flavor singlet tetraquark currents of $J^P = 0^-$ classified in subsection~\ref{subsec:singletpseudoscalar}.}
\begin{tabular}{c c c}
\hline\hline
Tetraquark Currents of $(\mathbf 1_F, J^P = 0^-)$ & Chiral Representations & Chirality
\\ \hline \hline
$\eta_1^{\rm PS,\mathbb{S}} - \eta_2^{\rm PS,\mathbb{S}}$, $\eta_4^{\rm PS,\mathbb{S}} - \eta_5^{\rm PS,\mathbb{S}}$ & $[({\mathbf 1}, {\mathbf 1}) + ({\mathbf 1}, {\mathbf 1})]$ & $L L \bar L \bar L + R R \bar R \bar R$
\\ \hline
$\eta_1^{\rm PS,\mathbb{S}} + \eta_2^{\rm PS,\mathbb{S}}$ & $[(\mathbf {\bar 3}, {\mathbf 3}) + ({\mathbf 3}, \mathbf {\bar 3})]$ & $L L \bar R \bar R + R R \bar L \bar L$
\\ \hline
$\eta_4^{\rm PS,\mathbb{S}} + \eta_5^{\rm PS,\mathbb{S}}$ & $[({\mathbf 6}, \mathbf {\bar 6}) + (\mathbf {\bar 6}, {\mathbf 6})]$ & $L L \bar R \bar R + R R \bar L \bar L$
\\ \hline
$\eta_3^{\rm PS,\mathbb{S}}$ & $[(\mathbf {\bar 3}, {\mathbf 3}) + ({\mathbf 3}, \mathbf {\bar 3})]$ & $L L \bar R \bar R + R R \bar L \bar L$
\\ \hline
$\eta_6^{\rm PS,\mathbb{S}}$ & $[({\mathbf 6}, \mathbf {\bar 6}) + (\mathbf {\bar 6}, {\mathbf 6})]$ & $L L \bar R \bar R + R R \bar L \bar L$
\\ \hline\hline
\end{tabular}
\label{tab:singletpseudoscalar}
\end{center}
\renewcommand{\arraystretch}{1}
\end{table}

\subsection{Scalar Tetraquark Currents of Flavor Octet}
\label{subsec:8scalar}

In this subsection we study flavor octet tetraquark currents of $J^P = 0^+$. There are altogether ten independent scalar currents as listed in the following:
\begin{eqnarray}
\nonumber \eta_{1,N}^{\rm S,\mathbb{O}} &=& \lambda_N^{DB} ( q_A^{aT} \mathbb{C} \gamma_5 q_B^b ) ( \bar{q}_A^a \gamma_5 \mathbb{C} \bar{q}_D^{bT} - \bar{q}_A^b \gamma_5 \mathbb{C} \bar{q}_D^{aT} ) \, ,
\\ \nonumber \eta_{2,N}^{\rm S,\mathbb{O}} &=& \lambda_N^{DB} ( q_A^{aT} \mathbb{C} q_B^b ) (\bar{q}_A^a \mathbb{C} \bar{q}_D^{bT} - \bar{q}_A^b \mathbb{C} \bar{q}_D^{aT}) \, ,
\\ \nonumber \eta_{3,N}^{\rm S,\mathbb{O}} &=& \lambda_N^{DB} ( q_A^{aT} \mathbb{C} \sigma_{\mu\nu} q_B^b ) ( \bar{q}_A^a \sigma^{\mu\nu} \mathbb{C} \bar{q}_D^{bT} + \bar{q}_A^b \sigma^{\mu\nu} \mathbb{C} \bar{q}_D^{aT} ) \, ,
\\ \nonumber \eta_{4,N}^{\rm S,\mathbb{O}} &=& \lambda_N^{DB} ( q_A^{aT} \mathbb{C} \gamma_5 q_B^b ) ( \bar{q}_A^a \gamma_5 \mathbb{C} \bar{q}_D^{bT} + \bar{q}_A^b \gamma_5 \mathbb{C} \bar{q}_D^{aT} ) \, ,
\\ \eta_{5,N}^{\rm S,\mathbb{O}} &=& \lambda_N^{DB} ( q_A^{aT} \mathbb{C} q_B^b ) (\bar{q}_A^a \mathbb{C} \bar{q}_D^{bT} + \bar{q}_A^b \mathbb{C} \bar{q}_D^{aT}) \, ,
\\ \nonumber \eta_{6,N}^{\rm S,\mathbb{O}} &=& \lambda_N^{DB} ( q_A^{aT} \mathbb{C} \sigma_{\mu\nu} q_B^b ) ( \bar{q}_A^a \sigma^{\mu\nu} \mathbb{C} \bar{q}_D^{bT} - \bar{q}_A^b \sigma^{\mu\nu} \mathbb{C} \bar{q}_D^{aT} ) \, ,
\\ \nonumber \eta_{7,N}^{\rm S,\mathbb{O}} &=& \lambda_N^{DB} ( q_A^{aT} \mathbb{C} \gamma_\mu \gamma_5 q_B^b ) ( \bar{q}_A^a \gamma^\mu \gamma_5 \mathbb{C} \bar{q}_D^{bT} - \bar{q}_A^b\gamma^\mu \gamma_5 \mathbb{C} \bar{q}_D^{aT}) \, ,
\\ \nonumber \eta_{8,N}^{\rm S,\mathbb{O}} &=& \lambda_N^{DB} ( q_A^{aT} \mathbb{C} \gamma_\mu q_B^b ) (\bar{q}_A^a \gamma^\mu \mathbb{C} \bar{q}_D^{bT} + \bar{q}_A^b \gamma^\mu \mathbb{C} \bar{q}_D^{aT}) \, ,
\\ \nonumber \eta_{9,N}^{\rm S,\mathbb{O}} &=& \lambda_N^{DB} ( q_A^{aT} \mathbb{C} \gamma_\mu \gamma_5 q_B^b ) ( \bar{q}_A^a \gamma^\mu \gamma_5 \mathbb{C} \bar{q}_D^{bT} + \bar{q}_A^b\gamma^\mu \gamma_5 \mathbb{C} \bar{q}_D^{aT}) \, ,
\\ \nonumber \eta_{10,N}^{\rm S,\mathbb{O}} &=& \lambda_N^{DB} ( q_A^{aT} \mathbb{C} \gamma_\mu q_B^b ) (\bar{q}_A^a \gamma^\mu \mathbb{C} \bar{q}_D^{bT} - \bar{q}_A^b \gamma^\mu \mathbb{C} \bar{q}_D^{aT}) \, .
\end{eqnarray}
The two superscripts S and $\mathbb{O}$ denote scalar and flavor octet, respectively. Five currents $\eta_{1,2,3,7,8}^{\rm S,\mathbb{O}}$ contain diquarks and antidiquarks having the antisymmetric flavor structure $\mathbf{\bar 3} \otimes \mathbf{3}$ and other five currents $\eta_{4,5,6,9,10}^{\rm S,\mathbb{O}}$ contain diquarks and antidiquarks having the symmetric flavor structure $\mathbf{6} \otimes \mathbf{\bar6}$. From the following combinations we can clearly see their chiral structure, where the left handed quark field $L_A^a \equiv q_{LA}^a = {1 - \gamma_5 \over 2} q_A^a$ and the right handed quark field $R_A^a \equiv q_{RA}^a = {1 + \gamma_5 \over 2} q_A^a$ are used:
\begin{eqnarray}
\nonumber \eta_{1,N}^{\rm S,\mathbb{O}} - \eta_{2,N}^{\rm S,\mathbb{O}} &=& - 2 \lambda_N^{DB} L_A^{aT} \mathbb{C} L_B^b (\bar{L}_A^a \mathbb{C} \bar{L}_D^{bT} - \bar{L}_A^b \mathbb{C} \bar{L}_D^{aT})
- 2 \lambda_N^{DB} R_A^{aT} \mathbb{C} R_B^b (\bar{R}_A^a \mathbb{C} \bar{R}_D^{bT} - \bar{R}_A^b \mathbb{C} \bar{R}_D^{aT}) \, ,
\\ \nonumber \eta_{1,N}^{\rm S,\mathbb{O}} + \eta_{2,N}^{\rm S,\mathbb{O}} &=& 2 \lambda_N^{DB} L_A^{aT} \mathbb{C} L_B^b (\bar{R}_A^a \mathbb{C} \bar{R}_D^{bT} - \bar{R}_A^b \mathbb{C} \bar{R}_D^{aT})
+ 2 \lambda_N^{DB} R_A^{aT} \mathbb{C} R_B^b (\bar{L}_A^a \mathbb{C} \bar{L}_D^{bT} - \bar{L}_A^b \mathbb{C} \bar{L}_D^{aT}) \, ,
\\ \nonumber \eta_{4,N}^{\rm S,\mathbb{O}} - \eta_{5,N}^{\rm S,\mathbb{O}} &=& - 2 \lambda_N^{DB} L_A^{aT} \mathbb{C} L_B^b (\bar{L}_A^a \mathbb{C} \bar{L}_D^{bT} + \bar{L}_A^b \mathbb{C} \bar{L}_D^{aT})
- 2 \lambda_N^{DB} R_A^{aT} \mathbb{C} R_B^b (\bar{R}_A^a \mathbb{C} \bar{R}_D^{bT} + \bar{R}_A^b \mathbb{C} \bar{R}_D^{aT}) \, ,
\\ \nonumber \eta_{4,N}^{\rm S,\mathbb{O}} + \eta_{5,N}^{\rm S,\mathbb{O}} &=& 2 \lambda_N^{DB} L_A^{aT} \mathbb{C} L_B^b (\bar{R}_A^a \mathbb{C} \bar{R}_D^{bT} + \bar{R}_A^b \mathbb{C} \bar{R}_D^{aT})
+ 2 \lambda_N^{DB} R_A^{aT} \mathbb{C} R_B^b (\bar{L}_A^a \mathbb{C} \bar{L}_D^{bT} + \bar{L}_A^b \mathbb{C} \bar{L}_D^{aT}) \, ,
\\ \nonumber \eta_{3,N}^{\rm S,\mathbb{O}} &=& \lambda_N^{DB} L_A^{aT} \mathbb{C} \sigma_{\mu\nu} L_B^b (\bar{R}_A^a \sigma_{\mu\nu} \mathbb{C} \bar{R}_D^{bT} + \bar{R}_A^b \sigma_{\mu\nu} \mathbb{C} \bar{R}_D^{aT})
+ \lambda_N^{DB} R_A^{aT} \mathbb{C} \sigma_{\mu\nu} R_B^b (\bar{L}_A^a \sigma_{\mu\nu} \mathbb{C} \bar{L}_D^{bT} + \bar{L}_A^b \sigma_{\mu\nu} \mathbb{C} \bar{L}_D^{aT}) \, ,
\\ \nonumber \eta_{6,N}^{\rm S,\mathbb{O}} &=& \lambda_N^{DB} L_A^{aT} \mathbb{C} \sigma_{\mu\nu} L_B^b (\bar{R}_A^a \sigma_{\mu\nu} \mathbb{C} \bar{R}_D^{bT} - \bar{R}_A^b \sigma_{\mu\nu} \mathbb{C} \bar{R}_D^{aT})
+ \lambda_N^{DB} R_A^{aT} \mathbb{C} \sigma_{\mu\nu} R_B^b (\bar{L}_A^a \sigma_{\mu\nu} \mathbb{C} \bar{L}_D^{bT} - \bar{L}_A^b \sigma_{\mu\nu} \mathbb{C} \bar{L}_D^{aT}) \, ,
\\ \nonumber \eta_{7,N}^{\rm S,\mathbb{O}} - \eta_{10,N}^{\rm S,\mathbb{O}} &=& - 2 \lambda_N^{DB} L_A^{aT} \mathbb{C} \gamma_\mu R_B^b ( \bar{L}_A^a \gamma^\mu \mathbb{C} \bar{R}_D^{bT} - \bar{L}_A^b \gamma^\mu \mathbb{C} \bar{R}_D^{aT})
- 2 \lambda_N^{DB} R_A^{aT} \mathbb{C} \gamma_\mu L_B^b ( \bar{R}_A^a \gamma^\mu \mathbb{C} \bar{L}_D^{bT} - \bar{R}_A^b \gamma^\mu \mathbb{C} \bar{L}_D^{aT})  \, ,
\\ 5 \eta_{7,N}^{\rm S,\mathbb{O}} + \eta_{10,N}^{\rm S,\mathbb{O}} &=& 6 d^{NMO} \lambda_M^{CA} \lambda_O^{DB} L_A^{aT} \mathbb{C} \gamma_\mu R_B^b ( \bar{L}_C^a \gamma^\mu \mathbb{C} \bar{R}_D^{bT} - \bar{L}_C^b \gamma^\mu \mathbb{C} \bar{R}_D^{aT}) \, ,
\\ \nonumber \eta_{8,N}^{\rm S,\mathbb{O}} - \eta_{9,N}^{\rm S,\mathbb{O}} &=& 2 \lambda_N^{DB} L_A^{aT} \mathbb{C} \gamma_\mu R_B^b ( \bar{L}_A^a \gamma^\mu \mathbb{C} \bar{R}_D^{bT} + \bar{L}_A^b \gamma^\mu \mathbb{C} \bar{R}_D^{aT})
+ 2 \lambda_N^{DB} R_A^{aT} \mathbb{C} \gamma_\mu L_B^b ( \bar{R}_A^a \gamma^\mu \mathbb{C} \bar{L}_D^{bT} + \bar{R}_A^b \gamma^\mu \mathbb{C} \bar{L}_D^{aT}) \, ,
\\ \nonumber 5 \eta_{8,N}^{\rm S,\mathbb{O}} + \eta_{9,N}^{\rm S,\mathbb{O}} &=& - 6 d^{NMO} \lambda_M^{CA} \lambda_O^{DB} L_A^{aT} \mathbb{C} \gamma_\mu R_B^b ( \bar{L}_C^a \gamma^\mu \mathbb{C} \bar{R}_D^{bT} + \bar{L}_C^b \gamma^\mu \mathbb{C} \bar{R}_D^{aT}) \, .
\end{eqnarray}
We list their chirality and chiral representations in Table~\ref{tab:8scalar}.
\begin{table}[!hbt]
\renewcommand{\arraystretch}{1.5}
\begin{center}
\caption{Flavor octet tetraquark currents of $J^P = 0^+$ classified in subsection~\ref{subsec:8scalar}.}
\begin{tabular}{c c c}
\hline\hline
Tetraquark Currents of $(\mathbf {8}_F, J^P = 0^+)$ & Chiral Representations & Chirality
\\ \hline \hline
$\eta_{1,N}^{\rm S,\mathbb{O}} - \eta_{2,N}^{\rm S,\mathbb{O}}$, $\eta_{4,N}^{\rm S,\mathbb{O}} - \eta_{5,N}^{\rm S,\mathbb{O}}$ & $[({\mathbf 8}, {\mathbf 1}) + ({\mathbf 1}, {\mathbf 8})]$ & $L L \bar L \bar L + R R \bar R \bar R$
\\ \hline
$\eta_{1,N}^{\rm S,\mathbb{O}} + \eta_{2,N}^{\rm S,\mathbb{O}}$ & $[(\mathbf {\bar 3}, {\mathbf 3}) + ({\mathbf 3}, \mathbf {\bar 3})]$ & $L L \bar R \bar R + R R \bar L \bar L$
\\ \hline
$\eta_{4,N}^{\rm S,\mathbb{O}} + \eta_{5,N}^{\rm S,\mathbb{O}}$ & $[({\mathbf 6}, \mathbf {\bar 6}) + (\mathbf {\bar 6}, {\mathbf 6})]$ & $L L \bar R \bar R + R R \bar L \bar L$
\\ \hline
$\eta_{3,N}^{\rm S,\mathbb{O}}$ & $[(\mathbf {\bar 3}, {\mathbf 3}) + ({\mathbf 3}, \mathbf {\bar 3})]$ & $L L \bar R \bar R + R R \bar L \bar L$
\\ \hline
$\eta_{6,N}^{\rm S,\mathbb{O}}$ & $[({\mathbf 6}, \mathbf {\bar 6}) + (\mathbf {\bar 6}, {\mathbf 6})]$ & $L L \bar R \bar R + R R \bar L \bar L$
\\ \hline
$\eta_{7,N}^{\rm S,\mathbb{O}} - \eta_{10,N}^{\rm S,\mathbb{O}}$, $\eta_{8,N}^{\rm S,\mathbb{O}} - \eta_{9,N}^{\rm S,\mathbb{O}}$ & $[({\mathbf 1}, {\mathbf 8}) + ({\mathbf 8}, {\mathbf 1})]$ & $L R \bar L \bar R + R L \bar R \bar L$
\\ \hline
$5 \eta_{7,N}^{\rm S,\mathbb{O}} + \eta_{10,N}^{\rm S,\mathbb{O}}$, $5 \eta_{8,N}^{\rm S,\mathbb{O}} + \eta_{9,N}^{\rm S,\mathbb{O}}$ & $[({\mathbf 8}, {\mathbf 8}) + ({\mathbf 8}, {\mathbf 8})]$ & $L R \bar L \bar R + R L \bar R \bar L$
\\ \hline\hline
\end{tabular}
\label{tab:8scalar}
\end{center}
\renewcommand{\arraystretch}{1}
\end{table}

\subsection{Pseudoscalar Tetraquark Currents of Flavor Octet}
\label{subsec:8pseudoscalar}

In this subsection we study flavor octet tetraquark currents of $J^P = 0^-$. There are altogether ten independent pseudoscalar currents as listed in the following:
\begin{eqnarray}
\nonumber \eta_{1,N}^{\rm PS,\mathbb{O}} &=& \lambda_N^{BD} ( q_A^{aT} \mathbb{C} q_B^b ) (\bar{q}_A^a \gamma_5 \mathbb{C} \bar{q}_D^{bT} - \bar{q}_A^b \gamma_5 \mathbb{C} \bar{q}_D^{aT}) \, ,
\\ \nonumber \eta_{2,N}^{\rm PS,\mathbb{O}} &=& \lambda_N^{BD} ( q_A^{aT} \mathbb{C} \gamma_5 q_B^b ) (\bar{q}_A^a \mathbb{C} \bar{q}_D^{bT} - \bar{q}_A^b \mathbb{C} \bar{q}_D^{aT}) \, ,
\\ \nonumber \eta_{3,N}^{\rm PS,\mathbb{O}} &=& \lambda_N^{BD} ( q_A^{aT} \mathbb{C} \sigma_{\mu\nu} q_B^b ) (\bar{q}_A^a \sigma^{\mu\nu} \gamma_5 \mathbb{C} \bar{q}_D^{bT} + \bar{q}_A^b \sigma^{\mu\nu} \gamma_5 \mathbb{C} \bar{q}_D^{aT}) \, ,
\\ \nonumber \eta_{4,N}^{\rm PS,\mathbb{O}} &=& \lambda_N^{BD} ( q_A^{aT} \mathbb{C} q_B^b ) (\bar{q}_A^a \gamma_5 \mathbb{C} \bar{q}_D^{bT} + \bar{q}_A^b \gamma_5 \mathbb{C} \bar{q}_D^{aT}) \, ,
\\ \eta_{5,N}^{\rm PS,\mathbb{O}} &=& \lambda_N^{BD} ( q_A^{aT} \mathbb{C} \gamma_5 q_B^b ) (\bar{q}_A^a \mathbb{C} \bar{q}_D^{bT} + \bar{q}_A^b \mathbb{C} \bar{q}_D^{aT}) \, ,
\\ \nonumber \eta_{6,N}^{\rm PS,\mathbb{O}} &=& \lambda_N^{BD} ( q_A^{aT} \mathbb{C} \sigma_{\mu\nu} q_B^b ) (\bar{q}_A^a \sigma^{\mu\nu} \gamma_5 \mathbb{C} \bar{q}_D^{bT} - \bar{q}_A^b \sigma^{\mu\nu} \gamma_5 \mathbb{C} \bar{q}_D^{aT}) \, ,
\\ \nonumber \eta_{7,N}^{\rm PS,\mathbb{O}} &=& \lambda_N^{BD} ( q_A^{aT} \mathbb{C} \gamma_\mu q_B^b ) (\bar{q}_A^a \gamma^\mu \gamma_5 \mathbb{C} \bar{q}_D^{bT} - \bar{q}_A^b \gamma^\mu \gamma_5 \mathbb{C} \bar{q}_D^{aT} ) \, ,
\\ \nonumber \eta_{8,N}^{\rm PS,\mathbb{O}} &=& \lambda_N^{BD} ( q_A^{aT} \mathbb{C} \gamma_\mu \gamma_5 q_B^b ) (\bar{q}_A^a \gamma^\mu \mathbb{C} \bar{q}_D^{bT} + \bar{q}_A^b \gamma^\mu \mathbb{C} \bar{q}_D^{aT} ) \, ,
\\ \nonumber \eta_{9,N}^{\rm PS,\mathbb{O}} &=& \lambda_N^{BD} ( q_A^{aT} \mathbb{C} \gamma_\mu q_B^b ) (\bar{q}_A^a \gamma^\mu \gamma_5 \mathbb{C} \bar{q}_D^{bT} + \bar{q}_A^b \gamma^\mu \gamma_5 \mathbb{C} \bar{q}_D^{aT} ) \, ,
\\ \nonumber \eta_{10,N}^{\rm PS,\mathbb{O}} &=& \lambda_N^{BD} ( q_A^{aT} \mathbb{C} \gamma_\mu \gamma_5 q_B^b ) (\bar{q}_A^a \gamma^\mu \mathbb{C} \bar{q}_D^{bT} - \bar{q}_A^b \gamma^\mu \mathbb{C} \bar{q}_D^{aT} ) \, .
\end{eqnarray}
The two superscripts PS and $\mathbb{O}$ denote pseudoscalar and flavor octet, respectively. Among these ten currents, $\eta_{1,2,3}^{\rm PS,\mathbb{O}}$ contain diquarks and antidiquarks having both the antisymmetric flavor structure $\mathbf{\bar 3} \otimes \mathbf{3}$; $\eta_{4,5,6}^{\rm PS,\mathbb{O}}$ contain diquarks and antidiquarks having both the symmetric flavor structure $\mathbf{6} \otimes \mathbf{\bar6}$; $\eta_{7,8}^{\rm PS,\mathbb{O}}$ contain diquarks having the symmetric flavor structure and antidiquarks the antisymmetric flavor structure $\mathbf{6} \otimes \mathbf{3}$; $\eta_{9,10}^{\rm PS,\mathbb{O}}$ contain diquarks having the antisymmetric flavor structure and antidiquarks the symmetric flavor structure $\mathbf{\bar3} \otimes \mathbf{\bar6}$. From the following combinations we can clearly see their chiral structure, where the left handed quark field $L_A^a \equiv q_{LA}^a = {1 - \gamma_5 \over 2} q_A^a$ and the right handed quark field $R_A^a \equiv q_{RA}^a = {1 + \gamma_5 \over 2} q_A^a$ are used:
\begin{eqnarray}
\nonumber \eta_{1,N}^{\rm PS,\mathbb{O}} - \eta_{2,N}^{\rm PS,\mathbb{O}} &=& 2 \lambda_N^{DB} L_A^{aT} \mathbb{C} L_B^b (\bar{L}_A^a \mathbb{C} \bar{L}_D^{bT} - \bar{L}_A^b \mathbb{C} \bar{L}_D^{aT})
- 2 \lambda_N^{DB} R_A^{aT} \mathbb{C} R_B^b (\bar{R}_A^a \mathbb{C} \bar{R}_D^{bT} - \bar{R}_A^b \mathbb{C} \bar{R}_D^{aT}) \, ,
\\ \nonumber \eta_{1,N}^{\rm PS,\mathbb{O}} + \eta_{2,N}^{\rm PS,\mathbb{O}} &=& - 2 \lambda_N^{DB} L_A^{aT} \mathbb{C} L_B^b (\bar{R}_A^a \mathbb{C} \bar{R}_D^{bT} - \bar{R}_A^b \mathbb{C} \bar{R}_D^{aT})
+ 2 \lambda_N^{DB} R_A^{aT} \mathbb{C} R_B^b (\bar{L}_A^a \mathbb{C} \bar{L}_D^{bT} - \bar{L}_A^b \mathbb{C} \bar{L}_D^{aT}) \, ,
\\ \nonumber \eta_{4,N}^{\rm PS,\mathbb{O}} - \eta_{5,N}^{\rm PS,\mathbb{O}} &=& 2 \lambda_N^{DB} L_A^{aT} \mathbb{C} L_B^b (\bar{L}_A^a \mathbb{C} \bar{L}_D^{bT} + \bar{L}_A^b \mathbb{C} \bar{L}_D^{aT})
- 2 \lambda_N^{DB} R_A^{aT} \mathbb{C} R_B^b (\bar{R}_A^a \mathbb{C} \bar{R}_D^{bT} + \bar{R}_A^b \mathbb{C} \bar{R}_D^{aT}) \, ,
\\ \eta_{4,N}^{\rm PS,\mathbb{O}} + \eta_{5,N}^{\rm PS,\mathbb{O}} &=& - 2 \lambda_N^{DB} L_A^{aT} \mathbb{C} L_B^b (\bar{R}_A^a \mathbb{C} \bar{R}_D^{bT} + \bar{R}_A^b \mathbb{C} \bar{R}_D^{aT})
+ 2 \lambda_N^{DB} R_A^{aT} \mathbb{C} R_B^b (\bar{L}_A^a \mathbb{C} \bar{L}_D^{bT} + \bar{L}_A^b \mathbb{C} \bar{L}_D^{aT}) \, ,
\\ \nonumber \eta_{3,N}^{\rm PS,\mathbb{O}} &=& - \lambda_N^{DB} L_A^{aT} \mathbb{C} \sigma_{\mu\nu} L_B^b (\bar{R}_A^a \sigma_{\mu\nu} \mathbb{C} \bar{R}_D^{bT} + \bar{R}_A^b \sigma_{\mu\nu} \mathbb{C} \bar{R}_D^{aT})
+ \lambda_N^{DB} R_A^{aT} \mathbb{C} \sigma_{\mu\nu} R_B^b (\bar{L}_A^a \sigma_{\mu\nu} \mathbb{C} \bar{L}_D^{bT} + \bar{L}_A^b \sigma_{\mu\nu} \mathbb{C} \bar{L}_D^{aT}) \, ,
\\ \nonumber \eta_{6,N}^{\rm PS,\mathbb{O}} &=& - \lambda_N^{DB} L_A^{aT} \mathbb{C} \sigma_{\mu\nu} L_B^b (\bar{R}_A^a \sigma_{\mu\nu} \mathbb{C} \bar{R}_D^{bT} - \bar{R}_A^b \sigma_{\mu\nu} \mathbb{C} \bar{R}_D^{aT})
+ \lambda_N^{DB} R_A^{aT} \mathbb{C} \sigma_{\mu\nu} R_B^b (\bar{L}_A^a \sigma_{\mu\nu} \mathbb{C} \bar{L}_D^{bT} - \bar{L}_A^b \sigma_{\mu\nu} \mathbb{C} \bar{L}_D^{aT}) \, ,
\\ \nonumber \eta_{8,N}^{\rm PS,\mathbb{O}} - \eta_{9,N}^{\rm PS,\mathbb{O}} &=& 2 \lambda_N^{DB} L_A^{aT} \mathbb{C} \gamma_\mu R_B^b ( \bar{L}_A^a \gamma^\mu \mathbb{C} \bar{R}_D^{bT} + \bar{L}_A^b \gamma^\mu \mathbb{C} \bar{R}_D^{aT})
- 2 \lambda_N^{DB} R_A^{aT} \mathbb{C} \gamma_\mu L_B^b ( \bar{R}_A^a \gamma^\mu \mathbb{C} \bar{L}_D^{bT} + \bar{R}_A^b \gamma^\mu \mathbb{C} \bar{L}_D^{aT}) \, ,
\\ \nonumber \eta_{8,N}^{\rm PS,\mathbb{O}} + \eta_{9,N}^{\rm PS,\mathbb{O}} &=& - 2 i f^{NMO} \lambda_M^{CA} \lambda_O^{DB} L_A^{aT} \mathbb{C} \gamma_\mu R_B^b ( \bar{L}_C^a \gamma^\mu \mathbb{C} \bar{R}_D^{bT} + \bar{L}_C^b \gamma^\mu \mathbb{C} \bar{R}_D^{aT}) \, ,
\\ \nonumber \eta_{7,N}^{\rm PS,\mathbb{O}} - \eta_{10,N}^{\rm PS,\mathbb{O}} &=& - 2 \lambda_N^{DB} L_A^{aT} \mathbb{C} \gamma_\mu R_B^b ( \bar{L}_A^a \gamma^\mu \mathbb{C} \bar{R}_D^{bT} - \bar{L}_A^b \gamma^\mu \mathbb{C} \bar{R}_D^{aT})
+ 2 \lambda_N^{DB} R_A^{aT} \mathbb{C} \gamma_\mu L_B^b ( \bar{R}_A^a \gamma^\mu \mathbb{C} \bar{L}_D^{bT} - \bar{R}_A^b \gamma^\mu \mathbb{C} \bar{L}_D^{aT}) \, ,
\\ \nonumber \eta_{7,N}^{\rm PS,\mathbb{O}} + \eta_{10,N}^{\rm PS,\mathbb{O}} &=& 2 i f^{NMO} \lambda_M^{CA} \lambda_O^{DB} L_A^{aT} \mathbb{C} \gamma_\mu R_B^b ( \bar{L}_C^a \gamma^\mu \mathbb{C} \bar{R}_D^{bT} - \bar{L}_C^b \gamma^\mu \mathbb{C} \bar{R}_D^{aT}) \, .
\end{eqnarray}
We list their chirality and chiral representations in Table~\ref{tab:8pseudoscalar}.
\begin{table}[!hbt]
\renewcommand{\arraystretch}{1.5}
\begin{center}
\caption{Flavor octet tetraquark currents of $J^P = 0^-$ classified in subsection~\ref{subsec:8pseudoscalar}.}
\begin{tabular}{c c c}
\hline\hline
Tetraquark Currents of $(\mathbf {8}_F, J^P = 0^-)$ & Chiral Representations & Chirality
\\ \hline \hline
$\eta_{1,N}^{\rm PS,\mathbb{O}} - \eta_{2,N}^{\rm PS,\mathbb{O}}$, $\eta_4^{\rm PS,\mathbb{O}} - \eta_5^{\rm PS,\mathbb{O}}$ & $[({\mathbf 8}, {\mathbf 1}) + ({\mathbf 1}, {\mathbf 8})]$ & $L L \bar L \bar L + R R \bar R \bar R$
\\ \hline
$\eta_{1,N}^{\rm PS,\mathbb{O}} + \eta_{2,N}^{\rm PS,\mathbb{O}}$ & $[(\mathbf {\bar 3}, {\mathbf 3}) + ({\mathbf 3}, \mathbf {\bar 3})]$ & $L L \bar R \bar R + R R \bar L \bar L$
\\ \hline
$\eta_{4,N}^{\rm PS,\mathbb{O}} + \eta_{5,N}^{\rm PS,\mathbb{O}}$ & $[({\mathbf 6}, \mathbf {\bar 6}) + (\mathbf {\bar 6}, {\mathbf 6})]$ & $L L \bar R \bar R + R R \bar L \bar L$
\\ \hline
$\eta_{3,N}^{\rm PS,\mathbb{O}}$ & $[(\mathbf {\bar 3}, {\mathbf 3}) + ({\mathbf 3}, \mathbf {\bar 3})]$ & $L L \bar R \bar R + R R \bar L \bar L$
\\ \hline
$\eta_{6,N}^{\rm PS,\mathbb{O}}$ & $[({\mathbf 6}, \mathbf {\bar 6}) + (\mathbf {\bar 6}, {\mathbf 6})]$ & $L L \bar R \bar R + R R \bar L \bar L$
\\ \hline
$\eta_{8,N}^{\rm PS,\mathbb{O}} - \eta_{9,N}^{\rm PS,\mathbb{O}}$, $\eta_{7,N}^{\rm PS,\mathbb{O}} - \eta_{10,N}^{\rm PS,\mathbb{O}}$ & $[({\mathbf 1}, {\mathbf 8}) + ({\mathbf 8}, {\mathbf 1})]$ & $L R \bar L \bar R + R L \bar R \bar L$
\\ \hline
$\eta_{8,N}^{\rm PS,\mathbb{O}} + \eta_{9,N}^{\rm PS,\mathbb{O}}$, $\eta_{7,N}^{\rm PS,\mathbb{O}} + \eta_{10,N}^{\rm PS,\mathbb{O}}$ & $[({\mathbf 8}, {\mathbf 8}) + ({\mathbf 8}, {\mathbf 8})]$ & $L R \bar L \bar R + R L \bar R \bar L$
\\ \hline\hline
\end{tabular}
\label{tab:8pseudoscalar}
\end{center}
\renewcommand{\arraystretch}{1}
\end{table}

\subsection{Scalar Tetraquark Currents of Flavor $\mathbf{27}_F$}
\label{subsec:27scalar}

In this subsection we study flavor $\mathbf{27}_F$ tetraquark currents of $J^P = 0^+$. There are altogether five independent scalar currents as listed in the following:
\begin{eqnarray}
\nonumber \eta_{1,U}^{\rm S, \mathbb{TS}} &=& S_U^{CD;AB} ( q_A^{aT} C \gamma_5 q_B^b ) ( \bar{q}_C^a \gamma_5 C \bar{q}_D^{bT} ) \, ,
\\ \nonumber \eta_{2,U}^{\rm S, \mathbb{TS}} &=& S_U^{CD;AB} ( q_A^{aT} C q_B^b ) ( \bar{q}_C^a C \bar{q}_D^{bT} ) \, ,
\\ \eta_{3,U}^{\rm S, \mathbb{TS}} &=& S_U^{CD;AB} ( q_A^{aT} C \sigma_{\mu\nu} q_B^b ) ( \bar{q}_C^a \sigma^{\mu\nu} C \bar{q}_D^{bT} ) \, ,
\\ \nonumber \eta_{4,U}^{\rm S, \mathbb{TS}} &=& S_U^{CD;AB} ( q_A^{aT} C \gamma_\mu \gamma_5 q_B^b ) ( \bar{q}_C^a \gamma^\mu \gamma_5 C \bar{q}_D^{bT} ) \, ,
\\ \nonumber \eta_{5,U}^{\rm S, \mathbb{TS}} &=& S_U^{CD;AB} ( q_A^{aT} C \gamma_\mu q_B^b ) ( \bar{q}_C^a \gamma^\mu C \bar{q}_D^{bT} ) \, .
\end{eqnarray}
All these five currents contain diquarks and antidiquarks having the symmetric flavor structure $\mathbf{6} \otimes \mathbf{\bar6}$. From the following combinations we can clearly see their chiral structure, where the left handed quark field $L_A^a \equiv q_{LA}^a = {1 - \gamma_5 \over 2} q_A^a$ and the right handed quark field $R_A^a \equiv q_{RA}^a = {1 + \gamma_5 \over 2} q_A^a$ are used:
\begin{eqnarray}
\nonumber \eta_{1,U}^{\rm S, \mathbb{TS}} - \eta_{2,U}^{\rm S, \mathbb{TS}} &=& - 2 S_U^{CD;AB} L_A^{aT} C L_B^b \bar{L}_C^a C \bar{L}_D^{bT} - 2 S_U^{CD;AB} R_A^{aT} C R_B^b \bar{R}_C^a C \bar{R}_D^{bT}  \, ,
\\ \nonumber \eta_{1,U}^{\rm S, \mathbb{TS}} + \eta_{2,U}^{\rm S, \mathbb{TS}} &=& 2 S_U^{CD;AB} L_A^{aT} C L_B^b \bar{R}_C^a C \bar{R}_D^{bT} + 2 S_U^{CD;AB} R_A^{aT} C R_B^b \bar{L}_C^a C \bar{L}_D^{bT}  \, ,
\\ \nonumber \eta_{3,U}^{\rm S, \mathbb{TS}} &=& S_U^{CD;AB} L_A^{aT} C \sigma_{\mu\nu} L_B^b \bar{R}_C^a \sigma_{\mu\nu} C \bar{R}_D^{bT} + S_U^{CD;AB} R_A^{aT} C \sigma_{\mu\nu} R_B^b \bar{L}_C^a \sigma_{\mu\nu} C \bar{L}_D^{bT} \, ,
\\ \eta_{4,U}^{\rm S, \mathbb{TS}} &=& - 2 S_U^{CD;AB} L_A^{aT} C \gamma_\mu R_B^b ( \bar{L}_C^a \gamma^\mu C \bar{R}_D^{bT} + \bar{L}_C^b \gamma^\mu C \bar{R}_D^{aT}) \, ,
\\ \nonumber \eta_{5,U}^{\rm S, \mathbb{TS}} &=& 2 S_U^{CD;AB} L_A^{aT} C \gamma_\mu R_B^b ( \bar{L}_C^a \gamma^\mu C \bar{R}_D^{bT} - \bar{L}_C^b \gamma^\mu C \bar{R}_D^{aT}) \, .
\end{eqnarray}
We list their chirality and chiral representations in Table~\ref{tab:27scalar}.
\begin{table}[!hbt]
\renewcommand{\arraystretch}{1.5}
\begin{center}
\caption{Flavor $\mathbf {27}_F$ tetraquark currents of $J^P = 0^+$ classified in subsection~\ref{subsec:27scalar}.}
\begin{tabular}{c c c}
\hline\hline
Tetraquark Currents of $(\mathbf {27}_F, J^P = 0^+)$ & Chiral Representations & Chirality
\\ \hline \hline
$\eta_{1,U}^{\rm S, \mathbb{TS}} - \eta_{2,U}^{\rm S, \mathbb{TS}}$ & $[({\mathbf {27}}, {\mathbf 1}) + ({\mathbf 1}, {\mathbf {27}})]$ & $L L \bar L \bar L + R R \bar R \bar R$
\\ \hline
$\eta_{1,U}^{\rm S, \mathbb{TS}} + \eta_{2,U}^{\rm S, \mathbb{TS}}$ & $[({\mathbf 6}, \mathbf {\bar 6}) + (\mathbf {\bar 6}, {\mathbf 6})]$ & $L L \bar R \bar R + R R \bar L \bar L$
\\ \hline
$\eta_{3,U}^{\rm S, \mathbb{TS}}$ & $[({\mathbf 6}, \mathbf{ \bar 6}) + (\mathbf {\bar 6}, {\mathbf 6})]$ & $L L \bar R \bar R + R R \bar L \bar L$
\\ \hline
$\eta_{4,U}^{\rm S, \mathbb{TS}}$, $\eta_{5,U}^{\rm S, \mathbb{TS}}$ & $[({\mathbf 8}, {\mathbf 8}) + ({\mathbf 8}, {\mathbf 8})]$ & $L R \bar L \bar R + R L \bar R \bar L$
\\ \hline\hline
\end{tabular}
\label{tab:27scalar}
\end{center}
\renewcommand{\arraystretch}{1}
\end{table}

\subsection{Pseudoscalar Tetraquark Currents of Flavor $\mathbf{27}_F$}
\label{subsec:27pseudoscalar}

In this subsection we study flavor $\mathbf{27}_F$ tetraquark currents of $J^P = 0^-$. There are altogether three independent pseudoscalar currents as listed in the following:
\begin{eqnarray}
\nonumber \eta_{1,U}^{\rm PS, \mathbb{TS}} &=& S_U^{CD;AB} ( q_A^{aT} C q_B^b ) ( \bar{q}_C^a \gamma_5 C \bar{q}_D^{bT} ) \, ,
\\ \eta_{2,U}^{\rm PS, \mathbb{TS}} &=& S_U^{CD;AB} ( q_A^{aT} C \gamma_5 q_B^b ) ( \bar{q}_C^a C \bar{q}_D^{bT} ) \, ,
\\ \nonumber \eta_{3,U}^{\rm PS, \mathbb{TS}} &=& S_U^{CD;AB} ( q_A^{aT} C \sigma_{\mu\nu} q_B^b ) ( \bar{q}_C^a \sigma^{\mu\nu} \gamma_5 C \bar{q}_D^{bT} ) \, .
\end{eqnarray}
All these three currents contain diquarks and antidiquarks having the symmetric flavor structure $\mathbf{6} \otimes \mathbf{\bar6}$. From the following combinations we can clearly see their chiral structure, where the left handed quark field $L_A^a \equiv q_{LA}^a = {1 - \gamma_5 \over 2} q_A^a$ and the right handed quark field $R_A^a \equiv q_{RA}^a = {1 + \gamma_5 \over 2} q_A^a$ are used:
\begin{eqnarray}
\nonumber \eta_{1,U}^{\rm PS, \mathbb{TS}} - \eta_{2,U}^{\rm PS, \mathbb{TS}} &=& 2 S_U^{CD;AB} L_A^{aT} C L_B^b \bar{L}_C^a C \bar{L}_D^{bT} - 2 S_U^{CD;AB} R_A^{aT} C R_B^b \bar{R}_C^a C \bar{R}_D^{bT}  \, ,
\\ \eta_{1,U}^{\rm PS, \mathbb{TS}} + \eta_{2,U}^{\rm PS, \mathbb{TS}} &=& - 2 S_U^{CD;AB} L_A^{aT} C L_B^b \bar{R}_C^a C \bar{R}_D^{bT} + 2 S_U^{CD;AB} R_A^{aT} C R_B^b \bar{L}_C^a C \bar{L}_D^{bT}  \, ,
\\ \nonumber \eta_{3,U}^{\rm PS, \mathbb{TS}} &=& - S_U^{CD;AB} L_A^{aT} C \sigma_{\mu\nu} L_B^b \bar{R}_C^a \sigma_{\mu\nu} C \bar{R}_D^{bT} + S_U^{CD;AB} R_A^{aT} C \sigma_{\mu\nu} R_B^b \bar{L}_C^a \sigma_{\mu\nu} C \bar{L}_D^{bT} \, .
\end{eqnarray}
We list their chirality and chiral representations in Table~\ref{tab:27pseudoscalar}.
\begin{table}[!hbt]
\renewcommand{\arraystretch}{1.5}
\begin{center}
\caption{Flavor $\mathbf {27}_F$ tetraquark currents of $J^P = 0^-$ classified in subsection~\ref{subsec:27pseudoscalar}.}
\begin{tabular}{c c c}
\hline\hline
Tetraquark Currents of $(\mathbf {27}_F, J^P = 0^-)$ & Chiral Representations & Chirality
\\ \hline \hline
$\eta_{1,U}^{\rm PS, \mathbb{TS}} - \eta_{2,U}^{\rm PS, \mathbb{TS}}$ & $[({\mathbf {27}}, {\mathbf 1}) + ({\mathbf 1}, {\mathbf {27}})]$ & $L L \bar L \bar L + R R \bar R \bar R$
\\ \hline
$\eta_{1,U}^{\rm PS, \mathbb{TS}} + \eta_{2,U}^{\rm PS, \mathbb{TS}}$ & $[({\mathbf 6}, \mathbf {\bar 6}) + (\mathbf {\bar 6}, {\mathbf 6})]$ & $L L \bar R \bar R + R R \bar L \bar L$
\\ \hline
$\eta_{3,U}^{\rm PS, \mathbb{TS}}$ & $[({\mathbf 6}, \mathbf {\bar 6}) + (\mathbf {\bar 6}, {\mathbf 6})]$ & $L L \bar R \bar R + R R \bar L \bar L$
\\ \hline\hline
\end{tabular}
\label{tab:27pseudoscalar}
\end{center}
\renewcommand{\arraystretch}{1}
\end{table}

\subsection{Pseudoscalar Tetraquark Currents of Flavor $\overline{\mathbf{10}}_F$}
\label{subsec:10barpseudoscalar}

In this subsection we study flavor $\overline{\mathbf{10}}_F$ tetraquark currents of $J^P = 0^-$. There are altogether two independent pseudoscalar currents as listed in the following:
\begin{eqnarray}
\eta_{1,P}^{\rm PS, \bar \mathbb{D}} &=& \epsilon^{ABE} S_P^{CDE} ( q_A^{aT} C \gamma_\mu q_B^b ) (\bar{q}_C^a \gamma^\mu \gamma_5 C \bar{q}_D^{bT} ) \, ,
\\ \nonumber \eta_{2,P}^{\rm PS, \bar \mathbb{D}} &=& \epsilon^{ABE} S_P^{CDE} ( q_A^{aT} C \gamma_\mu \gamma_5 q_B^b ) (\bar{q}_C^a \gamma^\mu C \bar{q}_D^{bT} ) \, .
\end{eqnarray}
These two currents both contain diquarks having the antisymmetric flavor structure and antidiquarks the symmetric flavor structure $\mathbf{\bar 3} \otimes \mathbf{\bar 6}$. From the following combinations we can clearly see their chiral structure, where the left handed quark field $L_A^a \equiv q_{LA}^a = {1 - \gamma_5 \over 2} q_A^a$ and the right handed quark field $R_A^a \equiv q_{RA}^a = {1 + \gamma_5 \over 2} q_A^a$ are used:
\begin{eqnarray}
\eta_{1,P}^{\rm PS, \bar \mathbb{D}} &=& -2 \epsilon^{ABE} S_P^{CDE} L_A^{aT} C \gamma_\mu R_B^b ( \bar{L}_C^a \gamma^\mu C \bar{R}_D^{bT} + \bar{L}_C^b \gamma^\mu C \bar{R}_D^{aT} ) \, ,
\\ \nonumber \eta_{2,P}^{\rm PS, \bar \mathbb{D}} &=& 2 \epsilon^{ABE} S_P^{CDE} L_A^{aT} C \gamma_\mu R_B^b ( \bar{L}_C^a \gamma^\mu C \bar{R}_D^{bT} - \bar{L}_C^b \gamma^\mu C \bar{R}_D^{aT} ) \, .
\end{eqnarray}
We find that these two currents both belong to the chiral representation $[({\mathbf 8}, {\mathbf 8}) + ({\mathbf 8}, {\mathbf 8})]$ and their chirality is $L R \bar L \bar R + R L \bar R \bar L$.

\subsection{Pseudoscalar Tetraquark Currents of Flavor ${\mathbf{10}}_F$}
\label{subsec:10pseudoscalar}

In this subsection we study flavor ${\mathbf{10}}_F$ tetraquark currents of $J^P = 0^-$. There are altogether two independent pseudoscalar currents as listed in the following:
\begin{eqnarray}
\eta_{1,P}^{\rm PS, \mathbb{D}} &=& S_P^{ABE} \epsilon^{CDE} ( q_A^{aT} C \gamma_\mu q_B^b ) (\bar{q}_C^a \gamma^\mu \gamma_5 C \bar{q}_D^{bT} ) \, ,
\\ \nonumber \eta_{2,P}^{\rm PS, \mathbb{D}} &=& S_P^{ABE} \epsilon^{CDE} ( q_A^{aT} C \gamma_\mu \gamma_5 q_B^b ) (\bar{q}_C^a \gamma^\mu C \bar{q}_D^{bT} ) \, .
\end{eqnarray}
These two currents both contain diquarks having the antisymmetric flavor structure and antidiquarks the symmetric flavor structure $\mathbf{6} \otimes \mathbf{3}$. From the following combinations we can clearly see their chiral structure, where the left handed quark field $L_A^a \equiv q_{LA}^a = {1 - \gamma_5 \over 2} q_A^a$ and the right handed quark field $R_A^a \equiv q_{RA}^a = {1 + \gamma_5 \over 2} q_A^a$ are used:
\begin{eqnarray}
\eta_{1,P}^{\rm PS, \mathbb{D}} &=& -2 S_P^{ABE} \epsilon^{CDE} L_A^{aT} C \gamma_\mu R_B^b ( \bar{L}_C^a \gamma^\mu C \bar{R}_D^{bT} - \bar{L}_C^b \gamma^\mu C \bar{R}_D^{aT} ) \, ,
\\ \nonumber \eta_{2,P}^{\rm PS, \mathbb{D}} &=& 2 S_P^{ABE} \epsilon^{CDE} L_A^{aT} C \gamma_\mu R_B^b ( \bar{L}_C^a \gamma^\mu C \bar{R}_D^{bT} + \bar{L}_C^b \gamma^\mu C \bar{R}_D^{aT} ) \, .
\end{eqnarray}
We find that these two currents both belong to the chiral representation $[({\mathbf 8}, {\mathbf 8}) + ({\mathbf 8}, {\mathbf 8})]$ and their chirality is $L R \bar L \bar R + R L \bar R \bar L$.

\section{Other Chiral Transformations}
\label{app:othertransform}

There are four $[({\mathbf 1},{\mathbf 1})]$ chiral multiplets, $\big ( \eta_{1}^{\rm S, \mathbb{S}} - \eta_{2}^{\rm S, \mathbb{S}}, \eta_{1}^{\rm PS, \mathbb{S}} - \eta_{2}^{\rm PS, \mathbb{S}} \big )$, $\big ( \eta_{4}^{\rm S, \mathbb{S}} - \eta_{5}^{\rm S, \mathbb{S}}, \eta_{4}^{\rm PS, \mathbb{S}} - \eta_{5}^{\rm PS, \mathbb{S}}\big )$, $\big ( \eta_{7}^{\rm S, \mathbb{S}} - \eta_{10}^{\rm S, \mathbb{S}}, \eta_{7}^{\rm PS, \mathbb{S}} - \eta_{10}^{\rm PS, \mathbb{S}} = 0 \big )$, $\big ( \eta_{8}^{\rm S, \mathbb{S}} - \eta_{9}^{\rm S, \mathbb{S}}, \eta_{8}^{\rm PS, \mathbb{S}} - \eta_{9}^{\rm PS, \mathbb{S}}  = 0 \big )$. We use $\big ( \eta_{(\mathbf{1},\mathbf{1})}^{\rm S, \mathbb{S}}, \eta_{(\mathbf{1},\mathbf{1})}^{\rm PS, \mathbb{S}} \big )$ to denote them, and their chiral transformation properties are
\begin{eqnarray}
\nonumber \delta_5 \eta_{(\mathbf{1},\mathbf{1})}^{\rm S(PS), \mathbb{S}} &=& 0 \, ,
\\ \nonumber \delta^{\vec a} \eta_{(\mathbf{1},\mathbf{1})}^{\rm S(PS), \mathbb{S}} &=& 0 \, ,
\\ \nonumber \delta_5^{\vec b} \eta_{(\mathbf{1},\mathbf{1})}^{\rm S(PS), \mathbb{S}} &=& 0 \, .
\end{eqnarray}
There are two $[({\mathbf 6},\mathbf {\bar 6}) \oplus (\mathbf {\bar 6},{\mathbf 6})]$ chiral multiplets, $\big ( \eta_{4}^{\rm S, \mathbb{S}} + \eta_{5}^{\rm S, \mathbb{S}}, \eta_{4}^{\rm PS, \mathbb{S}} + \eta_{5}^{\rm PS, \mathbb{S}}, \eta_{4,N}^{\rm S, \mathbb{O}} + \eta_{5,N}^{\rm S, \mathbb{O}}, \eta_{4,N}^{\rm PS, \mathbb{O}} + \eta_{5,N}^{\rm PS, \mathbb{O}}, \eta_{1,U}^{\rm S, \mathbb{TS}} + \eta_{2,U}^{\rm S, \mathbb{TS}}, \eta_{1,U}^{\rm PS, \mathbb{TS}} + \eta_{2,U}^{\rm PS, \mathbb{TS}} \big )$, $\big ( \eta_{6}^{\rm S, \mathbb{S}}, \eta_{6}^{\rm PS, \mathbb{S}}, \eta_{6,N}^{\rm S, \mathbb{O}}, \eta_{6,N}^{\rm PS, \mathbb{O}}, \eta_{3,U}^{\rm S, \mathbb{TS}}, \eta_{3,U}^{\rm PS, \mathbb{TS}} \big )$. We use $\big ( \eta_{(\mathbf{6}, \mathbf{\bar 6})}^{\rm S, \mathbb{S}}, \eta_{(\mathbf{6},\mathbf{\bar 6})}^{\rm PS, \mathbb{S}}, \eta_{(\mathbf{6},\mathbf{\bar 6}),N}^{\rm S, \mathbb{O}}, \eta_{(\mathbf{6}, \mathbf{\bar 6}),N}^{\rm PS, \mathbb{O}}, \eta_{(\mathbf{6},\mathbf{\bar 6})}^{\rm S, \mathbb{TS}}, \eta_{(\mathbf{6},\mathbf{\bar 6})}^{\rm PS, \mathbb{TS}} \big )$ to denote them, and their chiral transformation properties are
\begin{eqnarray}
\nonumber \delta_5 \eta_{(\mathbf{6},\mathbf{\bar 6})}^{\rm S(PS), \mathbb{S}} &=& 4 i b \eta_{(\mathbf{6},\mathbf{\bar 6})}^{\rm PS(S), \mathbb{S}} \, ,
\\ \nonumber \delta^{\vec a} \eta_{(\mathbf{6},\mathbf{\bar 6})}^{\rm S(PS), \mathbb{S}} &=& 0 \, ,
\\ \nonumber \delta_5^{\vec b} \eta_{(\mathbf{6},\mathbf{\bar 6})}^{\rm S(PS), \mathbb{S}} &=& 4 i b^N \eta_{(\mathbf{6},\mathbf{\bar 6}),N}^{\rm PS(S), \mathbb{O}} \, ,
\\ \nonumber \delta_5 \eta_{(\mathbf{6},\mathbf{\bar 6}),N}^{\rm S(PS), \mathbb{O}} &=& 4 i b \eta_{(\mathbf{6},\mathbf{\bar 6}),N}^{\rm PS(S), \mathbb{O}} \, ,
\\ \nonumber \delta^{\vec a} \eta_{(\mathbf{6},\mathbf{\bar 6}),N}^{\rm S(PS), \mathbb{O}} &=& 2 a^N f_{NMO} \eta_{(\mathbf{6},\mathbf{\bar 6}),O}^{\rm S(PS), \mathbb{O}} \, ,
\\ \nonumber \delta_5^{\vec b} \eta_{(\mathbf{6},\mathbf{\bar 6}),N}^{\rm S(PS), \mathbb{O}} &=& {5\over3} i b^M \eta_{(\mathbf{6},\mathbf{\bar 6})}^{\rm PS(S), \mathbb{S}} + {14\over5} i b^N d_{NMO} \eta_{(\mathbf{6},\mathbf{\bar 6}),O}^{\rm PS(S), \mathbb{O}} + 4 i b^N \big ( {\bf T}_{8\times27} \big )^N_{MU} \eta_{(\mathbf{6},\mathbf{\bar 6}),U}^{\rm PS(S), \mathbb{TS}} \, ,
\\ \nonumber \delta_5 \eta_{(\mathbf{6},\mathbf{\bar 6})}^{\rm S(PS), \mathbb{TS}} &=& 4 i b \eta_{(\mathbf{6},\mathbf{\bar 6}),U}^{\rm PS(S), \mathbb{TS}} \, ,
\\ \nonumber \delta^{\vec a} \eta_{(\mathbf{6},\mathbf{\bar 6})}^{\rm S(PS), \mathbb{TS}} &=& 2 i a^N \big ({\bf T^A}_{27\times27} - {\bf T^B}_{27\times27}\big )^N_{UV} \eta_{(\mathbf{6},\mathbf{\bar 6}),V}^{\rm S(PS), \mathbb{TS}} \, ,
\\ \nonumber \delta_5^{\vec b} \eta_{(\mathbf{6},\mathbf{\bar 6})}^{\rm S(PS), \mathbb{TS}} &=& {6\over5} i b^N \big ( {\bf T}^\dagger_{8\times27} \big )^N_{UO} \eta_{(\mathbf{6},\mathbf{\bar 6}),O}^{\rm PS(S), \mathbb{O}} + 2 i b^N \big ( {\bf T}^{\bf A}_{27\times27} + {\bf T}^{\bf B}_{27\times27} \big )^N_{UV} \eta_{(\mathbf{6},\mathbf{\bar 6}),V}^{\rm PS(S), \mathbb{TS}} \, .
\end{eqnarray}
There are two $[({\mathbf 8},{\mathbf 8}) \oplus ({\mathbf 8},{\mathbf 8})]$ chiral multiplets, $\big ( 2 \eta_{7}^{\rm S, \mathbb{S}} + \eta_{10}^{\rm S, \mathbb{S}}, 2 \eta_{7}^{\rm PS, \mathbb{S}} + \eta_{10}^{\rm PS, \mathbb{S}} = 0, 5 \eta_{7,N}^{\rm S, \mathbb{O}} + \eta_{10,N}^{\rm S, \mathbb{O}}, \eta_{7,N}^{\rm PS, \mathbb{O}} + \eta_{10,N}^{\rm PS, \mathbb{O}}, \eta_{5,U}^{\rm S, \mathbb{TS}}, \eta_{1,P}^{\rm PS, \mathbb{D}}, \eta_{2,P}^{\rm PS, \bar \mathbb{D}} \big )$, $\big ( 2 \eta_{8}^{\rm S, \mathbb{S}} + \eta_{9}^{\rm S, \mathbb{S}}, 2 \eta_{8}^{\rm PS, \mathbb{S}} + \eta_{9}^{\rm PS, \mathbb{S}} = 0, 5 \eta_{8,N}^{\rm S, \mathbb{O}} + \eta_{9,N}^{\rm S, \mathbb{O}}, \eta_{8,N}^{\rm PS, \mathbb{O}} + \eta_{9,N}^{\rm PS, \mathbb{O}}, \eta_{4,U}^{\rm S, \mathbb{TS}}, \eta_{2,P}^{\rm PS, \mathbb{D}}, \eta_{1,P}^{\rm PS,\bar \mathbb{D}} \big )$. We use $\big ( \eta_{({\mathbf 8},{\mathbf 8})}^{\rm S, \mathbb{S}}, \eta_{({\mathbf 8},{\mathbf 8})}^{\rm PS, \mathbb{S}}, \eta_{({\mathbf 8},{\mathbf 8}),N}^{\rm S, \mathbb{O}}, \eta_{({\mathbf 8},{\mathbf 8}),N}^{\rm PS, \mathbb{O}}, \eta_{({\mathbf 8},{\mathbf 8}),U}^{\rm S, \mathbb{TS}}, \eta_{({\mathbf 8},{\mathbf 8}),P}^{\rm PS, \mathbb{D}}, \eta_{({\mathbf 8},{\mathbf 8}),P}^{\rm PS, \bar \mathbb{D}} \big )$ to denote them, and their chiral transformation properties are
\begin{eqnarray}
\nonumber \delta_5 \eta_{({\mathbf 8},{\mathbf 8})}^{\rm S, \mathbb{S}} &=& 0 \, ,
\\ \nonumber \delta^{\vec a} \eta_{({\mathbf 8},{\mathbf 8})}^{\rm S, \mathbb{S}} &=& 0 \, ,
\\ \nonumber \delta_5^{\vec b} \eta_{({\mathbf 8},{\mathbf 8})}^{\rm S, \mathbb{S}} &=& 6 i b^N \eta_{({\mathbf 8},{\mathbf 8}),N}^{\rm PS, \mathbb{O}} \, ,
\\ \nonumber \delta_5 \eta_{({\mathbf 8},{\mathbf 8}),N}^{\rm S, \mathbb{O}} &=& 0 \, ,
\\ \nonumber \delta^{\vec a} \eta_{({\mathbf 8},{\mathbf 8}),N}^{\rm S, \mathbb{O}} &=& 2 a^N f_{NMO} \eta_{({\mathbf 8},{\mathbf 8}),O}^{\rm S, \mathbb{O}} \, ,
\\ \nonumber \delta_5^{\vec b} \eta_{({\mathbf 8},{\mathbf 8}),N}^{\rm S, \mathbb{O}} &=& 
6 i b^N d_{NMO} \eta_{({\mathbf 8},{\mathbf 8}),O}^{\rm PS, \mathbb{O}} - 12 i b^N \big ( {\bf T}_{8\times10} \big )^N_{MP} \eta_{({\mathbf 8},{\mathbf 8}),P}^{\rm PS, \mathbb{D}} - 12 i b^N \big ( {\bf T}^*_{8\times10} \big )^N_{MP} \eta_{({\mathbf 8},{\mathbf 8}),P}^{\rm PS, \bar \mathbb{D}} \, ,
\\ \nonumber \delta_5 \eta_{({\mathbf 8},{\mathbf 8}),N}^{\rm PS, \mathbb{O}} &=& 0 \, ,
\\ \nonumber \delta^{\vec a} \eta_{({\mathbf 8},{\mathbf 8}),N}^{\rm PS, \mathbb{O}} &=& 2 a^N f_{NMO} \eta_{({\mathbf 8},{\mathbf 8}),O}^{\rm PS, \mathbb{O}} \, ,
\\ \nonumber \delta_5^{\vec b} \eta_{({\mathbf 8},{\mathbf 8}),N}^{\rm PS, \mathbb{O}} &=& i b^M \eta_{({\mathbf 8},{\mathbf 8})}^{\rm S, \mathbb{S}} + {6\over5} i b^N d_{NMO} \eta_{({\mathbf 8},{\mathbf 8}),O}^{\rm S, \mathbb{O}} - 4 i b^N \big ( {\bf T}_{8\times27} \big )^N_{MU} \eta_{({\mathbf 8},{\mathbf 8}),U}^{\rm S, \mathbb{TS}} \, ,
\\ \nonumber \delta_5 \eta_{({\mathbf 8},{\mathbf 8}),U}^{\rm S, \mathbb{TS}} &=& 0 \, ,
\\ \nonumber \delta^{\vec a} \eta_{({\mathbf 8},{\mathbf 8}),U}^{\rm S, \mathbb{TS}} &=& 2 i a^N \big ({\bf T}^{\bf A}_{27\times27} - {\bf T}^{\bf B}_{27\times27}\big )^N_{UV} \eta_{({\mathbf 8},{\mathbf 8}),V}^{\rm S, \mathbb{TS}} \, ,
\\ \nonumber \delta_5^{\vec b} \eta_{({\mathbf 8},{\mathbf 8}),U}^{\rm S, \mathbb{TS}} &=& - i b^N \big ( {\bf T}^\dagger_{8\times27} \big )^N_{UO} \eta_{({\mathbf 8},{\mathbf 8}),O}^{\rm PS, \mathbb{O}} + 3 i b^N \big ( {\bf T}^{{\bf B}\dagger}_{10\times27} \big )^N_{UP} \eta_{({\mathbf 8},{\mathbf 8}),P}^{\rm PS, \bar \mathbb{D}} + 3 i b^N \big ( {\bf T}^{{\bf A}\dagger}_{10\times27} \big )^N_{UP} \eta_{({\mathbf 8},{\mathbf 8}),P}^{\rm PS, \mathbb{D}} \, ,
\\ \nonumber \delta_5 \eta_{({\mathbf 8},{\mathbf 8}),P}^{\rm PS, \mathbb{D}} &=& 0 \, ,
\\ \nonumber \delta^{\vec a} \eta_{({\mathbf 8},{\mathbf 8}),P}^{\rm PS, \mathbb{D}} &=& 3 i a^N \big ({\bf T}_{10\times10}\big )^N_{PQ} \eta_{({\mathbf 8},{\mathbf 8}),Q}^{\rm PS, \mathbb{D}} \, ,
\\ \nonumber \delta_5^{\vec b} \eta_{({\mathbf 8},{\mathbf 8}),P}^{\rm PS, \mathbb{D}} &=& - {2\over5} i b^N \big ( {\bf T}^\dagger_{8\times10} \big )^N_{PO} \eta_{({\mathbf 8},{\mathbf 8}),O}^{\rm S, \mathbb{O}} + 2 i b^N \big ( {\bf T}^{\bf A}_{10\times27} \big )^N_{PU} \eta_{({\mathbf 8},{\mathbf 8}),U}^{\rm S, \mathbb{TS}} \, ,
\\ \nonumber \delta_5 \eta_{({\mathbf 8},{\mathbf 8}),P}^{\rm PS, \bar \mathbb{D}} &=& 0 \, ,
\\ \nonumber \delta^{\vec a} \eta_{({\mathbf 8},{\mathbf 8}),P}^{\rm PS, \bar \mathbb{D}} &=& -3 i a^N \big ({\bf T}^*_{10\times10}\big )^N_{PQ} \eta_{({\mathbf 8},{\mathbf 8}),Q}^{\rm PS, \bar \mathbb{D}} \, ,
\\ \nonumber \delta_5^{\vec b} \eta_{({\mathbf 8},{\mathbf 8}),P}^{\rm PS, \bar \mathbb{D}} &=& - {2\over5} i b^N \big ( {\bf T}^{\dagger*}_{8\times10} \big )^N_{PO} \eta_{({\mathbf 8},{\mathbf 8}),O}^{\rm S, \mathbb{O}} + 2 i b^N \big ( {\bf T}^{\bf B}_{10\times27} \big )^N_{PU} \eta_{({\mathbf 8},{\mathbf 8}),U}^{\rm S, \mathbb{TS}} \, .
\end{eqnarray}
There are four $[({\mathbf 8},{\mathbf 1}) \oplus ({\mathbf 1},{\mathbf 8})]$ chiral multiplets, $\big ( \eta_{1,N}^{\rm S, \mathbb{O}} - \eta_{2,N}^{\rm S, \mathbb{O}}, \eta_{1,N}^{\rm PS, \mathbb{O}} - \eta_{2,N}^{\rm PS, \mathbb{O}} \big )$, $\big ( \eta_{4,N}^{\rm S, \mathbb{O}} - \eta_{5,N}^{\rm S, \mathbb{O}}, \eta_{4,N}^{\rm PS, \mathbb{O}} - \eta_{5,N}^{\rm PS, \mathbb{O}} \big )$, $\big ( \eta_{7,N}^{\rm S, \mathbb{O}} - \eta_{10,N}^{\rm S, \mathbb{O}}, \eta_{7,N}^{\rm PS, \mathbb{O}} - \eta_{10,N}^{\rm PS, \mathbb{O}} \big )$, $\big ( \eta_{8,N}^{\rm S, \mathbb{O}} - \eta_{9,N}^{\rm S, \mathbb{O}}, \eta_{8,N}^{\rm PS, \mathbb{O}} - \eta_{9,N}^{\rm PS, \mathbb{O}} \big )$. We use $\big ( \eta_{({\mathbf 8},{\mathbf 1}),N}^{\rm S, \mathbb{O}}, \eta_{({\mathbf 8},{\mathbf 1}),N}^{\rm PS, \mathbb{O}} \big )$ to denote them, and their chiral transformation properties are
\begin{eqnarray}
\nonumber \delta_5 \eta_{({\mathbf 8},{\mathbf 1}),N}^{\rm S(PS), \mathbb{O}} &=& 0 \, ,
\\ \nonumber \delta^{\vec a} \eta_{({\mathbf 8},{\mathbf 1}),N}^{\rm S(PS), \mathbb{O}} &=& 2 a^N f_{NMO} \eta_{({\mathbf 8},{\mathbf 1}),O}^{\rm S(PS), \mathbb{O}} \, ,
\\ \nonumber \delta_5^{\vec b} \eta_{({\mathbf 8},{\mathbf 1}),N}^{\rm S(PS), \mathbb{O}} &=& 2 b^N f_{NMO} \eta_{({\mathbf 8},{\mathbf 1}),O}^{\rm PS(S), \mathbb{O}} \, .
\end{eqnarray}
There is only one $[(\mathbf{27},{\mathbf 1}) \oplus ({\mathbf 1},\mathbf{27})]$ chiral multiplet, $\big ( \eta_{1,U}^{\rm S, \mathbb{TS}} - \eta_{2,U}^{\rm S, \mathbb{TS}}, \eta_{1,U}^{\rm PS, \mathbb{TS}} - \eta_{2,U}^{\rm PS, \mathbb{TS}} \big )$. Its chiral transformation properties are
\begin{eqnarray}
\nonumber \delta_5 \eta_{1-2,U}^{\rm S(PS), \mathbb{TS}} &=& 0 \, ,
\\ \nonumber \delta^{\vec a} \eta_{1-2,U}^{\rm S(PS) \mathbb{TS}} &=& 2 i a^N \big ({\bf T}^{\bf A}_{27\times27} - {\bf T}^{\bf B}_{27\times27}\big )^N_{UV} \eta_{1-2,V}^{\rm S(PS), \mathbb{TS}} \, ,
\\ \nonumber \delta_5^{\vec b} \eta_{1-2,U}^{\rm S(PS), \mathbb{TS}} &=& 2 i a^N \big ({\bf T}^{\bf A}_{27\times27} - {\bf T}^{\bf B}_{27\times27}\big )^N_{UV} \eta_{1-2,V}^{\rm PS(S), \mathbb{TS}} \, .
\end{eqnarray}

\section{Transition Matrices}
\label{app:matrices}

The transition matrices ${\bf T}_{10\times10}^N$ are
\begin{eqnarray}
\nonumber {\bf T}_{10\times10}^1 &=& \left(
\begin{array}{cccccccccc}
 0 & \frac{1}{\sqrt{3}} & 0 & 0 & 0 & 0 & 0 & 0 & 0 & 0 \\
 \frac{1}{\sqrt{3}} & 0 & \frac{2}{3} & 0 & 0 & 0 & 0 & 0 & 0 & 0 \\
 0 & \frac{2}{3} & 0 & \frac{1}{\sqrt{3}} & 0 & 0 & 0 & 0 & 0 & 0 \\
 0 & 0 & \frac{1}{\sqrt{3}} & 0 & 0 & 0 & 0 & 0 & 0 & 0 \\
 0 & 0 & 0 & 0 & 0 & \frac{\sqrt{2}}{3} & 0 & 0 & 0 & 0 \\
 0 & 0 & 0 & 0 & \frac{\sqrt{2}}{3} & 0 & \frac{\sqrt{2}}{3} & 0 & 0 & 0 \\
 0 & 0 & 0 & 0 & 0 & \frac{\sqrt{2}}{3} & 0 & 0 & 0 & 0 \\
 0 & 0 & 0 & 0 & 0 & 0 & 0 & 0 & \frac{1}{3} & 0 \\
 0 & 0 & 0 & 0 & 0 & 0 & 0 & \frac{1}{3} & 0 & 0 \\
 0 & 0 & 0 & 0 & 0 & 0 & 0 & 0 & 0 & 0
\end{array}
\right) \, ,
\\ \nonumber {\bf T}_{10\times10}^2 &=& \left(
\begin{array}{cccccccccc}
 0 & -\frac{i}{\sqrt{3}} & 0 & 0 & 0 & 0 & 0 & 0 & 0 & 0 \\
 \frac{i}{\sqrt{3}} & 0 & -\frac{2 i}{3} & 0 & 0 & 0 & 0 & 0 & 0 & 0 \\
 0 & \frac{2 i}{3} & 0 & -\frac{i}{\sqrt{3}} & 0 & 0 & 0 & 0 & 0 & 0 \\
 0 & 0 & \frac{i}{\sqrt{3}} & 0 & 0 & 0 & 0 & 0 & 0 & 0 \\
 0 & 0 & 0 & 0 & 0 & -\frac{i \sqrt{2}}{3} & 0 & 0 & 0 & 0 \\
 0 & 0 & 0 & 0 & \frac{i \sqrt{2}}{3} & 0 & -\frac{i \sqrt{2}}{3} & 0 & 0 & 0 \\
 0 & 0 & 0 & 0 & 0 & \frac{i \sqrt{2}}{3} & 0 & 0 & 0 & 0 \\
 0 & 0 & 0 & 0 & 0 & 0 & 0 & 0 & -\frac{i}{3} & 0 \\
 0 & 0 & 0 & 0 & 0 & 0 & 0 & \frac{i}{3} & 0 & 0 \\
 0 & 0 & 0 & 0 & 0 & 0 & 0 & 0 & 0 & 0
\end{array}
\right) \, ,
\\ \nonumber {\bf T}_{10\times10}^3 &=& \left(
\begin{array}{cccccccccc}
 1 & 0 & 0 & 0 & 0 & 0 & 0 & 0 & 0 & 0 \\
 0 & \frac{1}{3} & 0 & 0 & 0 & 0 & 0 & 0 & 0 & 0 \\
 0 & 0 & -\frac{1}{3} & 0 & 0 & 0 & 0 & 0 & 0 & 0 \\
 0 & 0 & 0 & -1 & 0 & 0 & 0 & 0 & 0 & 0 \\
 0 & 0 & 0 & 0 & \frac{2}{3} & 0 & 0 & 0 & 0 & 0 \\
 0 & 0 & 0 & 0 & 0 & 0 & 0 & 0 & 0 & 0 \\
 0 & 0 & 0 & 0 & 0 & 0 & -\frac{2}{3} & 0 & 0 & 0 \\
 0 & 0 & 0 & 0 & 0 & 0 & 0 & \frac{1}{3} & 0 & 0 \\
 0 & 0 & 0 & 0 & 0 & 0 & 0 & 0 & -\frac{1}{3} & 0 \\
 0 & 0 & 0 & 0 & 0 & 0 & 0 & 0 & 0 & 0
\end{array}
\right) \, ,
\\ \nonumber {\bf T}_{10\times10}^4 &=& \left(
\begin{array}{cccccccccc}
 0 & 0 & 0 & 0 & \frac{1}{\sqrt{3}} & 0 & 0 & 0 & 0 & 0 \\
 0 & 0 & 0 & 0 & 0 & \frac{\sqrt{2}}{3} & 0 & 0 & 0 & 0 \\
 0 & 0 & 0 & 0 & 0 & 0 & \frac{1}{3} & 0 & 0 & 0 \\
 0 & 0 & 0 & 0 & 0 & 0 & 0 & 0 & 0 & 0 \\
 \frac{1}{\sqrt{3}} & 0 & 0 & 0 & 0 & 0 & 0 & \frac{2}{3} & 0 & 0 \\
 0 & \frac{\sqrt{2}}{3} & 0 & 0 & 0 & 0 & 0 & 0 & \frac{\sqrt{2}}{3} & 0 \\
 0 & 0 & \frac{1}{3} & 0 & 0 & 0 & 0 & 0 & 0 & 0 \\
 0 & 0 & 0 & 0 & \frac{2}{3} & 0 & 0 & 0 & 0 & \frac{1}{\sqrt{3}} \\
 0 & 0 & 0 & 0 & 0 & \frac{\sqrt{2}}{3} & 0 & 0 & 0 & 0 \\
 0 & 0 & 0 & 0 & 0 & 0 & 0 & \frac{1}{\sqrt{3}} & 0 & 0
\end{array}
\right) \, ,
\\ \nonumber {\bf T}_{10\times10}^5 &=& \left(
\begin{array}{cccccccccc}
 0 & 0 & 0 & 0 & -\frac{i}{\sqrt{3}} & 0 & 0 & 0 & 0 & 0 \\
 0 & 0 & 0 & 0 & 0 & -\frac{i \sqrt{2}}{3} & 0 & 0 & 0 & 0 \\
 0 & 0 & 0 & 0 & 0 & 0 & -\frac{i}{3} & 0 & 0 & 0 \\
 0 & 0 & 0 & 0 & 0 & 0 & 0 & 0 & 0 & 0 \\
 \frac{i}{\sqrt{3}} & 0 & 0 & 0 & 0 & 0 & 0 & -\frac{2 i}{3} & 0 & 0 \\
 0 & \frac{i \sqrt{2}}{3} & 0 & 0 & 0 & 0 & 0 & 0 & -\frac{i \sqrt{2}}{3} & 0 \\
 0 & 0 & \frac{i}{3} & 0 & 0 & 0 & 0 & 0 & 0 & 0 \\
 0 & 0 & 0 & 0 & \frac{2 i}{3} & 0 & 0 & 0 & 0 & -\frac{i}{\sqrt{3}} \\
 0 & 0 & 0 & 0 & 0 & \frac{i \sqrt{2}}{3} & 0 & 0 & 0 & 0 \\
 0 & 0 & 0 & 0 & 0 & 0 & 0 & \frac{i}{\sqrt{3}} & 0 & 0
\end{array}
\right) \, ,
\\ \nonumber {\bf T}_{10\times10}^6 &=& \left(
\begin{array}{cccccccccc}
 0 & 0 & 0 & 0 & 0 & 0 & 0 & 0 & 0 & 0 \\
 0 & 0 & 0 & 0 & \frac{1}{3} & 0 & 0 & 0 & 0 & 0 \\
 0 & 0 & 0 & 0 & 0 & \frac{\sqrt{2}}{3} & 0 & 0 & 0 & 0 \\
 0 & 0 & 0 & 0 & 0 & 0 & \frac{1}{\sqrt{3}} & 0 & 0 & 0 \\
 0 & \frac{1}{3} & 0 & 0 & 0 & 0 & 0 & 0 & 0 & 0 \\
 0 & 0 & \frac{\sqrt{2}}{3} & 0 & 0 & 0 & 0 & \frac{\sqrt{2}}{3} & 0 & 0 \\
 0 & 0 & 0 & \frac{1}{\sqrt{3}} & 0 & 0 & 0 & 0 & \frac{2}{3} & 0 \\
 0 & 0 & 0 & 0 & 0 & \frac{\sqrt{2}}{3} & 0 & 0 & 0 & 0 \\
 0 & 0 & 0 & 0 & 0 & 0 & \frac{2}{3} & 0 & 0 & \frac{1}{\sqrt{3}} \\
 0 & 0 & 0 & 0 & 0 & 0 & 0 & 0 & \frac{1}{\sqrt{3}} & 0
\end{array}
\right) \, ,
\\ \nonumber {\bf T}_{10\times10}^7 &=& \left(
\begin{array}{cccccccccc}
 0 & 0 & 0 & 0 & 0 & 0 & 0 & 0 & 0 & 0 \\
 0 & 0 & 0 & 0 & -\frac{i}{3} & 0 & 0 & 0 & 0 & 0 \\
 0 & 0 & 0 & 0 & 0 & -\frac{i \sqrt{2}}{3} & 0 & 0 & 0 & 0 \\
 0 & 0 & 0 & 0 & 0 & 0 & -\frac{i}{\sqrt{3}} & 0 & 0 & 0 \\
 0 & \frac{i}{3} & 0 & 0 & 0 & 0 & 0 & 0 & 0 & 0 \\
 0 & 0 & \frac{i \sqrt{2}}{3} & 0 & 0 & 0 & 0 & -\frac{i \sqrt{2}}{3} & 0 & 0 \\
 0 & 0 & 0 & \frac{i}{\sqrt{3}} & 0 & 0 & 0 & 0 & -\frac{2 i}{3} & 0 \\
 0 & 0 & 0 & 0 & 0 & \frac{i \sqrt{2}}{3} & 0 & 0 & 0 & 0 \\
 0 & 0 & 0 & 0 & 0 & 0 & \frac{2 i}{3} & 0 & 0 & -\frac{i}{\sqrt{3}} \\
 0 & 0 & 0 & 0 & 0 & 0 & 0 & 0 & \frac{i}{\sqrt{3}} & 0
\end{array}
\right) \, ,
\\ \nonumber {\bf T}_{10\times10}^8 &=& \left(
\begin{array}{cccccccccc}
 \frac{1}{\sqrt{3}} & 0 & 0 & 0 & 0 & 0 & 0 & 0 & 0 & 0 \\
 0 & \frac{1}{\sqrt{3}} & 0 & 0 & 0 & 0 & 0 & 0 & 0 & 0 \\
 0 & 0 & \frac{1}{\sqrt{3}} & 0 & 0 & 0 & 0 & 0 & 0 & 0 \\
 0 & 0 & 0 & \frac{1}{\sqrt{3}} & 0 & 0 & 0 & 0 & 0 & 0 \\
 0 & 0 & 0 & 0 & 0 & 0 & 0 & 0 & 0 & 0 \\
 0 & 0 & 0 & 0 & 0 & 0 & 0 & 0 & 0 & 0 \\
 0 & 0 & 0 & 0 & 0 & 0 & 0 & 0 & 0 & 0 \\
 0 & 0 & 0 & 0 & 0 & 0 & 0 & -\frac{1}{\sqrt{3}} & 0 & 0 \\
 0 & 0 & 0 & 0 & 0 & 0 & 0 & 0 & -\frac{1}{\sqrt{3}} & 0 \\
 0 & 0 & 0 & 0 & 0 & 0 & 0 & 0 & 0 & -\frac{2}{\sqrt{3}}
\end{array}
\right) \, .
\end{eqnarray}

The transition matrices $({\bf T}^A_{10\times27})^{N\dagger}$ are
\begin{eqnarray}
\nonumber ({\bf T}^A_{10\times27})^{1\dagger} &=&
\left(
\begin{array}{cccccccccc}
 0 & 0 & 0 & 0 & \frac{2}{3} \sqrt{\frac{2}{5}} & 0 & -\frac{2}{3} \sqrt{\frac{2}{5}} & 0 & 0 & 0 \\
 0 & 0 & 0 & 0 & 0 & \sqrt{\frac{3}{10}} & 0 & 0 & 0 & 0 \\
 0 & 0 & 0 & 0 & \sqrt{\frac{3}{10}} & 0 & \sqrt{\frac{3}{10}} & 0 & 0 & 0 \\
 0 & 0 & 0 & 0 & 0 & -\sqrt{\frac{3}{10}} & 0 & 0 & 0 & 0 \\
 0 & 0 & 0 & 0 & 0 & 0 & 0 & \frac{7}{3 \sqrt{10}} & 0 & 0 \\
 0 & 0 & 0 & 0 & 0 & 0 & 0 & 0 & -\frac{7}{3 \sqrt{10}} & 0 \\
 0 & -\frac{1}{3}\sqrt{\frac{2}{5}} & 0 & \sqrt{\frac{2}{15}} & 0 & 0 & 0 & 0 & 0 & 0 \\
 -\sqrt{\frac{2}{15}} & 0 & \frac{1}{3}\sqrt{\frac{2}{5}} & 0 & 0 & 0 & 0 & 0 & 0 & 0 \\
 0 & 0 & 0 & 0 & 0 & 0 & \frac{1}{\sqrt{3}} & 0 & 0 & 0 \\
 0 & 0 & 0 & 0 & 0 & \frac{1}{\sqrt{6}} & 0 & 0 & 0 & 0 \\
 0 & 0 & 0 & 0 & \frac{1}{3 \sqrt{2}} & 0 & -\frac{1}{3 \sqrt{2}} & 0 & 0 & 0 \\
 0 & 0 & 0 & 0 & 0 & \frac{1}{\sqrt{6}} & 0 & 0 & 0 & 0 \\
 0 & 0 & 0 & 0 & -\frac{1}{\sqrt{3}} & 0 & 0 & 0 & 0 & 0 \\
 0 & 0 & 0 & 0 & 0 & 0 & 0 & 0 & \sqrt{\frac{2}{3}} & 0 \\
 0 & 0 & 0 & 0 & 0 & 0 & 0 & \frac{\sqrt{2}}{3} & 0 & 0 \\
 0 & 0 & 0 & 0 & 0 & 0 & 0 & 0 & \frac{\sqrt{2}}{3} & 0 \\
 0 & 0 & 0 & 0 & 0 & 0 & 0 & -\sqrt{\frac{2}{3}} & 0 & 0 \\
 0 & 0 & -\frac{1}{\sqrt{6}} & 0 & 0 & 0 & 0 & 0 & 0 & 0 \\
 0 & -\frac{\sqrt{2}}{3} & 0 & -\frac{1}{\sqrt{6}} & 0 & 0 & 0 & 0 & 0 & 0 \\
 -\frac{1}{\sqrt{6}} & 0 & -\frac{\sqrt{2}}{3} & 0 & 0 & 0 & 0 & 0 & 0 & 0 \\
 0 & \frac{1}{\sqrt{6}} & 0 & 0 & 0 & 0 & 0 & 0 & 0 & 0 \\
 0 & 0 & 0 & 0 & 0 & 0 & 0 & 0 & 0 & 1 \\
 0 & 0 & 0 & 0 & 0 & 0 & 0 & 0 & 0 & 0 \\
 0 & 0 & 0 & 0 & 0 & 0 & 0 & 0 & 0 & -1 \\
 0 & 0 & 0 & 0 & 0 & 0 & 0 & 0 & 0 & 0 \\
 0 & 0 & 0 & 0 & 0 & 0 & 0 & 0 & 0 & 0 \\
 0 & 0 & 0 & 0 & 0 & 0 & 0 & 0 & 0 & 0
\end{array}
\right) \, ,
\\ \nonumber ({\bf T}^A_{10\times27})^{2\dagger} &=&
\left(
\begin{array}{cccccccccc}
 0 & 0 & 0 & 0 & \frac{2}{3} i \sqrt{\frac{2}{5}} & 0 & \frac{2}{3} i \sqrt{\frac{2}{5}} & 0 & 0 & 0 \\
 0 & 0 & 0 & 0 & 0 & i \sqrt{\frac{3}{10}} & 0 & 0 & 0 & 0 \\
 0 & 0 & 0 & 0 & i \sqrt{\frac{3}{10}} & 0 & -i \sqrt{\frac{3}{10}} & 0 & 0 & 0 \\
 0 & 0 & 0 & 0 & 0 & i \sqrt{\frac{3}{10}} & 0 & 0 & 0 & 0 \\
 0 & 0 & 0 & 0 & 0 & 0 & 0 & \frac{7 i}{3 \sqrt{10}} & 0 & 0 \\
 0 & 0 & 0 & 0 & 0 & 0 & 0 & 0 & \frac{7 i}{3 \sqrt{10}} & 0 \\
 0 & -\frac{1}{3} i \sqrt{\frac{2}{5}} & 0 & -i \sqrt{\frac{2}{15}} & 0 & 0 & 0 & 0 & 0 & 0 \\
 -i \sqrt{\frac{2}{15}} & 0 & -\frac{1}{3} i \sqrt{\frac{2}{5}} & 0 & 0 & 0 & 0 & 0 & 0 & 0 \\
 0 & 0 & 0 & 0 & 0 & 0 & \frac{i}{\sqrt{3}} & 0 & 0 & 0 \\
 0 & 0 & 0 & 0 & 0 & \frac{i}{\sqrt{6}} & 0 & 0 & 0 & 0 \\
 0 & 0 & 0 & 0 & \frac{i}{3 \sqrt{2}} & 0 & \frac{i}{3 \sqrt{2}} & 0 & 0 & 0 \\
 0 & 0 & 0 & 0 & 0 & -\frac{i}{\sqrt{6}} & 0 & 0 & 0 & 0 \\
 0 & 0 & 0 & 0 & \frac{i}{\sqrt{3}} & 0 & 0 & 0 & 0 & 0 \\
 0 & 0 & 0 & 0 & 0 & 0 & 0 & 0 & i \sqrt{\frac{2}{3}} & 0 \\
 0 & 0 & 0 & 0 & 0 & 0 & 0 & \frac{i \sqrt{2}}{3} & 0 & 0 \\
 0 & 0 & 0 & 0 & 0 & 0 & 0 & 0 & -\frac{i \sqrt{2}}{3} & 0 \\
 0 & 0 & 0 & 0 & 0 & 0 & 0 & i \sqrt{\frac{2}{3}} & 0 & 0 \\
 0 & 0 & -\frac{i}{\sqrt{6}} & 0 & 0 & 0 & 0 & 0 & 0 & 0 \\
 0 & -\frac{i \sqrt{2}}{3} & 0 & \frac{i}{\sqrt{6}} & 0 & 0 & 0 & 0 & 0 & 0 \\
 -\frac{i}{\sqrt{6}} & 0 & \frac{i \sqrt{2}}{3} & 0 & 0 & 0 & 0 & 0 & 0 & 0 \\
 0 & -\frac{i}{\sqrt{6}} & 0 & 0 & 0 & 0 & 0 & 0 & 0 & 0 \\
 0 & 0 & 0 & 0 & 0 & 0 & 0 & 0 & 0 & i \\
 0 & 0 & 0 & 0 & 0 & 0 & 0 & 0 & 0 & 0 \\
 0 & 0 & 0 & 0 & 0 & 0 & 0 & 0 & 0 & i \\
 0 & 0 & 0 & 0 & 0 & 0 & 0 & 0 & 0 & 0 \\
 0 & 0 & 0 & 0 & 0 & 0 & 0 & 0 & 0 & 0 \\
 0 & 0 & 0 & 0 & 0 & 0 & 0 & 0 & 0 & 0
\end{array}
\right) \, ,
\\ \nonumber ({\bf T}^A_{10\times27})^{3\dagger} &=&
\left(
\begin{array}{cccccccccc}
 0 & 0 & 0 & 0 & 0 & -\frac{4}{3 \sqrt{5}} & 0 & 0 & 0 & 0 \\
 0 & 0 & 0 & 0 & 0 & 0 & -\sqrt{\frac{3}{5}} & 0 & 0 & 0 \\
 0 & 0 & 0 & 0 & 0 & 0 & 0 & 0 & 0 & 0 \\
 0 & 0 & 0 & 0 & -\sqrt{\frac{3}{5}} & 0 & 0 & 0 & 0 & 0 \\
 0 & 0 & 0 & 0 & 0 & 0 & 0 & 0 & -\frac{7}{3 \sqrt{10}} & 0 \\
 0 & 0 & 0 & 0 & 0 & 0 & 0 & -\frac{7}{3 \sqrt{10}} & 0 & 0 \\
 0 & 0 & \frac{2}{3} \sqrt{\frac{2}{5}} & 0 & 0 & 0 & 0 & 0 & 0 & 0 \\
 0 & \frac{2}{3} \sqrt{\frac{2}{5}} & 0 & 0 & 0 & 0 & 0 & 0 & 0 & 0 \\
 0 & 0 & 0 & 0 & 0 & 0 & 0 & 0 & 0 & 0 \\
 0 & 0 & 0 & 0 & 0 & 0 & \frac{1}{\sqrt{3}} & 0 & 0 & 0 \\
 0 & 0 & 0 & 0 & 0 & \frac{2}{3} & 0 & 0 & 0 & 0 \\
 0 & 0 & 0 & 0 & -\frac{1}{\sqrt{3}} & 0 & 0 & 0 & 0 & 0 \\
 0 & 0 & 0 & 0 & 0 & 0 & 0 & 0 & 0 & 0 \\
 0 & 0 & 0 & 0 & 0 & 0 & 0 & 0 & 0 & 0 \\
 0 & 0 & 0 & 0 & 0 & 0 & 0 & 0 & \frac{2 \sqrt{2}}{3} & 0 \\
 0 & 0 & 0 & 0 & 0 & 0 & 0 & -\frac{2 \sqrt{2}}{3} & 0 & 0 \\
 0 & 0 & 0 & 0 & 0 & 0 & 0 & 0 & 0 & 0 \\
 0 & 0 & 0 & \frac{1}{\sqrt{2}} & 0 & 0 & 0 & 0 & 0 & 0 \\
 0 & 0 & \frac{1}{3 \sqrt{2}} & 0 & 0 & 0 & 0 & 0 & 0 & 0 \\
 0 & -\frac{1}{3 \sqrt{2}} & 0 & 0 & 0 & 0 & 0 & 0 & 0 & 0 \\
 \frac{1}{\sqrt{2}} & 0 & 0 & 0 & 0 & 0 & 0 & 0 & 0 & 0 \\
 0 & 0 & 0 & 0 & 0 & 0 & 0 & 0 & 0 & 0 \\
 0 & 0 & 0 & 0 & 0 & 0 & 0 & 0 & 0 & -\sqrt{2} \\
 0 & 0 & 0 & 0 & 0 & 0 & 0 & 0 & 0 & 0 \\
 0 & 0 & 0 & 0 & 0 & 0 & 0 & 0 & 0 & 0 \\
 0 & 0 & 0 & 0 & 0 & 0 & 0 & 0 & 0 & 0 \\
 0 & 0 & 0 & 0 & 0 & 0 & 0 & 0 & 0 & 0
\end{array}
\right) \, ,
\\ \nonumber ({\bf T}^A_{10\times27})^{4\dagger} &=&
\left(
\begin{array}{cccccccccc}
 0 & 0 & 0 & 0 & 0 & 0 & 0 & 0 & \sqrt{\frac{2}{5}} & 0 \\
 0 & 0 & -\frac{1}{2 \sqrt{15}} & 0 & 0 & 0 & 0 & 0 & 0 & 0 \\
 0 & -\frac{1}{\sqrt{30}} & 0 & 0 & 0 & 0 & 0 & 0 & -\sqrt{\frac{2}{15}} & 0 \\
 \frac{1}{2 \sqrt{5}} & 0 & 0 & 0 & 0 & 0 & 0 & \frac{2}{\sqrt{15}} & 0 & 0 \\
 0 & 0 & 0 & 0 & 0 & -\frac{1}{3 \sqrt{5}} & 0 & 0 & 0 & 0 \\
 0 & 0 & 0 & 0 & \frac{1}{3}\sqrt{\frac{2}{5}} & 0 & 0 & 0 & 0 & \sqrt{\frac{3}{10}} \\
 0 & 0 & 0 & 0 & 0 & 0 & -\frac{4}{3} \sqrt{\frac{2}{5}} & 0 & 0 & 0 \\
 0 & 0 & 0 & 0 & 0 & -\frac{4}{3 \sqrt{5}} & 0 & 0 & 0 & 0 \\
 0 & 0 & 0 & -1 & 0 & 0 & 0 & 0 & 0 & 0 \\
 0 & 0 & -\frac{\sqrt{3}}{2} & 0 & 0 & 0 & 0 & 0 & 0 & 0 \\
 0 & -\frac{1}{\sqrt{2}} & 0 & 0 & 0 & 0 & 0 & 0 & 0 & 0 \\
 \frac{1}{2} & 0 & 0 & 0 & 0 & 0 & 0 & 0 & 0 & 0 \\
 0 & 0 & 0 & 0 & 0 & 0 & 0 & 0 & 0 & 0 \\
 0 & 0 & 0 & 0 & 0 & 0 & -\sqrt{\frac{2}{3}} & 0 & 0 & 0 \\
 0 & 0 & 0 & 0 & 0 & -\frac{2}{3} & 0 & 0 & 0 & 0 \\
 0 & 0 & 0 & 0 & \frac{\sqrt{2}}{3} & 0 & 0 & 0 & 0 & 0 \\
 0 & 0 & 0 & 0 & 0 & 0 & 0 & 0 & 0 & 0 \\
 0 & 0 & 0 & 0 & 0 & 0 & 0 & 0 & 0 & 0 \\
 0 & 0 & 0 & 0 & 0 & 0 & \frac{1}{3 \sqrt{2}} & 0 & 0 & 0 \\
 0 & 0 & 0 & 0 & 0 & \frac{1}{3} & 0 & 0 & 0 & 0 \\
 0 & 0 & 0 & 0 & -\frac{1}{\sqrt{6}} & 0 & 0 & 0 & 0 & 0 \\
 0 & 0 & 0 & 0 & 0 & 0 & 0 & 0 & -\frac{1}{\sqrt{3}} & 0 \\
 0 & 0 & 0 & 0 & 0 & 0 & 0 & \frac{1}{\sqrt{6}} & 0 & 0 \\
 0 & 0 & 0 & 0 & 0 & 0 & 0 & 0 & 0 & 0 \\
 0 & 0 & 0 & 1 & 0 & 0 & 0 & 0 & 0 & 0 \\
 0 & 0 & \sqrt{\frac{2}{3}} & 0 & 0 & 0 & 0 & 0 & 0 & 0 \\
 0 & \frac{1}{\sqrt{3}} & 0 & 0 & 0 & 0 & 0 & 0 & 0 & 0
\end{array}
\right) \, ,
\\ \nonumber ({\bf T}^A_{10\times27})^{5\dagger} &=&
\left(
\begin{array}{cccccccccc}
 0 & 0 & 0 & 0 & 0 & 0 & 0 & 0 & -i \sqrt{\frac{2}{5}} & 0 \\
 0 & 0 & -\frac{i}{2 \sqrt{15}} & 0 & 0 & 0 & 0 & 0 & 0 & 0 \\
 0 & -\frac{i}{\sqrt{30}} & 0 & 0 & 0 & 0 & 0 & 0 & i \sqrt{\frac{2}{15}} & 0 \\
 \frac{i}{2 \sqrt{5}} & 0 & 0 & 0 & 0 & 0 & 0 & -\frac{2 i}{\sqrt{15}} & 0 & 0 \\
 0 & 0 & 0 & 0 & 0 & -\frac{i}{3 \sqrt{5}} & 0 & 0 & 0 & 0 \\
 0 & 0 & 0 & 0 & \frac{1}{3} i \sqrt{\frac{2}{5}} & 0 & 0 & 0 & 0 & -i \sqrt{\frac{3}{10}} \\
 0 & 0 & 0 & 0 & 0 & 0 & \frac{4}{3} i \sqrt{\frac{2}{5}} & 0 & 0 & 0 \\
 0 & 0 & 0 & 0 & 0 & \frac{4 i}{3 \sqrt{5}} & 0 & 0 & 0 & 0 \\
 0 & 0 & 0 & -i & 0 & 0 & 0 & 0 & 0 & 0 \\
 0 & 0 & -\frac{i \sqrt{3}}{2} & 0 & 0 & 0 & 0 & 0 & 0 & 0 \\
 0 & -\frac{i}{\sqrt{2}} & 0 & 0 & 0 & 0 & 0 & 0 & 0 & 0 \\
 \frac{i}{2} & 0 & 0 & 0 & 0 & 0 & 0 & 0 & 0 & 0 \\
 0 & 0 & 0 & 0 & 0 & 0 & 0 & 0 & 0 & 0 \\
 0 & 0 & 0 & 0 & 0 & 0 & -i \sqrt{\frac{2}{3}} & 0 & 0 & 0 \\
 0 & 0 & 0 & 0 & 0 & -\frac{2 i}{3} & 0 & 0 & 0 & 0 \\
 0 & 0 & 0 & 0 & \frac{i \sqrt{2}}{3} & 0 & 0 & 0 & 0 & 0 \\
 0 & 0 & 0 & 0 & 0 & 0 & 0 & 0 & 0 & 0 \\
 0 & 0 & 0 & 0 & 0 & 0 & 0 & 0 & 0 & 0 \\
 0 & 0 & 0 & 0 & 0 & 0 & -\frac{i}{3 \sqrt{2}} & 0 & 0 & 0 \\
 0 & 0 & 0 & 0 & 0 & -\frac{i}{3} & 0 & 0 & 0 & 0 \\
 0 & 0 & 0 & 0 & \frac{i}{\sqrt{6}} & 0 & 0 & 0 & 0 & 0 \\
 0 & 0 & 0 & 0 & 0 & 0 & 0 & 0 & -\frac{i}{\sqrt{3}} & 0 \\
 0 & 0 & 0 & 0 & 0 & 0 & 0 & \frac{i}{\sqrt{6}} & 0 & 0 \\
 0 & 0 & 0 & 0 & 0 & 0 & 0 & 0 & 0 & 0 \\
 0 & 0 & 0 & -i & 0 & 0 & 0 & 0 & 0 & 0 \\
 0 & 0 & -i \sqrt{\frac{2}{3}} & 0 & 0 & 0 & 0 & 0 & 0 & 0 \\
 0 & -\frac{i}{\sqrt{3}} & 0 & 0 & 0 & 0 & 0 & 0 & 0 & 0
\end{array}
\right) \, ,
\\ \nonumber ({\bf T}^A_{10\times27})^{6\dagger} &=&
\left(
\begin{array}{cccccccccc}
 0 & 0 & 0 & 0 & 0 & 0 & 0 & -\sqrt{\frac{2}{5}} & 0 & 0 \\
 0 & 0 & 0 & -\frac{1}{2 \sqrt{5}} & 0 & 0 & 0 & 0 & -\frac{2}{\sqrt{15}} & 0 \\
 0 & 0 & -\frac{1}{\sqrt{30}} & 0 & 0 & 0 & 0 & -\sqrt{\frac{2}{15}} & 0 & 0 \\
 0 & \frac{1}{2 \sqrt{15}} & 0 & 0 & 0 & 0 & 0 & 0 & 0 & 0 \\
 0 & 0 & 0 & 0 & 0 & 0 & -\frac{1}{3}\sqrt{\frac{2}{5}} & 0 & 0 & -\sqrt{\frac{3}{10}} \\
 0 & 0 & 0 & 0 & 0 & \frac{1}{3 \sqrt{5}} & 0 & 0 & 0 & 0 \\
 0 & 0 & 0 & 0 & 0 & \frac{4}{3 \sqrt{5}} & 0 & 0 & 0 & 0 \\
 0 & 0 & 0 & 0 & \frac{4}{3} \sqrt{\frac{2}{5}} & 0 & 0 & 0 & 0 & 0 \\
 0 & 0 & 0 & 0 & 0 & 0 & 0 & 0 & 0 & 0 \\
 0 & 0 & 0 & \frac{1}{2} & 0 & 0 & 0 & 0 & 0 & 0 \\
 0 & 0 & \frac{1}{\sqrt{2}} & 0 & 0 & 0 & 0 & 0 & 0 & 0 \\
 0 & -\frac{\sqrt{3}}{2} & 0 & 0 & 0 & 0 & 0 & 0 & 0 & 0 \\
 1 & 0 & 0 & 0 & 0 & 0 & 0 & 0 & 0 & 0 \\
 0 & 0 & 0 & 0 & 0 & 0 & 0 & 0 & 0 & 0 \\
 0 & 0 & 0 & 0 & 0 & 0 & \frac{\sqrt{2}}{3} & 0 & 0 & 0 \\
 0 & 0 & 0 & 0 & 0 & -\frac{2}{3} & 0 & 0 & 0 & 0 \\
 0 & 0 & 0 & 0 & \sqrt{\frac{2}{3}} & 0 & 0 & 0 & 0 & 0 \\
 0 & 0 & 0 & 0 & 0 & 0 & \frac{1}{\sqrt{6}} & 0 & 0 & 0 \\
 0 & 0 & 0 & 0 & 0 & \frac{1}{3} & 0 & 0 & 0 & 0 \\
 0 & 0 & 0 & 0 & \frac{1}{3 \sqrt{2}} & 0 & 0 & 0 & 0 & 0 \\
 0 & 0 & 0 & 0 & 0 & 0 & 0 & 0 & 0 & 0 \\
 0 & 0 & 0 & 0 & 0 & 0 & 0 & 0 & 0 & 0 \\
 0 & 0 & 0 & 0 & 0 & 0 & 0 & 0 & -\frac{1}{\sqrt{6}} & 0 \\
 0 & 0 & 0 & 0 & 0 & 0 & 0 & \frac{1}{\sqrt{3}} & 0 & 0 \\
 0 & 0 & -\frac{1}{\sqrt{3}} & 0 & 0 & 0 & 0 & 0 & 0 & 0 \\
 0 & -\sqrt{\frac{2}{3}} & 0 & 0 & 0 & 0 & 0 & 0 & 0 & 0 \\
 -1 & 0 & 0 & 0 & 0 & 0 & 0 & 0 & 0 & 0
\end{array}
\right) \, ,
\\ \nonumber ({\bf T}^A_{10\times27})^{7\dagger} &=&
\left(
\begin{array}{cccccccccc}
 0 & 0 & 0 & 0 & 0 & 0 & 0 & i \sqrt{\frac{2}{5}} & 0 & 0 \\
 0 & 0 & 0 & -\frac{i}{2 \sqrt{5}} & 0 & 0 & 0 & 0 & \frac{2 i}{\sqrt{15}} & 0 \\
 0 & 0 & -\frac{i}{\sqrt{30}} & 0 & 0 & 0 & 0 & i \sqrt{\frac{2}{15}} & 0 & 0 \\
 0 & \frac{i}{2 \sqrt{15}} & 0 & 0 & 0 & 0 & 0 & 0 & 0 & 0 \\
 0 & 0 & 0 & 0 & 0 & 0 & -\frac{1}{3} i \sqrt{\frac{2}{5}} & 0 & 0 & i \sqrt{\frac{3}{10}} \\
 0 & 0 & 0 & 0 & 0 & \frac{i}{3 \sqrt{5}} & 0 & 0 & 0 & 0 \\
 0 & 0 & 0 & 0 & 0 & -\frac{4 i}{3 \sqrt{5}} & 0 & 0 & 0 & 0 \\
 0 & 0 & 0 & 0 & -\frac{4}{3} i \sqrt{\frac{2}{5}} & 0 & 0 & 0 & 0 & 0 \\
 0 & 0 & 0 & 0 & 0 & 0 & 0 & 0 & 0 & 0 \\
 0 & 0 & 0 & \frac{i}{2} & 0 & 0 & 0 & 0 & 0 & 0 \\
 0 & 0 & \frac{i}{\sqrt{2}} & 0 & 0 & 0 & 0 & 0 & 0 & 0 \\
 0 & -\frac{i \sqrt{3}}{2} & 0 & 0 & 0 & 0 & 0 & 0 & 0 & 0 \\
 i & 0 & 0 & 0 & 0 & 0 & 0 & 0 & 0 & 0 \\
 0 & 0 & 0 & 0 & 0 & 0 & 0 & 0 & 0 & 0 \\
 0 & 0 & 0 & 0 & 0 & 0 & \frac{i \sqrt{2}}{3} & 0 & 0 & 0 \\
 0 & 0 & 0 & 0 & 0 & -\frac{2 i}{3} & 0 & 0 & 0 & 0 \\
 0 & 0 & 0 & 0 & i \sqrt{\frac{2}{3}} & 0 & 0 & 0 & 0 & 0 \\
 0 & 0 & 0 & 0 & 0 & 0 & -\frac{i}{\sqrt{6}} & 0 & 0 & 0 \\
 0 & 0 & 0 & 0 & 0 & -\frac{i}{3} & 0 & 0 & 0 & 0 \\
 0 & 0 & 0 & 0 & -\frac{i}{3 \sqrt{2}} & 0 & 0 & 0 & 0 & 0 \\
 0 & 0 & 0 & 0 & 0 & 0 & 0 & 0 & 0 & 0 \\
 0 & 0 & 0 & 0 & 0 & 0 & 0 & 0 & 0 & 0 \\
 0 & 0 & 0 & 0 & 0 & 0 & 0 & 0 & -\frac{i}{\sqrt{6}} & 0 \\
 0 & 0 & 0 & 0 & 0 & 0 & 0 & \frac{i}{\sqrt{3}} & 0 & 0 \\
 0 & 0 & \frac{i}{\sqrt{3}} & 0 & 0 & 0 & 0 & 0 & 0 & 0 \\
 0 & i \sqrt{\frac{2}{3}} & 0 & 0 & 0 & 0 & 0 & 0 & 0 & 0 \\
 i & 0 & 0 & 0 & 0 & 0 & 0 & 0 & 0 & 0
\end{array}
\right) \, ,
\\ \nonumber ({\bf T}^A_{10\times27})^{8\dagger} &=&
\left(
\begin{array}{cccccccccc}
 0 & 0 & 0 & 0 & 0 & 0 & 0 & 0 & 0 & 0 \\
 0 & 0 & 0 & 0 & 0 & 0 & -\frac{2}{\sqrt{5}} & 0 & 0 & 0 \\
 0 & 0 & 0 & 0 & 0 & -\frac{2}{\sqrt{5}} & 0 & 0 & 0 & 0 \\
 0 & 0 & 0 & 0 & \frac{2}{\sqrt{5}} & 0 & 0 & 0 & 0 & 0 \\
 0 & 0 & 0 & 0 & 0 & 0 & 0 & 0 & -\sqrt{\frac{3}{10}} & 0 \\
 0 & 0 & 0 & 0 & 0 & 0 & 0 & \sqrt{\frac{3}{10}} & 0 & 0 \\
 0 & 0 & 0 & 0 & 0 & 0 & 0 & 0 & 0 & 0 \\
 0 & 0 & 0 & 0 & 0 & 0 & 0 & 0 & 0 & 0 \\
 0 & 0 & 0 & 0 & 0 & 0 & 0 & 0 & 0 & 0 \\
 0 & 0 & 0 & 0 & 0 & 0 & 0 & 0 & 0 & 0 \\
 0 & 0 & 0 & 0 & 0 & 0 & 0 & 0 & 0 & 0 \\
 0 & 0 & 0 & 0 & 0 & 0 & 0 & 0 & 0 & 0 \\
 0 & 0 & 0 & 0 & 0 & 0 & 0 & 0 & 0 & 0 \\
 0 & 0 & 0 & 0 & 0 & 0 & 0 & 0 & 0 & 0 \\
 0 & 0 & 0 & 0 & 0 & 0 & 0 & 0 & 0 & 0 \\
 0 & 0 & 0 & 0 & 0 & 0 & 0 & 0 & 0 & 0 \\
 0 & 0 & 0 & 0 & 0 & 0 & 0 & 0 & 0 & 0 \\
 0 & 0 & 0 & \sqrt{\frac{3}{2}} & 0 & 0 & 0 & 0 & 0 & 0 \\
 0 & 0 & \sqrt{\frac{3}{2}} & 0 & 0 & 0 & 0 & 0 & 0 & 0 \\
 0 & \sqrt{\frac{3}{2}} & 0 & 0 & 0 & 0 & 0 & 0 & 0 & 0 \\
 -\sqrt{\frac{3}{2}} & 0 & 0 & 0 & 0 & 0 & 0 & 0 & 0 & 0 \\
 0 & 0 & 0 & 0 & 0 & 0 & 0 & 0 & 0 & 0 \\
 0 & 0 & 0 & 0 & 0 & 0 & 0 & 0 & 0 & 0 \\
 0 & 0 & 0 & 0 & 0 & 0 & 0 & 0 & 0 & 0 \\
 0 & 0 & 0 & 0 & 0 & 0 & 0 & 0 & 0 & 0 \\
 0 & 0 & 0 & 0 & 0 & 0 & 0 & 0 & 0 & 0 \\
 0 & 0 & 0 & 0 & 0 & 0 & 0 & 0 & 0 & 0
\end{array}
\right) \, .
\end{eqnarray}

The transition matrices $({\bf T}^B_{10\times27})^{N\dagger}$ are
\begin{eqnarray}
\nonumber ({\bf T}^B_{10\times27})^{1\dagger} &=&
\left(
\begin{array}{cccccccccc}
 0 & 0 & 0 & 0 & \frac{2}{3} \sqrt{\frac{2}{5}} & 0 & -\frac{2}{3} \sqrt{\frac{2}{5}} & 0 & 0 & 0 \\
 0 & 0 & 0 & 0 & 0 & -\sqrt{\frac{3}{10}} & 0 & 0 & 0 & 0 \\
 0 & 0 & 0 & 0 & \sqrt{\frac{3}{10}} & 0 & \sqrt{\frac{3}{10}} & 0 & 0 & 0 \\
 0 & 0 & 0 & 0 & 0 & \sqrt{\frac{3}{10}} & 0 & 0 & 0 & 0 \\
 -\sqrt{\frac{2}{15}} & 0 & \frac{\sqrt{\frac{2}{5}}}{3} & 0 & 0 & 0 & 0 & 0 & 0 & 0 \\
 0 & -\frac{\sqrt{\frac{2}{5}}}{3} & 0 & \sqrt{\frac{2}{15}} & 0 & 0 & 0 & 0 & 0 & 0 \\
 0 & 0 & 0 & 0 & 0 & 0 & 0 & 0 & -\frac{7}{3 \sqrt{10}} & 0 \\
 0 & 0 & 0 & 0 & 0 & 0 & 0 & \frac{7}{3 \sqrt{10}} & 0 & 0 \\
 0 & 0 & 0 & 0 & -\frac{1}{\sqrt{3}} & 0 & 0 & 0 & 0 & 0 \\
 0 & 0 & 0 & 0 & 0 & \frac{1}{\sqrt{6}} & 0 & 0 & 0 & 0 \\
 0 & 0 & 0 & 0 & \frac{1}{3 \sqrt{2}} & 0 & -\frac{1}{3 \sqrt{2}} & 0 & 0 & 0 \\
 0 & 0 & 0 & 0 & 0 & \frac{1}{\sqrt{6}} & 0 & 0 & 0 & 0 \\
 0 & 0 & 0 & 0 & 0 & 0 & \frac{1}{\sqrt{3}} & 0 & 0 & 0 \\
 0 & \frac{1}{\sqrt{6}} & 0 & 0 & 0 & 0 & 0 & 0 & 0 & 0 \\
 -\frac{1}{\sqrt{6}} & 0 & -\frac{\sqrt{2}}{3} & 0 & 0 & 0 & 0 & 0 & 0 & 0 \\
 0 & -\frac{\sqrt{2}}{3} & 0 & -\frac{1}{\sqrt{6}} & 0 & 0 & 0 & 0 & 0 & 0 \\
 0 & 0 & -\frac{1}{\sqrt{6}} & 0 & 0 & 0 & 0 & 0 & 0 & 0 \\
 0 & 0 & 0 & 0 & 0 & 0 & 0 & -\sqrt{\frac{2}{3}} & 0 & 0 \\
 0 & 0 & 0 & 0 & 0 & 0 & 0 & 0 & \frac{\sqrt{2}}{3} & 0 \\
 0 & 0 & 0 & 0 & 0 & 0 & 0 & \frac{\sqrt{2}}{3} & 0 & 0 \\
 0 & 0 & 0 & 0 & 0 & 0 & 0 & 0 & \sqrt{\frac{2}{3}} & 0 \\
 0 & 0 & 0 & 0 & 0 & 0 & 0 & 0 & 0 & 0 \\
 0 & 0 & 0 & 0 & 0 & 0 & 0 & 0 & 0 & 0 \\
 0 & 0 & 0 & 0 & 0 & 0 & 0 & 0 & 0 & 0 \\
 0 & 0 & 0 & 0 & 0 & 0 & 0 & 0 & 0 & -1 \\
 0 & 0 & 0 & 0 & 0 & 0 & 0 & 0 & 0 & 0 \\
 0 & 0 & 0 & 0 & 0 & 0 & 0 & 0 & 0 & 1
\end{array}
\right) \, ,
\\ \nonumber ({\bf T}^B_{10\times27})^{2\dagger} &=&
\left(
\begin{array}{cccccccccc}
 0 & 0 & 0 & 0 & -\frac{2}{3} i \sqrt{\frac{2}{5}} & 0 & -\frac{2}{3} i \sqrt{\frac{2}{5}} & 0 & 0 & 0 \\
 0 & 0 & 0 & 0 & 0 & -i \sqrt{\frac{3}{10}} & 0 & 0 & 0 & 0 \\
 0 & 0 & 0 & 0 & -i \sqrt{\frac{3}{10}} & 0 & i \sqrt{\frac{3}{10}} & 0 & 0 & 0 \\
 0 & 0 & 0 & 0 & 0 & -i \sqrt{\frac{3}{10}} & 0 & 0 & 0 & 0 \\
 i \sqrt{\frac{2}{15}} & 0 & \frac{1}{3} i \sqrt{\frac{2}{5}} & 0 & 0 & 0 & 0 & 0 & 0 & 0 \\
 0 & \frac{1}{3} i \sqrt{\frac{2}{5}} & 0 & i \sqrt{\frac{2}{15}} & 0 & 0 & 0 & 0 & 0 & 0 \\
 0 & 0 & 0 & 0 & 0 & 0 & 0 & 0 & -\frac{7 i}{3 \sqrt{10}} & 0 \\
 0 & 0 & 0 & 0 & 0 & 0 & 0 & -\frac{7 i}{3 \sqrt{10}} & 0 & 0 \\
 0 & 0 & 0 & 0 & -\frac{i}{\sqrt{3}} & 0 & 0 & 0 & 0 & 0 \\
 0 & 0 & 0 & 0 & 0 & \frac{i}{\sqrt{6}} & 0 & 0 & 0 & 0 \\
 0 & 0 & 0 & 0 & -\frac{i}{3 \sqrt{2}} & 0 & -\frac{i}{3 \sqrt{2}} & 0 & 0 & 0 \\
 0 & 0 & 0 & 0 & 0 & -\frac{i}{\sqrt{6}} & 0 & 0 & 0 & 0 \\
 0 & 0 & 0 & 0 & 0 & 0 & -\frac{i}{\sqrt{3}} & 0 & 0 & 0 \\
 0 & \frac{i}{\sqrt{6}} & 0 & 0 & 0 & 0 & 0 & 0 & 0 & 0 \\
 \frac{i}{\sqrt{6}} & 0 & -\frac{i \sqrt{2}}{3} & 0 & 0 & 0 & 0 & 0 & 0 & 0 \\
 0 & \frac{i \sqrt{2}}{3} & 0 & -\frac{i}{\sqrt{6}} & 0 & 0 & 0 & 0 & 0 & 0 \\
 0 & 0 & \frac{i}{\sqrt{6}} & 0 & 0 & 0 & 0 & 0 & 0 & 0 \\
 0 & 0 & 0 & 0 & 0 & 0 & 0 & -i \sqrt{\frac{2}{3}} & 0 & 0 \\
 0 & 0 & 0 & 0 & 0 & 0 & 0 & 0 & \frac{i \sqrt{2}}{3} & 0 \\
 0 & 0 & 0 & 0 & 0 & 0 & 0 & -\frac{i \sqrt{2}}{3} & 0 & 0 \\
 0 & 0 & 0 & 0 & 0 & 0 & 0 & 0 & -i \sqrt{\frac{2}{3}} & 0 \\
 0 & 0 & 0 & 0 & 0 & 0 & 0 & 0 & 0 & 0 \\
 0 & 0 & 0 & 0 & 0 & 0 & 0 & 0 & 0 & 0 \\
 0 & 0 & 0 & 0 & 0 & 0 & 0 & 0 & 0 & 0 \\
 0 & 0 & 0 & 0 & 0 & 0 & 0 & 0 & 0 & -i \\
 0 & 0 & 0 & 0 & 0 & 0 & 0 & 0 & 0 & 0 \\
 0 & 0 & 0 & 0 & 0 & 0 & 0 & 0 & 0 & -i
\end{array}
\right) \, ,
\\ \nonumber ({\bf T}^B_{10\times27})^{3\dagger} &=&
\left(
\begin{array}{cccccccccc}
 0 & 0 & 0 & 0 & 0 & -\frac{4}{3 \sqrt{5}} & 0 & 0 & 0 & 0 \\
 0 & 0 & 0 & 0 & -\sqrt{\frac{3}{5}} & 0 & 0 & 0 & 0 & 0 \\
 0 & 0 & 0 & 0 & 0 & 0 & 0 & 0 & 0 & 0 \\
 0 & 0 & 0 & 0 & 0 & 0 & -\sqrt{\frac{3}{5}} & 0 & 0 & 0 \\
 0 & \frac{2}{3} \sqrt{\frac{2}{5}} & 0 & 0 & 0 & 0 & 0 & 0 & 0 & 0 \\
 0 & 0 & \frac{2}{3} \sqrt{\frac{2}{5}} & 0 & 0 & 0 & 0 & 0 & 0 & 0 \\
 0 & 0 & 0 & 0 & 0 & 0 & 0 & -\frac{7}{3 \sqrt{10}} & 0 & 0 \\
 0 & 0 & 0 & 0 & 0 & 0 & 0 & 0 & -\frac{7}{3 \sqrt{10}} & 0 \\
 0 & 0 & 0 & 0 & 0 & 0 & 0 & 0 & 0 & 0 \\
 0 & 0 & 0 & 0 & -\frac{1}{\sqrt{3}} & 0 & 0 & 0 & 0 & 0 \\
 0 & 0 & 0 & 0 & 0 & \frac{2}{3} & 0 & 0 & 0 & 0 \\
 0 & 0 & 0 & 0 & 0 & 0 & \frac{1}{\sqrt{3}} & 0 & 0 & 0 \\
 0 & 0 & 0 & 0 & 0 & 0 & 0 & 0 & 0 & 0 \\
 \frac{1}{\sqrt{2}} & 0 & 0 & 0 & 0 & 0 & 0 & 0 & 0 & 0 \\
 0 & -\frac{1}{3 \sqrt{2}} & 0 & 0 & 0 & 0 & 0 & 0 & 0 & 0 \\
 0 & 0 & \frac{1}{3 \sqrt{2}} & 0 & 0 & 0 & 0 & 0 & 0 & 0 \\
 0 & 0 & 0 & \frac{1}{\sqrt{2}} & 0 & 0 & 0 & 0 & 0 & 0 \\
 0 & 0 & 0 & 0 & 0 & 0 & 0 & 0 & 0 & 0 \\
 0 & 0 & 0 & 0 & 0 & 0 & 0 & -\frac{2 \sqrt{2}}{3} & 0 & 0 \\
 0 & 0 & 0 & 0 & 0 & 0 & 0 & 0 & \frac{2 \sqrt{2}}{3} & 0 \\
 0 & 0 & 0 & 0 & 0 & 0 & 0 & 0 & 0 & 0 \\
 0 & 0 & 0 & 0 & 0 & 0 & 0 & 0 & 0 & 0 \\
 0 & 0 & 0 & 0 & 0 & 0 & 0 & 0 & 0 & 0 \\
 0 & 0 & 0 & 0 & 0 & 0 & 0 & 0 & 0 & 0 \\
 0 & 0 & 0 & 0 & 0 & 0 & 0 & 0 & 0 & 0 \\
 0 & 0 & 0 & 0 & 0 & 0 & 0 & 0 & 0 & -\sqrt{2} \\
 0 & 0 & 0 & 0 & 0 & 0 & 0 & 0 & 0 & 0
\end{array}
\right) \, ,
\\ \nonumber ({\bf T}^B_{10\times27})^{4\dagger} &=&
\left(
\begin{array}{cccccccccc}
 0 & 0 & 0 & 0 & 0 & 0 & 0 & 0 & \sqrt{\frac{2}{5}} & 0 \\
 \frac{1}{2 \sqrt{5}} & 0 & 0 & 0 & 0 & 0 & 0 & \frac{2}{\sqrt{15}} & 0 & 0 \\
 0 & -\frac{1}{\sqrt{30}} & 0 & 0 & 0 & 0 & 0 & 0 & -\sqrt{\frac{2}{15}} & 0 \\
 0 & 0 & -\frac{1}{2 \sqrt{15}} & 0 & 0 & 0 & 0 & 0 & 0 & 0 \\
 0 & 0 & 0 & 0 & 0 & -\frac{4}{3 \sqrt{5}} & 0 & 0 & 0 & 0 \\
 0 & 0 & 0 & 0 & 0 & 0 & -\frac{4}{3} \sqrt{\frac{2}{5}} & 0 & 0 & 0 \\
 0 & 0 & 0 & 0 & \frac{1}{3}\sqrt{\frac{2}{5}} & 0 & 0 & 0 & 0 & \sqrt{\frac{3}{10}} \\
 0 & 0 & 0 & 0 & 0 & -\frac{1}{3 \sqrt{5}} & 0 & 0 & 0 & 0 \\
 0 & 0 & 0 & 0 & 0 & 0 & 0 & 0 & 0 & 0 \\
 \frac{1}{2} & 0 & 0 & 0 & 0 & 0 & 0 & 0 & 0 & 0 \\
 0 & -\frac{1}{\sqrt{2}} & 0 & 0 & 0 & 0 & 0 & 0 & 0 & 0 \\
 0 & 0 & -\frac{\sqrt{3}}{2} & 0 & 0 & 0 & 0 & 0 & 0 & 0 \\
 0 & 0 & 0 & -1 & 0 & 0 & 0 & 0 & 0 & 0 \\
 0 & 0 & 0 & 0 & -\frac{1}{\sqrt{6}} & 0 & 0 & 0 & 0 & 0 \\
 0 & 0 & 0 & 0 & 0 & \frac{1}{3} & 0 & 0 & 0 & 0 \\
 0 & 0 & 0 & 0 & 0 & 0 & \frac{1}{3 \sqrt{2}} & 0 & 0 & 0 \\
 0 & 0 & 0 & 0 & 0 & 0 & 0 & 0 & 0 & 0 \\
 0 & 0 & 0 & 0 & 0 & 0 & 0 & 0 & 0 & 0 \\
 0 & 0 & 0 & 0 & \frac{\sqrt{2}}{3} & 0 & 0 & 0 & 0 & 0 \\
 0 & 0 & 0 & 0 & 0 & -\frac{2}{3} & 0 & 0 & 0 & 0 \\
 0 & 0 & 0 & 0 & 0 & 0 & -\sqrt{\frac{2}{3}} & 0 & 0 & 0 \\
 0 & \frac{1}{\sqrt{3}} & 0 & 0 & 0 & 0 & 0 & 0 & 0 & 0 \\
 0 & 0 & \sqrt{\frac{2}{3}} & 0 & 0 & 0 & 0 & 0 & 0 & 0 \\
 0 & 0 & 0 & 1 & 0 & 0 & 0 & 0 & 0 & 0 \\
 0 & 0 & 0 & 0 & 0 & 0 & 0 & 0 & 0 & 0 \\
 0 & 0 & 0 & 0 & 0 & 0 & 0 & \frac{1}{\sqrt{6}} & 0 & 0 \\
 0 & 0 & 0 & 0 & 0 & 0 & 0 & 0 & -\frac{1}{\sqrt{3}} & 0
\end{array}
\right) \, ,
\\ \nonumber ({\bf T}^B_{10\times27})^{5\dagger} &=&
\left(
\begin{array}{cccccccccc}
 0 & 0 & 0 & 0 & 0 & 0 & 0 & 0 & i \sqrt{\frac{2}{5}} & 0 \\
 -\frac{i}{2 \sqrt{5}} & 0 & 0 & 0 & 0 & 0 & 0 & \frac{2 i}{\sqrt{15}} & 0 & 0 \\
 0 & \frac{i}{\sqrt{30}} & 0 & 0 & 0 & 0 & 0 & 0 & -i \sqrt{\frac{2}{15}} & 0 \\
 0 & 0 & \frac{i}{2 \sqrt{15}} & 0 & 0 & 0 & 0 & 0 & 0 & 0 \\
 0 & 0 & 0 & 0 & 0 & -\frac{4 i}{3 \sqrt{5}} & 0 & 0 & 0 & 0 \\
 0 & 0 & 0 & 0 & 0 & 0 & -\frac{4}{3} i \sqrt{\frac{2}{5}} & 0 & 0 & 0 \\
 0 & 0 & 0 & 0 & -\frac{1}{3} i \sqrt{\frac{2}{5}} & 0 & 0 & 0 & 0 & i \sqrt{\frac{3}{10}} \\
 0 & 0 & 0 & 0 & 0 & \frac{i}{3 \sqrt{5}} & 0 & 0 & 0 & 0 \\
 0 & 0 & 0 & 0 & 0 & 0 & 0 & 0 & 0 & 0 \\
 -\frac{i}{2} & 0 & 0 & 0 & 0 & 0 & 0 & 0 & 0 & 0 \\
 0 & \frac{i}{\sqrt{2}} & 0 & 0 & 0 & 0 & 0 & 0 & 0 & 0 \\
 0 & 0 & \frac{i \sqrt{3}}{2} & 0 & 0 & 0 & 0 & 0 & 0 & 0 \\
 0 & 0 & 0 & i & 0 & 0 & 0 & 0 & 0 & 0 \\
 0 & 0 & 0 & 0 & -\frac{i}{\sqrt{6}} & 0 & 0 & 0 & 0 & 0 \\
 0 & 0 & 0 & 0 & 0 & \frac{i}{3} & 0 & 0 & 0 & 0 \\
 0 & 0 & 0 & 0 & 0 & 0 & \frac{i}{3 \sqrt{2}} & 0 & 0 & 0 \\
 0 & 0 & 0 & 0 & 0 & 0 & 0 & 0 & 0 & 0 \\
 0 & 0 & 0 & 0 & 0 & 0 & 0 & 0 & 0 & 0 \\
 0 & 0 & 0 & 0 & -\frac{i \sqrt{2}}{3} & 0 & 0 & 0 & 0 & 0 \\
 0 & 0 & 0 & 0 & 0 & \frac{2 i}{3} & 0 & 0 & 0 & 0 \\
 0 & 0 & 0 & 0 & 0 & 0 & i \sqrt{\frac{2}{3}} & 0 & 0 & 0 \\
 0 & \frac{i}{\sqrt{3}} & 0 & 0 & 0 & 0 & 0 & 0 & 0 & 0 \\
 0 & 0 & i \sqrt{\frac{2}{3}} & 0 & 0 & 0 & 0 & 0 & 0 & 0 \\
 0 & 0 & 0 & i & 0 & 0 & 0 & 0 & 0 & 0 \\
 0 & 0 & 0 & 0 & 0 & 0 & 0 & 0 & 0 & 0 \\
 0 & 0 & 0 & 0 & 0 & 0 & 0 & -\frac{i}{\sqrt{6}} & 0 & 0 \\
 0 & 0 & 0 & 0 & 0 & 0 & 0 & 0 & \frac{i}{\sqrt{3}} & 0
\end{array}
\right) \, ,
\\ \nonumber ({\bf T}^B_{10\times27})^{6\dagger} &=&
\left(
\begin{array}{cccccccccc}
 0 & 0 & 0 & 0 & 0 & 0 & 0 & -\sqrt{\frac{2}{5}} & 0 & 0 \\
 0 & \frac{1}{2 \sqrt{15}} & 0 & 0 & 0 & 0 & 0 & 0 & 0 & 0 \\
 0 & 0 & -\frac{1}{\sqrt{30}} & 0 & 0 & 0 & 0 & -\sqrt{\frac{2}{15}} & 0 & 0 \\
 0 & 0 & 0 & -\frac{1}{2 \sqrt{5}} & 0 & 0 & 0 & 0 & -\frac{2}{\sqrt{15}} & 0 \\
 0 & 0 & 0 & 0 & \frac{4}{3} \sqrt{\frac{2}{5}} & 0 & 0 & 0 & 0 & 0 \\
 0 & 0 & 0 & 0 & 0 & \frac{4}{3 \sqrt{5}} & 0 & 0 & 0 & 0 \\
 0 & 0 & 0 & 0 & 0 & \frac{1}{3 \sqrt{5}} & 0 & 0 & 0 & 0 \\
 0 & 0 & 0 & 0 & 0 & 0 & -\frac{1}{3}\sqrt{\frac{2}{5}} & 0 & 0 & -\sqrt{\frac{3}{10}} \\
 1 & 0 & 0 & 0 & 0 & 0 & 0 & 0 & 0 & 0 \\
 0 & -\frac{\sqrt{3}}{2} & 0 & 0 & 0 & 0 & 0 & 0 & 0 & 0 \\
 0 & 0 & \frac{1}{\sqrt{2}} & 0 & 0 & 0 & 0 & 0 & 0 & 0 \\
 0 & 0 & 0 & \frac{1}{2} & 0 & 0 & 0 & 0 & 0 & 0 \\
 0 & 0 & 0 & 0 & 0 & 0 & 0 & 0 & 0 & 0 \\
 0 & 0 & 0 & 0 & 0 & 0 & 0 & 0 & 0 & 0 \\
 0 & 0 & 0 & 0 & \frac{1}{3 \sqrt{2}} & 0 & 0 & 0 & 0 & 0 \\
 0 & 0 & 0 & 0 & 0 & \frac{1}{3} & 0 & 0 & 0 & 0 \\
 0 & 0 & 0 & 0 & 0 & 0 & \frac{1}{\sqrt{6}} & 0 & 0 & 0 \\
 0 & 0 & 0 & 0 & \sqrt{\frac{2}{3}} & 0 & 0 & 0 & 0 & 0 \\
 0 & 0 & 0 & 0 & 0 & -\frac{2}{3} & 0 & 0 & 0 & 0 \\
 0 & 0 & 0 & 0 & 0 & 0 & \frac{\sqrt{2}}{3} & 0 & 0 & 0 \\
 0 & 0 & 0 & 0 & 0 & 0 & 0 & 0 & 0 & 0 \\
 -1 & 0 & 0 & 0 & 0 & 0 & 0 & 0 & 0 & 0 \\
 0 & -\sqrt{\frac{2}{3}} & 0 & 0 & 0 & 0 & 0 & 0 & 0 & 0 \\
 0 & 0 & -\frac{1}{\sqrt{3}} & 0 & 0 & 0 & 0 & 0 & 0 & 0 \\
 0 & 0 & 0 & 0 & 0 & 0 & 0 & \frac{1}{\sqrt{3}} & 0 & 0 \\
 0 & 0 & 0 & 0 & 0 & 0 & 0 & 0 & -\frac{1}{\sqrt{6}} & 0 \\
 0 & 0 & 0 & 0 & 0 & 0 & 0 & 0 & 0 & 0
\end{array}
\right) \, ,
\\ \nonumber ({\bf T}^B_{10\times27})^{7\dagger} &=&
\left(
\begin{array}{cccccccccc}
 0 & 0 & 0 & 0 & 0 & 0 & 0 & -i \sqrt{\frac{2}{5}} & 0 & 0 \\
 0 & -\frac{i}{2 \sqrt{15}} & 0 & 0 & 0 & 0 & 0 & 0 & 0 & 0 \\
 0 & 0 & \frac{i}{\sqrt{30}} & 0 & 0 & 0 & 0 & -i \sqrt{\frac{2}{15}} & 0 & 0 \\
 0 & 0 & 0 & \frac{i}{2 \sqrt{5}} & 0 & 0 & 0 & 0 & -\frac{2 i}{\sqrt{15}} & 0 \\
 0 & 0 & 0 & 0 & \frac{4}{3} i \sqrt{\frac{2}{5}} & 0 & 0 & 0 & 0 & 0 \\
 0 & 0 & 0 & 0 & 0 & \frac{4 i}{3 \sqrt{5}} & 0 & 0 & 0 & 0 \\
 0 & 0 & 0 & 0 & 0 & -\frac{i}{3 \sqrt{5}} & 0 & 0 & 0 & 0 \\
 0 & 0 & 0 & 0 & 0 & 0 & \frac{1}{3} i \sqrt{\frac{2}{5}} & 0 & 0 & -i \sqrt{\frac{3}{10}} \\
 -i & 0 & 0 & 0 & 0 & 0 & 0 & 0 & 0 & 0 \\
 0 & \frac{i \sqrt{3}}{2} & 0 & 0 & 0 & 0 & 0 & 0 & 0 & 0 \\
 0 & 0 & -\frac{i}{\sqrt{2}} & 0 & 0 & 0 & 0 & 0 & 0 & 0 \\
 0 & 0 & 0 & -\frac{i}{2} & 0 & 0 & 0 & 0 & 0 & 0 \\
 0 & 0 & 0 & 0 & 0 & 0 & 0 & 0 & 0 & 0 \\
 0 & 0 & 0 & 0 & 0 & 0 & 0 & 0 & 0 & 0 \\
 0 & 0 & 0 & 0 & \frac{i}{3 \sqrt{2}} & 0 & 0 & 0 & 0 & 0 \\
 0 & 0 & 0 & 0 & 0 & \frac{i}{3} & 0 & 0 & 0 & 0 \\
 0 & 0 & 0 & 0 & 0 & 0 & \frac{i}{\sqrt{6}} & 0 & 0 & 0 \\
 0 & 0 & 0 & 0 & -i \sqrt{\frac{2}{3}} & 0 & 0 & 0 & 0 & 0 \\
 0 & 0 & 0 & 0 & 0 & \frac{2 i}{3} & 0 & 0 & 0 & 0 \\
 0 & 0 & 0 & 0 & 0 & 0 & -\frac{i \sqrt{2}}{3} & 0 & 0 & 0 \\
 0 & 0 & 0 & 0 & 0 & 0 & 0 & 0 & 0 & 0 \\
 -i & 0 & 0 & 0 & 0 & 0 & 0 & 0 & 0 & 0 \\
 0 & -i \sqrt{\frac{2}{3}} & 0 & 0 & 0 & 0 & 0 & 0 & 0 & 0 \\
 0 & 0 & -\frac{i}{\sqrt{3}} & 0 & 0 & 0 & 0 & 0 & 0 & 0 \\
 0 & 0 & 0 & 0 & 0 & 0 & 0 & -\frac{i}{\sqrt{3}} & 0 & 0 \\
 0 & 0 & 0 & 0 & 0 & 0 & 0 & 0 & \frac{i}{\sqrt{6}} & 0 \\
 0 & 0 & 0 & 0 & 0 & 0 & 0 & 0 & 0 & 0
\end{array}
\right) \, ,
\\ \nonumber ({\bf T}^B_{10\times27})^{8\dagger} &=&
\left(
\begin{array}{cccccccccc}
 0 & 0 & 0 & 0 & 0 & 0 & 0 & 0 & 0 & 0 \\
 0 & 0 & 0 & 0 & \frac{2}{\sqrt{5}} & 0 & 0 & 0 & 0 & 0 \\
 0 & 0 & 0 & 0 & 0 & -\frac{2}{\sqrt{5}} & 0 & 0 & 0 & 0 \\
 0 & 0 & 0 & 0 & 0 & 0 & -\frac{2}{\sqrt{5}} & 0 & 0 & 0 \\
 0 & 0 & 0 & 0 & 0 & 0 & 0 & 0 & 0 & 0 \\
 0 & 0 & 0 & 0 & 0 & 0 & 0 & 0 & 0 & 0 \\
 0 & 0 & 0 & 0 & 0 & 0 & 0 & \sqrt{\frac{3}{10}} & 0 & 0 \\
 0 & 0 & 0 & 0 & 0 & 0 & 0 & 0 & -\sqrt{\frac{3}{10}} & 0 \\
 0 & 0 & 0 & 0 & 0 & 0 & 0 & 0 & 0 & 0 \\
 0 & 0 & 0 & 0 & 0 & 0 & 0 & 0 & 0 & 0 \\
 0 & 0 & 0 & 0 & 0 & 0 & 0 & 0 & 0 & 0 \\
 0 & 0 & 0 & 0 & 0 & 0 & 0 & 0 & 0 & 0 \\
 0 & 0 & 0 & 0 & 0 & 0 & 0 & 0 & 0 & 0 \\
 -\sqrt{\frac{3}{2}} & 0 & 0 & 0 & 0 & 0 & 0 & 0 & 0 & 0 \\
 0 & \sqrt{\frac{3}{2}} & 0 & 0 & 0 & 0 & 0 & 0 & 0 & 0 \\
 0 & 0 & \sqrt{\frac{3}{2}} & 0 & 0 & 0 & 0 & 0 & 0 & 0 \\
 0 & 0 & 0 & \sqrt{\frac{3}{2}} & 0 & 0 & 0 & 0 & 0 & 0 \\
 0 & 0 & 0 & 0 & 0 & 0 & 0 & 0 & 0 & 0 \\
 0 & 0 & 0 & 0 & 0 & 0 & 0 & 0 & 0 & 0 \\
 0 & 0 & 0 & 0 & 0 & 0 & 0 & 0 & 0 & 0 \\
 0 & 0 & 0 & 0 & 0 & 0 & 0 & 0 & 0 & 0 \\
 0 & 0 & 0 & 0 & 0 & 0 & 0 & 0 & 0 & 0 \\
 0 & 0 & 0 & 0 & 0 & 0 & 0 & 0 & 0 & 0 \\
 0 & 0 & 0 & 0 & 0 & 0 & 0 & 0 & 0 & 0 \\
 0 & 0 & 0 & 0 & 0 & 0 & 0 & 0 & 0 & 0 \\
 0 & 0 & 0 & 0 & 0 & 0 & 0 & 0 & 0 & 0 \\
 0 & 0 & 0 & 0 & 0 & 0 & 0 & 0 & 0 & 0
\end{array}
\right) \, .
\end{eqnarray}

\section{$(\bar q q)(\bar q q)$ Tetraquark Currents}
\label{app:mesoniccurrents}

In Sec.~\ref{sec:currents} and Appendix.~\ref{app:othercurrents} we have investigate the chiral structure of local scalar and pseudoscalar tetraquark currents constructed using diquarks and antidiquarks. While they can also be constructed using two quark-antiquark pairs. These two different constructions can be related to each other through Firez transformations, and so they can equally describe the full space of local tetraquark currents. In this appendix we show these relations. We note that some of these relations have been obtained in Ref.~\cite{Chen:2006hy,Chen:2007xr,Jiao:2009ra}.

We shall separately investigate scalar and pseudoscalar in the following subsections. Since Firez transformations can only change the Lorentz structure and not change the flavor symmetry and the color symmetry of diquarks and antidiquarks, we shall fix these two symmetries in the following discussions, and separately study tetraquark currents having the antisymmetric flavor structure $\mathbf{\bar 3}_f (qq)\otimes\mathbf{3}_f(\bar q \bar q)$, the symmetric flavor structure $\mathbf{6}_f (qq)\otimes\mathbf{\bar 6}_f(\bar q \bar q)$ and the mixed flavor structure $\mathbf{\bar 3}_f (qq)\otimes\mathbf{\bar 6}_f(\bar q \bar q)$. The other case of mixed flavor structure $\mathbf{6}_f (qq)\otimes \mathbf{3}_f(\bar q \bar q)$ can be similarly investigated.

\subsection{Scalar Tetraquark Currents}

\subsubsection{Flavor Structure $\mathbf{{\bar 3}}_f(qq)\otimes\mathbf{3}_f(\bar q \bar q)$}

In this subsection we study the scalar tetraquark currents which contain diquarks and antidiquarks having both the antisymmetric flavor structure $\mathbf{\bar 3} \otimes \mathbf{3}$. There are altogether five independent scalar tetraquark currents constructed using diquarks and antidiquarks:
\begin{eqnarray}
\nonumber \eta_1 &=& q_A^{aT} \mathbb{C} \gamma_5 q_B^b (\bar{q}_C^a \gamma_5 \mathbb{C} \bar{q}_D^{bT} - \bar{q}_C^b \gamma_5 \mathbb{C} \bar{q}_D^{aT}) \, ,
\\ \nonumber \eta_2 &=& q_A^{aT} \mathbb{C} \gamma_\mu \gamma_5 q_B^b (\bar{q}_C^a \gamma^\mu \gamma_5 \mathbb{C} \bar{q}_D^{bT} - \bar{q}_C^b \gamma^\mu \gamma_5 \mathbb{C} \bar{q}_D^{aT}) \, ,
\\ \nonumber \eta_3 &=& q_A^{aT} \mathbb{C} \sigma_{\mu\nu} q_B^b (\bar{q}_C^a \sigma^{\mu\nu} \mathbb{C} \bar{q}_D^{bT} + \bar{q}_C^b \sigma^{\mu\nu} \mathbb{C} \bar{q}_D^{aT}) \, ,
\\ \nonumber \eta_4 &=& q_A^{aT} \mathbb{C} \gamma_\mu q_B^b (\bar{q}_C^a \gamma^\mu \mathbb{C} \bar{q}_D^{bT} + \bar{q}_C^b \gamma^\mu \mathbb{C} \bar{q}_D^{aT}) \, ,
\\ \nonumber \eta_5 &=& q_A^{aT} \mathbb{C} q_B^b (\bar{q}_C^a \mathbb{C} \bar{q}_D^{bT} - \bar{q}_C^b \mathbb{C} \bar{q}_D^{aT}) \, .
\end{eqnarray}
There are altogether ten scalar tetraquark currents constructed using quark-antiquark pairs:
\begin{eqnarray}
\nonumber \psi_1 &=& (\bar{q}_C^a q_A^a)(\bar q_D^b q_B^b) - (\bar{q}_C^a q_B^a)(\bar{q}_D^b q_A^b) \, ,
\\ \nonumber \psi_2 &=& (\bar{q}_C^a \gamma_\mu q_A^a)(\bar q_D^b \gamma^\mu q_B^b) - (\bar{q}_C^a \gamma_\mu q_B^a)(\bar{q}_D^b \gamma^\mu q_A^b) \, ,
\\ \nonumber \psi_3 &=& (\bar{q}_C^a \sigma_{\mu\nu} q_A^a)(\bar q_D^b \sigma^{\mu\nu} q_B^b) - (\bar{q}_C^a \sigma_{\mu\nu} q_B^a)(\bar{q}_D^b \sigma^{\mu\nu} q_A^b) \, ,
\\ \nonumber \psi_4 &=& (\bar{q}_C^a \gamma_\mu \gamma_5 q_A^a)(\bar q_D^b \gamma^\mu \gamma_5 q_B^b) - (\bar{q}_C^a \gamma_\mu \gamma_5 q_B^a)(\bar{q}_D^b \gamma^\mu \gamma_5 q_A^b) \, ,
\\ \nonumber \psi_5 &=& (\bar{q}_C^a \gamma_5 q_A^a)(\bar q_D^b \gamma_5 q_B^b) - (\bar{q}_C^a \gamma_5 q_B^a)(\bar{q}_D^b \gamma_5 q_A^b) \, ,
\\ \nonumber \psi_6 &=& (\bar{q}_C^a {\lambda_{ab}} q_A^b)(\bar q_D^c {\lambda_{cd}} q_B^d) - (\bar{q}_C^a {\lambda_{ab}} q_B^b)(\bar{q}_D^c {\lambda_{cd}} q_A^d) \, ,
\\ \nonumber \psi_7 &=& (\bar{q}_C^a \gamma_\mu {\lambda_{ab}} q_A^b)(\bar q_D^c \gamma^\mu {\lambda_{cd}} q_B^d) - (\bar{q}_C^a \gamma_\mu {\lambda_{ab}} q_B^b)(\bar{q}_D^c \gamma^\mu {\lambda_{cd}} q_A^d) \, ,
\\ \nonumber \psi_8 &=& (\bar{q}_C^a \sigma_{\mu\nu} {\lambda_{ab}} q_A^b)(\bar q_D^c \sigma^{\mu\nu} {\lambda_{cd}} q_B^d) - (\bar{q}_C^a \sigma_{\mu\nu} {\lambda_{ab}} q_B^b)(\bar{q}_D^c \sigma^{\mu\nu} {\lambda_{cd}} q_A^d) \, ,
\\ \nonumber \psi_9 &=& (\bar{q}_C^a \gamma_\mu{}\gamma_5 {\lambda_{ab}} q_A^b)(\bar q_D^c \gamma^\mu{}\gamma_5 {\lambda_{cd}} q_B^d) - (\bar{q}_C^a \gamma_\mu{}\gamma_5 {\lambda_{ab}} q_B^b)(\bar{q}_D^c \gamma^\mu{}\gamma_5 {\lambda_{cd}} q_A^d) \, ,
\\ \nonumber \psi_{10} &=& (\bar{q}_C^a \gamma_5 {\lambda_{ab}} q_A^b)(\bar q_D^c \gamma_5 {\lambda_{cd}} q_B^d) - (\bar{q}_C^a \gamma_5 {\lambda_{ab}} q_B^b)(\bar{q}_D^c \gamma_5 {\lambda_{cd}} q_A^d) \, .
\end{eqnarray}
Among these ten currents only five are independent. We can verify the following relations:
\begin{eqnarray}
\nonumber \eta_1 &=& -\frac{1}{4} \psi_1 - \frac{1}{4} \psi_2 + \frac{1}{8} \psi_3 - \frac{1}{4} \psi_4 - \frac{1}{4} \psi_5 \, ,
\\ \nonumber \eta_2 &=& \psi_1 - \frac{1}{2} \psi_2 + \frac{1}{2} \psi_4 - \psi_5 \, ,
\\ \nonumber \eta_3 &=& 3 \psi_1 + \frac{1}{2} \psi_3 + 3 \psi_5 \, ,
\\ \nonumber \eta_4 &=& \psi_1 + \frac{1}{2} \psi_2 - \frac{1}{2} \psi_4 - \psi_5 \, ,
\\ \nonumber \eta_5 &=& -\frac{1}{4} \psi_1 + \frac{1}{4} \psi_2 + \frac{1}{8} \psi_3 + \frac{1}{4} \psi_4 - \frac{1}{4} \psi_5 \, ,
\\ \nonumber \psi_6 &=&  -\frac{1}{6} \psi_1 + \frac{1}{2} \psi_2 + \frac{1}{4} \psi_3 - \frac{1}{2} \psi_4 + \frac{1}{2} \psi_5 \, ,
\\ \nonumber \psi_7 &=& 2 \psi_1 - \frac{5}{3} \psi_2 - \psi_4 - 2 \psi_5 \, ,
\\ \nonumber \psi_8 &=& 6 \psi_1 - \frac{5}{3} \psi_3 + 6 \psi_5 \, ,
\\ \nonumber \psi_9 &=& -2 \psi_1 - \psi_2 - \frac{5}{3} \psi_4 + 2 \psi_5 \, ,
\\ \nonumber \psi_{10} &=& \frac{1}{2} \psi_1 - \frac{1}{2} \psi_2 + \frac{1}{4} \psi_3 + \frac{1}{2} \psi_4 - \frac{1}{6} \psi_5 \, .
\end{eqnarray}

\subsubsection{Flavor Structure $\mathbf{6}_f(qq) \otimes \mathbf{\bar 6}_f(\bar q \bar q)$}

In this subsection we study the scalar tetraquark currents which contain diquarks and antidiquarks having both the symmetric flavor structure $\mathbf{6} \otimes \mathbf{\bar 6}$. There are altogether five independent scalar tetraquark currents constructed using diquarks and antidiquarks:
\begin{eqnarray}
\nonumber \eta_1 &=& q_A^{aT} \mathbb{C} \gamma_5 q_B^b (\bar{q}_C^a \gamma_5 \mathbb{C} \bar{q}_D^{bT} + \bar{q}_C^b \gamma_5 \mathbb{C} \bar{q}_D^{aT}) \, ,
\\ \nonumber \eta_2 &=& q_A^{aT} \mathbb{C} \gamma_\mu \gamma_5 q_B^b (\bar{q}_C^a \gamma^\mu \gamma_5 \mathbb{C} \bar{q}_D^{bT} + \bar{q}_C^b \gamma^\mu \gamma_5 \mathbb{C} \bar{q}_D^{aT}) \, ,
\\ \nonumber \eta_3 &=& q_A^{aT} \mathbb{C} \sigma_{\mu\nu} q_B^b (\bar{q}_C^a \sigma^{\mu\nu} \mathbb{C} \bar{q}_D^{bT} - \bar{q}_C^b \sigma^{\mu\nu} \mathbb{C} \bar{q}_D^{aT}) \, ,
\\ \nonumber \eta_4 &=& q_A^{aT} \mathbb{C} \gamma_\mu q_B^b (\bar{q}_C^a \gamma^\mu \mathbb{C} \bar{q}_D^{bT} - \bar{q}_C^b \gamma^\mu \mathbb{C} \bar{q}_D^{aT}) \, ,
\\ \nonumber \eta_5 &=& q_A^{aT} \mathbb{C} q_B^b (\bar{q}_C^a \mathbb{C} \bar{q}_D^{bT} + \bar{q}_C^b \mathbb{C} \bar{q}_D^{aT}) \, .
\end{eqnarray}
There are altogether ten scalar tetraquark currents constructed using quark-antiquark pairs:
\begin{eqnarray}
\nonumber \psi_1 &=& (\bar{q}_C^a q_A^a)(\bar q_D^b q_B^b) + (\bar{q}_C^a q_B^a)(\bar{q}_D^b q_A^b) \, ,
\\ \nonumber \psi_2 &=& (\bar{q}_C^a \gamma_\mu q_A^a)(\bar q_D^b \gamma^\mu q_B^b) + (\bar{q}_C^a \gamma_\mu q_B^a)(\bar{q}_D^b \gamma^\mu q_A^b) \, ,
\\ \nonumber \psi_3 &=& (\bar{q}_C^a \sigma_{\mu\nu} q_A^a)(\bar q_D^b \sigma^{\mu\nu} q_B^b) + (\bar{q}_C^a \sigma_{\mu\nu} q_B^a)(\bar{q}_D^b \sigma^{\mu\nu} q_A^b) \, ,
\\ \nonumber \psi_4 &=& (\bar{q}_C^a \gamma_\mu \gamma_5 q_A^a)(\bar q_D^b \gamma^\mu \gamma_5 q_B^b) + (\bar{q}_C^a \gamma_\mu \gamma_5 q_B^a)(\bar{q}_D^b \gamma^\mu \gamma_5 q_A^b) \, ,
\\ \nonumber \psi_5 &=& (\bar{q}_C^a \gamma_5 q_A^a)(\bar q_D^b \gamma_5 q_B^b) + (\bar{q}_C^a \gamma_5 q_B^a)(\bar{q}_D^b \gamma_5 q_A^b) \, ,
\\ \nonumber \psi_6 &=& (\bar{q}_C^a {\lambda_{ab}} q_A^b)(\bar q_D^c {\lambda_{cd}} q_B^d) + (\bar{q}_C^a {\lambda_{ab}} q_B^b)(\bar{q}_D^c {\lambda_{cd}} q_A^d) \, ,
\\ \nonumber \psi_7 &=& (\bar{q}_C^a \gamma_\mu {\lambda_{ab}} q_A^b)(\bar q_D^c \gamma^\mu {\lambda_{cd}} q_B^d) + (\bar{q}_C^a \gamma_\mu {\lambda_{ab}} q_B^b)(\bar{q}_D^c \gamma^\mu {\lambda_{cd}} q_A^d) \, ,
\\ \nonumber \psi_8 &=& (\bar{q}_C^a \sigma_{\mu\nu} {\lambda_{ab}} q_A^b)(\bar q_D^c \sigma^{\mu\nu} {\lambda_{cd}} q_B^d) + (\bar{q}_C^a \sigma_{\mu\nu} {\lambda_{ab}} q_B^b)(\bar{q}_D^c \sigma^{\mu\nu} {\lambda_{cd}} q_A^d) \, ,
\\ \nonumber \psi_9 &=& (\bar{q}_C^a \gamma_\mu{}\gamma_5 {\lambda_{ab}} q_A^b)(\bar q_D^c \gamma^\mu{}\gamma_5 {\lambda_{cd}} q_B^d) + (\bar{q}_C^a \gamma_\mu{}\gamma_5 {\lambda_{ab}} q_B^b)(\bar{q}_D^c \gamma^\mu{}\gamma_5 {\lambda_{cd}} q_A^d) \, ,
\\ \nonumber \psi_{10} &=& (\bar{q}_C^a \gamma_5 {\lambda_{ab}} q_A^b)(\bar q_D^c \gamma_5 {\lambda_{cd}} q_B^d) + (\bar{q}_C^a \gamma_5 {\lambda_{ab}} q_B^b)(\bar{q}_D^c \gamma_5 {\lambda_{cd}} q_A^d) \, .
\end{eqnarray}
Among these ten currents only five are independent. We can verify the following relations:
\begin{eqnarray}
\nonumber \eta_1 &=& -\frac{1}{4} \psi_1 - \frac{1}{4} \psi_2 + \frac{1}{8} \psi_3 - \frac{1}{4} \psi_4 - \frac{1}{4} \psi_5 \, ,
\\ \nonumber \eta_2 &=& \psi_1 - \frac{1}{2} \psi_2 + \frac{1}{2} \psi_4 - \psi_5 \, ,
\\ \nonumber \eta_3 &=& 3 \psi_1 + \frac{1}{2} \psi_3 + 3 \psi_5 \, ,
\\ \nonumber \eta_4 &=& \psi_1 + \frac{1}{2} \psi_2 - \frac{1}{2} \psi_4 - \psi_5 \, ,
\\ \nonumber \eta_5 &=& -\frac{1}{4} \psi_1 + \frac{1}{4} \psi_2 + \frac{1}{8} \psi_3 + \frac{1}{4} \psi_4 - \frac{1}{4} \psi_5 \, ,
\\ \nonumber \psi_6 &=&  -\frac{7}{6} \psi_1 - \frac{1}{2} \psi_2 - \frac{1}{4} \psi_3 + \frac{1}{2} \psi_4 - \frac{1}{2} \psi_5 \, ,
\\ \nonumber \psi_7 &=& - 2 \psi_1 + \frac{1}{3} \psi_2 + \psi_4 + 2 \psi_5 \, ,
\\ \nonumber \psi_8 &=& - 6 \psi_1 + \frac{1}{3} \psi_3 - 6 \psi_5 \, ,
\\ \nonumber \psi_9 &=&  2 \psi_1 + \psi_2 + \frac{1}{3} \psi_4 - 2 \psi_5 \, ,
\\ \nonumber \psi_{10} &=& - \frac{1}{2} \psi_1 + \frac{1}{2} \psi_2 - \frac{1}{4} \psi_3 - \frac{1}{2} \psi_4 - \frac{7}{6} \psi_5 \, .
\end{eqnarray}

\subsection{Pseudoscalar Tetraquark Currents}

\subsubsection{Flavor Structure $\mathbf{{\bar 3}}_f(qq)\otimes\mathbf{3}_f(\bar q \bar q)$}

In this subsection we study pseudoscalar tetraquark currents which contain diquarks and antidiquarks having both the antisymmetric flavor structure $\mathbf{\bar 3} \otimes \mathbf{3}$. There are altogether three independent pseudoscalar tetraquark currents constructed using diquarks and antidiquarks:
\begin{eqnarray}
\nonumber \eta_1 &=& q_A^{aT} \mathbb{C}
q_B^b (\bar{q}_C^a \gamma_5 \mathbb{C} \bar{q}_D^{bT} - \bar{q}_C^b \gamma_5 C \bar{q}_D^{aT}) \, ,
\\ \nonumber \eta_2 &=& q_A^{aT} C \gamma_5 q_B^b (\bar{q}_C^a \mathbb{C}
\bar{q}_D^{bT} - \bar{q}_C^b \mathbb{C} \bar{q}_D^{aT}) \, ,
\\ \nonumber \eta_3 &=& q_A^{aT} \mathbb{C} \sigma_{\mu\nu} q_B^b (\bar{q}_C^a \sigma^{\mu\nu} \gamma_5 \mathbb{C} \bar{q}_D^{bT} + \bar{q}_C^b \sigma^{\mu\nu} \gamma_5 \mathbb{C} \bar{q}_D^{aT}) \, .
\end{eqnarray}
There are altogether six pseudoscalar tetraquark currents constructed using quark-antiquark pairs:
\begin{eqnarray}
\nonumber \psi_1 &=& (\bar{q}_C^a q_A^a)(\bar q_D^b \gamma_5 q_B^b) + (\bar{q}_C^a \gamma_5 q_A^a)(\bar q_D^b q_B^b)
- (\bar{q}_C^a q_B^a)(\bar q_D^b \gamma_5 q_A^b) - (\bar{q}_C^a \gamma_5 q_B^a)(\bar q_D^b q_A^b) \, ,
\\ \nonumber \psi_2 &=& (\bar{q}_C^a \gamma_\mu q_A^a)(\bar q_D^b \gamma^\mu\gamma_5 q_B^b) + (\bar{q}_C^a \gamma_\mu\gamma_5 q_A^a)(\bar q_D^b \gamma^\mu q_B^b)
- (\bar{q}_C^a \gamma_\mu q_B^a)(\bar q_D^b \gamma^\mu\gamma_5 q_A^b) - (\bar{q}_C^a \gamma_\mu\gamma_5 q_B^a)(\bar q_D^b \gamma^\mu q_A^b) \, ,
\\ \nonumber \psi_3 &=& (\bar{q}_C^a \sigma_{\mu\nu} q_A^a)(\bar q_D^b \sigma^{\mu\nu} \gamma_5 q_B^b) - (\bar{q}_C^a \sigma_{\mu\nu} q_B^a)(\bar q_D^b \sigma^{\mu\nu} \gamma_5 q_A^b) \, ,
\\ \nonumber \psi_4 &=& {\lambda_{ab}}{\lambda_{cd}}\{(\bar{q}_C^a q_A^b)(\bar q_D^c \gamma_5 q_B^d) + (\bar{q}_C^a \gamma_5 q_A^b)(\bar q_D^c q_B^d)
- (\bar{q}_C^a q_B^b)(\bar q_D^c \gamma_5 q_A^d) - (\bar{q}_C^a \gamma_5 q_B^b)(\bar q_D^c q_A^d)\} \, ,
\\ \nonumber \psi_5 &=& {\lambda_{ab}}{\lambda_{cd}}\{(\bar{q}_C^a \gamma_\mu q_A^b)(\bar q_D^c \gamma^\mu\gamma_5 q_B^d) + (\bar{q}_C^a \gamma_\mu\gamma_5 q_A^b)(\bar q_D^c \gamma^\mu q_B^d)
- (\bar{q}_C^a \gamma_\mu q_B^b)(\bar q_D^c \gamma^\mu\gamma_5 q_A^d) - (\bar{q}_C^a \gamma_\mu\gamma_5 q_B^b)(\bar q_D^c \gamma^\mu q_A^d)\} \, ,
\\ \nonumber \psi_6 &=& {\lambda_{ab}}{\lambda_{cd}}\{ (\bar{q}_C^a \sigma_{\mu\nu} q_A^b)(\bar{q}_D^c \sigma^{\mu\nu} \gamma_5 q_B^d) - (\bar{q}_C^a \sigma_{\mu\nu} q_B^b)(\bar q_D^c \sigma^{\mu\nu} \gamma_5 q_A^d)\} \, .
\end{eqnarray}
Among these six currents only three are independent. We can verify the following relations:
\begin{eqnarray}
\nonumber \eta_1 &=&  -\frac{1}{4} \psi_1 - \frac{1}{4} \psi_2 + \frac{1}{8} \psi_3 \, ,
\\ \nonumber \eta_2 &=& -\frac{1}{4} \psi_1 + \frac{1}{4} \psi_2 + \frac{1}{8} \psi_3 \, ,
\\ \nonumber \eta_3 &=& 3 \psi_1  + \frac{1}{2} \psi_3 \, ,
\\ \nonumber \psi_4 &=& \frac{1}{3} \psi_1 + \frac{1}{2} \psi_3 \, ,
\\ \nonumber \psi_5 &=& - \frac{8}{3} \psi_2 \, ,
\\ \nonumber \psi_6 &=& 6 \psi_1 - \frac{5}{3} \psi_3 \, .
\end{eqnarray}

\subsubsection{Flavor Structure $\mathbf{6}_f(qq) \otimes \mathbf{\bar 6}_f(\bar q \bar q)$}

In this subsection we study pseudoscalar tetraquark currents which contain diquarks and antidiquarks having both the symmetric flavor structure $\mathbf{6} \otimes \mathbf{\bar 6}$. There are altogether three independent pseudoscalar tetraquark currents constructed using diquarks and antidiquarks:
\begin{eqnarray}
\nonumber \eta_1 &=& q_A^{aT} \mathbb{C} q_B^b (\bar{q}_C^a \gamma_5 \mathbb{C} \bar{q}_D^{bT} + \bar{q}_C^b \gamma_5 \mathbb{C} \bar{q}_D^{aT}) \, ,
\\ \nonumber \eta_2 &=& q_A^{aT} \mathbb{C} \gamma_5 q_B^b (\bar{q}_C^a \mathbb{C} \bar{q}_D^{bT} + \bar{q}_C^b \mathbb{C} \bar{q}_D^{aT}) \, ,
\\ \nonumber \eta_3 &=& q_A^{aT} \mathbb{C} \sigma_{\mu\nu} q_B^b (\bar{q}_C^a \sigma^{\mu\nu} \gamma_5 \mathbb{C} \bar{q}_D^{bT} - \bar{q}_C^b \sigma^{\mu\nu} \gamma_5 \mathbb{C} \bar{q}_D^{aT}) \, .
\end{eqnarray}
There are altogether six pseudoscalar currents constructed using quark-antiquark pairs:
\begin{eqnarray}
\nonumber \psi_1 &=& (\bar{q}_C^a q_A^a)(\bar q_D^b \gamma_5 q_B^b) + (\bar{q}_C^a \gamma_5 q_A^a)(\bar q_D^b q_B^b)
+ (\bar{q}_C^a q_B^a)(\bar q_D^b \gamma_5 q_A^b) + (\bar{q}_C^a \gamma_5 q_B^a)(\bar q_D^b q_A^b) \, ,
\\ \nonumber \psi_2 &=& (\bar{q}_C^a \gamma_\mu q_A^a)(\bar q_D^b \gamma^\mu\gamma_5 q_B^b) + (\bar{q}_C^a \gamma_\mu\gamma_5 q_A^a)(\bar q_D^b \gamma^\mu q_B^b)
+ (\bar{q}_C^a \gamma_\mu q_B^a)(\bar q_D^b \gamma^\mu\gamma_5 q_A^b) + (\bar{q}_C^a \gamma_\mu\gamma_5 q_B^a)(\bar q_D^b \gamma^\mu q_A^b) \, ,
\\ \nonumber \psi_3 &=& (\bar{q}_C^a \sigma_{\mu\nu} q_A^a)(\bar q_D^b \sigma^{\mu\nu} \gamma_5 q_B^b) + (\bar{q}_C^a \sigma_{\mu\nu} q_B^a)(\bar q_D^b \sigma^{\mu\nu} \gamma_5 q_A^b) \, ,
\\ \nonumber \psi_4 &=& {\lambda_{ab}}{\lambda_{cd}}\{(\bar{q}_C^a q_A^b)(\bar q_D^c \gamma_5 q_B^d) + (\bar{q}_C^a \gamma_5 q_A^b)(\bar q_D^c q_B^d)
+ (\bar{q}_C^a q_B^b)(\bar q_D^c \gamma_5 q_A^d) + (\bar{q}_C^a \gamma_5 q_B^b)(\bar q_D^c q_A^d)\} \, ,
\\ \nonumber \psi_5 &=& {\lambda_{ab}}{\lambda_{cd}}\{(\bar{q}_C^a \gamma_\mu q_A^b)(\bar q_D^c \gamma^\mu\gamma_5 q_B^d) + (\bar{q}_C^a \gamma_\mu\gamma_5 q_A^b)(\bar q_D^c \gamma^\mu q_B^d)
+ (\bar{q}_C^a \gamma_\mu q_B^b)(\bar q_D^c \gamma^\mu\gamma_5 q_A^d) + (\bar{q}_C^a \gamma_\mu\gamma_5 q_B^b)(\bar q_D^c \gamma^\mu q_A^d)\} \, ,
\\ \nonumber \psi_6 &=& {\lambda_{ab}}{\lambda_{cd}}\{ (\bar{q}_C^a \sigma_{\mu\nu} q_A^b)(\bar{q}_D^c \sigma^{\mu\nu} \gamma_5 q_B^d) + (\bar{q}_C^a \sigma_{\mu\nu} q_B^b)(\bar q_D^c \sigma^{\mu\nu} \gamma_5 q_A^d)\} \, .
\end{eqnarray}
Among these six currents only three are independent. We can verify the following relations:
\begin{eqnarray}
\nonumber \eta_1 &=&  -\frac{1}{4} \psi_1 - \frac{1}{4} \psi_2 + \frac{1}{8} \psi_3 \, ,
\\ \nonumber \eta_2 &=& -\frac{1}{4} \psi_1 + \frac{1}{4} \psi_2 + \frac{1}{8} \psi_3 \, ,
\\ \nonumber \eta_3 &=& 3 \psi_1 + \frac{1}{2} \psi_3 \, ,
\\ \nonumber \psi_4 &=&  -\frac{5}{3} \psi_1 - \frac{1}{2} \psi_3 \, ,
\\ \nonumber \psi_5 &=&  \frac{4}{3} \psi_2 \, ,
\\ \nonumber \psi_6 &=& - 6 \psi_1  + \frac{1}{3} \psi_3 \, .
\end{eqnarray}

\subsubsection{Flavor Structure $\mathbf{\bar 3}_f(qq)\otimes\mathbf{\bar 6}_f(\bar q \bar q)$}

In this subsection we study pseudoscalar tetraquark currents which contain diquarks having the antisymmetric flavor structure and antidiquarks the symmetric flavor structure $\mathbf{\bar3} \otimes \mathbf{\bar6}$. There are altogether two independent pseudoscalar tetraquark currents constructed using diquarks and antidiquarks:
\begin{eqnarray}
\nonumber \eta_1 &=& q_A^{aT} \mathbb{C} \gamma_\mu q_B^b (\bar{q}_C^a \gamma^\mu \gamma_5 \mathbb{C} \bar{q}_D^{bT} + \bar{q}_C^b \gamma^\mu \gamma_5 \mathbb{C} \bar{q}_D^{aT}) \, ,
\\ \nonumber \eta_2 &=& q_A^{aT} \mathbb{C} \gamma_\mu \gamma_5 q_B^b (\bar{q}_C^a \gamma^\mu \mathbb{C} \bar{q}_D^{bT} - \bar{q}_C^b \gamma^\mu \mathbb{C} \bar{q}_D^{aT}) \, .
\end{eqnarray}
There are altogether four pseudoscalar tetraquark currents constructed using quark-antiquark pairs:
\begin{eqnarray}
\nonumber \psi_1 &=& (\bar{q}_C^a q_A^a)(\bar q_D^b \gamma_5 q_B^b) - (\bar{q}_C^a \gamma_5 q_A^a)(\bar q_D^b q_B^b) - (\bar{q}_C^a q_B^a)(\bar q_D^b \gamma_5 q_A^b) + (\bar{q}_C^a \gamma_5 q_B^a)(\bar q_D^b q_A^b) \, ,
\\ \nonumber \psi_2 &=& (\bar{q}_C^a \gamma_\mu q_A^a)(\bar q_D^b \gamma^\mu\gamma_5 q_B^b) - (\bar{q}_C^a \gamma_\mu\gamma_5 q_A^a)(\bar q_D^b \gamma^\mu q_B^b)
- (\bar{q}_C^a \gamma_\mu q_B^a)(\bar q_D^b \gamma^\mu\gamma_5 q_A^b) + (\bar{q}_C^a \gamma_\mu\gamma_5 q_B^a)(\bar q_D^b \gamma^\mu q_A^b)  \, ,
\\ \nonumber \psi_3 &=& {\lambda_{ab}}{\lambda_{cd}}\{(\bar{q}_C^a q_A^b)(\bar q_D^c \gamma_5 q_B^d) - (\bar{q}_C^a \gamma_5 q_A^b)(\bar q_D^c q_B^d) - (\bar{q}_C^a q_B^b)(\bar q_D^c \gamma_5 q_A^d) + (\bar{q}_C^a \gamma_5 q_B^b)(\bar q_D^c q_A^d)\} \, ,
\\ \nonumber \psi_4 &=& {\lambda_{ab}}{\lambda_{cd}}\{(\bar{q}_C^a \gamma_\mu q_A^b)(\bar q_D^c \gamma^\mu\gamma_5 q_B^d) - (\bar{q}_C^a \gamma_\mu\gamma_5 q_A^b)(\bar q_D^c \gamma^\mu q_B^d)
- (\bar{q}_C^a \gamma_\mu q_B^b)(\bar q_D^c \gamma^\mu\gamma_5 q_A^d) + (\bar{q}_C^a \gamma_\mu\gamma_5 q_B^b)(\bar q_D^c \gamma^\mu q_A^d)\} \, .
\end{eqnarray}
Among these four currents only two are independent. We can verify the following relations:
\begin{eqnarray}
\nonumber \eta_1 &=& \psi_1 - \frac{1}{2} \psi_2 \, ,
\\ \nonumber \eta_2 &=& \psi_1 + \frac{1}{2} \psi_2 \, ,
\\ \nonumber \psi_3 &=&  - \frac{2}{3} \psi_1 - \psi_2 \, ,
\\ \nonumber \psi_4 &=& -4 \psi_1 - \frac{2}{3} \psi_2 \, .
\end{eqnarray}

\bibliographystyle{model1a-num-names}

\end{document}